\documentclass[twocolumn]{aastex631}

\graphicspath{{/Users/ahamanowicz/Desktop/METAL-Z/METAL-Z-I/paper-figures/}}

\shorttitle{METAL-Z}
\shortauthors{Hamanowicz et al.}
\graphicspath{{./}{figures/}}
\usepackage{graphicx}	
\usepackage{amsmath}	
\usepackage{amssymb}	
\usepackage{longtable}
\usepackage{rotating}

\begin{document}

\title{METAL-Z: Measuring dust depletion in low metalicity dwarf galaxies}

\correspondingauthor{Aleksandra Hamanowicz}
\email{ahamanowicz@stsci.edu}

\author[0000-0002-4646-7509]{Aleksandra Hamanowicz}
\affiliation{Space Telescope Science Institute, 3700 San Martin Drive, Baltimore, MD 21218, USA}

\author[0000-0003-0789-9939]{Kirill Tchernyshyov}
\affiliation{Department of Astronomy, Box 351580, University of Washington, Seattle, WA 98195, USA}

\author[0000-0001-6326-7069]{Julia Roman-Duval}
\affiliation{Space Telescope Science Institute, 3700 San Martin Drive, Baltimore, MD 21218, USA}

\author[0000-0003-1892-4423]{Edward B. Jenkins}
\affiliation{Department of Astrophysical Sciences, Princeton University, Princeton, NJ 08544-1001, USA}

\author[0000-0002-9946-4731]{Marc Rafelski}
\affiliation{Space Telescope Science Institute, 3700 San Martin Drive, Baltimore, MD 21218, USA}

\author[0000-0001-5340-6774]{Karl D.\ Gordon}
\affiliation{Space Telescope Science Institute, 3700 San Martin Drive, Baltimore, MD 21218, USA}

\author[0000-0003-4158-5116]{Yong Zheng}
\affiliation{Department of Physics, Applied Physics, and Astronomy, Rensselaer Polytechnic Institute, 110 8th St, Troy, NY 12180}

\author[0000-0003-0316-1208]{Miriam Garcia}
\affiliation{Centro de Astrobiologia, CSIC-INTA, Dpto. de Astrofisica, Instituto Nacional de Tecnica Aeroespacial, Ctra. de Torrejon a Ajalvir, km 4, 28850 Torrejon de Ardoz (Madrid), Spain}

\author[0000-0002-0355-0134]{Jessica Werk}
\affiliation{Department of Astronomy, Box 351580, University of Washington, Seattle, WA 98195, USA}

\begin{abstract}

The cycling of metals between interstellar gas and dust is a critical aspect of the baryon cycle of galaxies, yet our understanding of this process is limited. This study focuses on understanding dust depletion effects in the low metallicity regime ($< 20\% Z_{\odot}$) typical of cosmic noon. Using medium-resolution UV spectroscopy from the COS onboard the {\it Hubble Space Telescope}, gas-phase abundances and depletions of iron and sulfur were derived toward 18 sightlines in local dwarf galaxies IC 1613 and Sextans A. The results show that the depletion of Fe and S is consistent with that found in the Milky Way, LMC and SMC. The depletion level of Fe increases with gas column density, indicating dust growth in the interstellar medium (ISM). The level of Fe depletion decreases with decreasing metallicity, resulting in the fraction of iron in gas ranging from 3\% in the MW to 9\% in IC 1613 and $\sim$19\% in Sextans A. The dust-to-gas and dust-to-metal ratios (D/G, D/M) for these dwarf galaxies were estimated based on the MW relations between the depletion of Fe and other elements. The study finds that D/G decreases only slightly sub-linearly with metallicity, with D/M decreasing from 0.41 $\pm$ 0.05 in the MW to 0.11 $\pm$ 0.11 at 0.10 $Z_{\odot}$ (at $\log$ N(H) = 21 cm$^{-2}$). The trend of D/G vs. metallicity using depletion in local systems is similar to that inferred in Damped Ly-$\alpha$ systems from abundance ratios but lies higher than the trend inferred from FIR measurements in nearby galaxies.

\end{abstract}



\section{Introduction} \label{sec:intro}

\indent Despite amounting to only a small fraction of the Interstellar Medium (ISM) mass, metals play a significant role in the evolution of galaxies. Formed in stellar interiors and explosions, metals gradually enrich the ISM and influence galaxies' properties and evolution through their effects on heating and cooling, radiative transfer, and chemistry. In particular, interstellar dust, the main constituents of which are carbon, oxygen, silicon, magnesium, and iron, absorbs stellar radiation in the optical and UV and re-emits it in the FIR, affecting galaxies' Spectral Energy Distribution (SED). The opacity of dust versus wavelength, in turn, depends on the composition and size of dust grains \citep[e.g.,][]{gordon2003, demyk2017a, demyk2017b, ysard2018}. As a result, we need to understand the dust abundance and properties to correctly ``de-redden" galaxy SEDs, infer their star formation histories and stellar populations, as well as to convert FIR emission into dust and gas masses, a common and very effective way to trace the ISM at all redshifts \citep{hildebrand1983, bolatto2011, eales2012, schruba2012, rowlands2012, rowlands2014}.
  
\indent The abundance of dust and the fraction of metals locked onto dust grains is described by the dust-to-gas (D/G) and dust-to-metal (D/M) ratios (D/G = D/M  $\times$ Z, where Z is the metallicity of the system). These parameters are expected to vary with environment, especially with metallicity and density \citep{zhukovska2016}. At low metallicity, fewer metals and dust grains are available, which means their collisions are rarer. As a result, the timescale for ISM dust growth is expected to be inversely proportional to metallicity, and similarly, to gas density \citep{asano2013, feldmann2015, zhukovska2016}.

\indent The dust-to-gas ratio can be measured through two methods. The most common and efficient approach is to measure the dust mass (or surface density for resolved observations) of galaxies through their FIR dust emission, and their gas mass (or surface density) through 21 cm to trace the atomic gas, and CO rotational emission to estimate the molecular gas content \citep{remyruyer2014, devis2019}. However, the conversion of FIR flux to dust mass relies on modeling. While the dust temperature can be estimated through multiband FIR photometry, the approach still relies a purely theoretical estimate of the FIR opacity, which is known to vary with dust composition, shape, and size, and is observationally very poorly constrained \citep{stepnik2003, demyk2017, clark2019}.  In turn, gas mass estimates suffer from uncertainties due to CO-dark gas and optically thick HI \citep{romanduval2014}. 

\indent The other method to estimate D/G and D/G in galaxies relies on UV spectroscopy of interstellar absorption lines to measure gas-phase abundances and depletions. The depletion of a given element X (e.g., Fe, Si, Mg, etc.) corresponds to the logarithm of the fraction of that element in the gas phase, and is expressed as:

\begin{equation}
\label{eq:dep}
\delta(X) = \rm log_{10}(X/H)_{\rm gas} - \rm log_{10}(X/H)_{\rm total} ,
\end{equation}

where $\rm \log_{10}(X/H)_{gas}$ is the abundance of element X in the gas phase and $\rm \log_{10}(X/H)_{total}$ is the total abundance of element X (in gas and dust). In the Milky Way (MW) and nearby galaxies, gas phase abundances are measured through interstellar UV absorption lines towards UV-bright O and B stars. Total elemental abundances in the ISM are assumed to equal the photospheric abundances of young stars that recently formed out of interstellar gas and dust \citep{jenkins2009}. By summing the depletions of all major constituents of dust, one can estimate D/G and D/M \citep[e.g.,][]{romanduval2022a}.

\indent Understanding how depletions vary with environment (e.g., metallicity, gas density) is key to measuring the chemical enrichment of the Universe through Damped Lyman-$\alpha$ systems \citep[DLAs, e.g.][]{rafelski202, decia2016}. DLAs are quasar absorbers with high hydrogen column density ($\log$ N(\ion{H}{1}) $>$ 20.3 cm$^{-2}$), and their detectability depends solely on the brightness of the background object. This makes DLAs a powerful tool for tracing the chemical enrichment of the Universe over the past nine billion years ($z < 5.3$). However, to infer the total metallicities from gas-phase metallicities in DLAs, it is necessary to correct the measurements for depletion effects, particularly for refractory elements such as Fe. Given that stellar abundances cannot be used as a proxy for total ISM abundances in those systems, depletion corrections are estimated from abundance ratios such as [S/Fe] or [Zn/Fe] tied to the calibration of Zn depletion versus [Zn/Fe] in the Milky Way \citep{decia2016}. Such depletion corrections have only been compared to and calibrated against depletion measurements based on the comparison of gas-phase and stellar abundances in the MW, LMC and SMC, down to 20\% solar metallicity \citep{romanduval2022b}, but depletion measurements in nearby galaxies of lower metallicities are still lacking. 

\indent Indeed, while we have a good understanding of the depletion rate of gas-phase metals in the Milky Way \citep{jenkins2009}, depletion measurements remain sparse in low metallicity environments ($Z <$ 20\% solar) because the observations, which require deep medium-high resolution UV spectroscopy, are challenging. Relatively large samples of depletion measurements for some key constituents of dust (Fe, Mg, Si, Ni, Cr) and volatile elements (S, Zn) have recently been obtained in the LMC at 50\% solar metallicity \citep{tchernyshyov2015, romanduval2021} and SMC at 20\% solar metallicity \citep{tchernyshyov2015, jenkins2017}. These studies have shown that the metals are more depleted as the gas density and metallicity increases, resulting in variations of the dust-to-metal (D/M) and dust-to-gas (D/G) ratios with both gas density and metallicity \citep{romanduval2022a}. However, there have been no depletion measurements of this kind (i.e., toward individual stars in nearby galaxies) at metallicities lower than 20\% solar to date. 

\indent Yet, depletion measurements from UV spectroscopy can provide estimates of the  D/G, D/M, and dust composition, and offer a valuable tool for investigating the complex relationship between metallicity, gas density, and dust content in the ISM. In particular, a tension between FIR-based and depletion-based estimates of D/G and D/M has arisen in recent years. For galaxies with metallicities at or less than 20\% solar, D/G measured through FIR emission is much lower than D/G derived from rest-frame UV spectroscopy in DLAs \citep{galliano2018, romanduval2022a, romanduval2022b, popping2022}. To understand dust properties and abundance below 20\% solar metallicity and resolve this tension, we need to measure gas-phase abundances, elemental depletions, and subsequently, D/G and D/M in nearby low metallicity systems where stellar and gas-phase abundances can be measured and compared.

\indent In this work, we present the METAL-Z (``Metal Evolution, Transport, and Abundance at Low Metallicity (Z)") large program with the {\it Hubble Space Telescope} (77 orbits, GO-15880), designed to measure the gas-phase abundance and depletions of iron and sulfur (Fe and S), and infer D/G as a function of metallicity down to 0.1 $Z_{\odot}$. The program comprises medium-resolution UV spectra obtained with the {\it Cosmic Origins Spectrograph} (COS) onboard the {\it Hubble Space Telescope} (HST) toward 14 massive O and early B stars in two local galaxies with metallicity $<$ 20\% solar: IC 1613 (10\% solar metallicity in oxygen, 20\% solar metallicity in iron, distance 730 kpc ) and Sextans A (10\% solar metallicty, distance 1.3 Mpc). 

\indent This paper is organized as follows. In Section \ref{sec:survey}, we present the survey design and data reduction. Section \ref{sec:colden} describes the derivation of column densities of hydrogen, iron, and sulfur. Section \ref{sec:depl} examines the variations of the S and Fe depletions with gas density and metallicity. In Section \ref{sec:dg}, we present the derivation of D/G based on the depletion measurements in IC 1613 and Sextans A, combined with prior knowledge of depletion patterns obtained in the Milky Way and Magellanic Clouds. Section \ref{sec:concl} provides a summary of this work.

\begin{deluxetable*}{llccccc}
\tablecaption{Spectroscopic targets and their stellar parameters.  \label{tab:star}}

\tablehead{\colhead{Target} &\colhead{SIMBAD resolved name } & \colhead{RA} & \colhead{Dec} & \colhead{SpT} & \colhead{V} & \colhead{E(B-V)\tablenotemark{b}}\\
& & \colhead{h} &\colhead{deg}  & & \colhead{mag} & \colhead{mag}  }
\startdata
IC1613-61331 & [GHV2009] Star 61331 & 01:05:00.200 & +02:09:13.10 & O9.7 II & 19.14 & 0.05 \\ 
IC1613-62024 & [GHV2009] Star 62024 & 01:05:00.646 & +02:08:49.26 & O6.5 IIIf & 19.60 & 0.11 \\ 
IC1613-64066 & [GHV2009] Star 64066 & 01:05:20.700 & +02:09:28.10 & O3 III((f)) & 19.03 & 0.07 \\ 
IC1613-67559 & [GHV2009] Star 67559 & 01:05:04.767 & +02:09:23.19 & O8.5 III((f)) & 19.24 & 0.07 \\ 
IC1613-67684 & [GHV2009] Star 67684 & 01:05:04.900 & +02:09:32.60 & O8.5 I & 19.02 & 0.05 \\ 
IC1613-A13\tablenotemark{a} & [BUG2007] A 13 & 01:05:06.250 & +02:10:43.00 & O3-4 V((f)) & 18.96 & 0.05 \\ 
IC1613-B11\tablenotemark{a} & [BUG2007] B 11 & 01:04:43.800 & +02:06:44.75 & O9.5 I & 18.62 & 0.13 \\ 
IC1613-B2 & [BUG2007] B 2 & 01:05:03.068 & +02:10:04.54 & O7.5 III-V((f)) & 19.62 & 0.07 \\ 
IC1613-B3 & [BUG2007] B 3 & 01:05:06.370 & +02:09:31.34 & B0 Ia & 17.69 & 0.10 \\ 
IC1613-B7 & [BUG2007] B 7 & 01:05:01.970 & +02:08:05.10 & O9 II & 18.96 & 0.05 \\ 
SEXTANS-A-s050 & LGGS J101100.66-044044.3 & 10:11:00.660 & -04:40:44.30 & O9.7 I & 19.61 & 0.01 \\ 
SEXTANS-A-s014 & LGGS J101053.81-044113.0 & 10:10:53.800 & -04:41:13.00 & O7.5 III((f)) & 20.69 & 0.01 \\ 
SEXTANS-A-s022 & [VPW98] 451 & 10:11:05.380 & -04:42:40.10 & O8 V & 19.46 & 0.01 \\ 
SEXTANS-A-s038 & LGGS J101106.05-044211.4 & 10:11:06.047 & -04:42:11.37 & O9.7 I((f)) & 19.49 & 0.03 \\ 
SEXTANS-A-s029\tablenotemark{a} & [VPW98] 1744 & 10:10:58.190 & -04:43:18.40 & O8.5 III & 20.80 & 0.02 \\ 
SEXTANS-A-s037\tablenotemark{a} & LGGS J101104.78-044224.1 & 10:11:04.770 & -04:42:24.24 & O9 I & 20.68 & 0.03 \\ 
SEXTANS-A-SA2\tablenotemark{a} & LGGS J101056.86-044040.8 & 10:10:56.845 & -04:40:40.90 & O & 20.41 & 0.00 \\ 
SEXTANS-A-s021\tablenotemark{a} & LGGS J101104.79-044220.9 & 10:11:04.790 & -04:42:20.96 & O8 V & 20.60 & -0.20 \\ 
\enddata
\tablecomments{Names of stars in Sextans A are from \citet{Lorenzo2022}, except for Sextans A SA2, which is not included in that catalog.}
\tablenotetext{a}{Sightlines with archival data.}
\tablenotetext{b}{E(B-V) reddening  from the ULLYSES program.} 
\tablerefs{The SpT (spectral type) and V magnitudes of stars come from: \citet{bresolin2007,garcia2009, garcia-herrero2013, garcia2014} - IC1613, \citet{camacho2016,Lorenzo2022} - Sextans A.}
\end{deluxetable*}

\section{Survey Design}
\label{sec:survey}

\indent Measuring interstellar gas-phase abundances requires medium-resolution UV spectroscopy of bright background massive stars of type B0 or earlier (stars later than B0 have complex stellar continua with many weak lines unsuitable for ISM abundance determinations). Keeping the exposure time reasonable restricts the target pool to low-metallicity galaxies within $\sim$1.5 Mpc. The METAL-Z (GO-15880) large program with HST/COS observed several massive stars ($V<$ 19.5) in IC 1613 \citep[12 + $\log$ O/H = 7.86, or 13\% solar][]{skillman1989} and Sextans A \citep[12 + $\log$ O/H = 7.54, or 7\% solar][]{kniazev2005}. Those data are supplemented with archival observations of seven O and early B stars in Sextans A and IC 1613.

\begin{figure*}

	\includegraphics[width=2\columnwidth]{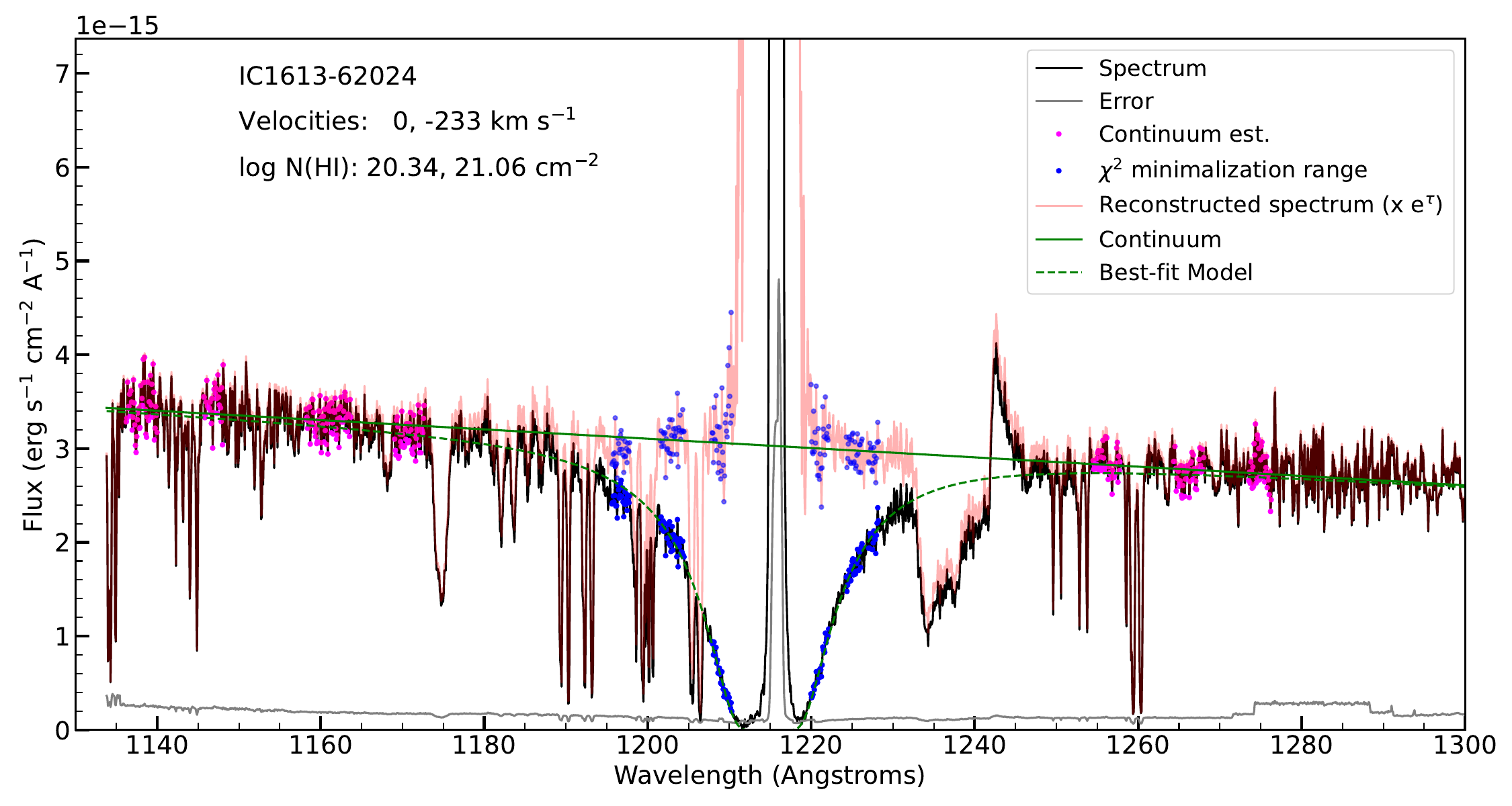}

    \caption{Illustration of the Ly-$\alpha$ absorption line profile fitting procedure for sightline IC1613-62024. The spectrum is zoomed at $\pm 80$ \AA~ around the rest frame Ly-$\alpha$ $\lambda$1216 \AA~ line. Error per wavelength is shown in grey. The central Ly-$\alpha$ absorption feature comes from the Milky Way and is blended with Ly-$\alpha$ from the targeted galaxy (here, IC1613). In black, we show the smoothed data; blue points mark the regions included in the profile fit, while magenta points mark the continuum fit windows. Continuum fit is marked with a straight green line, and the fitted line profile is marked with a green dashed line. For reference, we also show the reconstructed spectrum in red. In the top left, we show best-fit model parameters: heliocentric velocities of the centers of the line components and corresponding \ion{H}{1} column densities. Thanks to the significant velocity difference between Milky Way and METAL-Z galaxies, we can disentangle their \ion{H}{1} profiles from the line asymmetry.}
    \label{fig:hi}
\end{figure*}

\subsection{Target Sample}

\indent The sample is comprised of massive stars with spectral type B0 and earlier and with $V<$ 19.5 mag, taken from OB star catalogs in IC 1613 \citep{bresolin2007, garcia2014} and Sextans A \citep{camacho2016}. All targets are listed in Table \ref{tab:star}. O and early B stars have rotational velocities in excess of $\sim$ 70 km s$^{-1}$ \citep{garcia2014,tramper2014}, minimizing the risk of confusion between ISM and stellar photospheric absorption lines. Indeed, the typical velocity dispersion in ISM is 10-20 kms$^{-1}$ \citep{tamburro2009, choudhuri2019}. 

\indent Most targets were observed under {\it Hubble} program GO-15880 (PI: Roman-Duval). Additionally, we supplemented the GO-15880 sample with spectra of stars available in the archive: A13, B11, and B7 in IC 1613 (GO-12867, GO-15156) and s3, SA1, SA2, SA3 in Sextans A (GO-15967, GO-16920, and GO-16767). Note that the brightness of Sextans A star s3 falls below the original threshold V $<$ 19.5 mag; however, due to a limited number of available stars in that galaxy and the availability of archival observations, we decided to include it. 

\indent Both galaxies (IC 1613, Sextans A) have high-resolution ($\sim$8") VLA 21 cm data taken as part of the LITTLE THINGS \citep{hunter2012} or VLA-ANGST \citep{ott2012} surveys, providing spatially resolved information about the environment of the selected stellar sources and the estimates of the local N(\ion{H}{1}) values. 

\begin{figure*}
\centering

    \includegraphics[width=8.25cm]{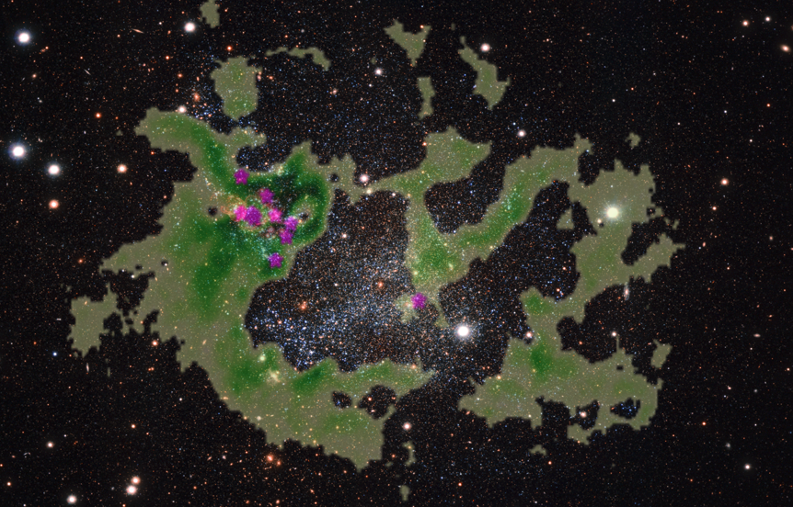}
    \includegraphics[width=7.75cm]{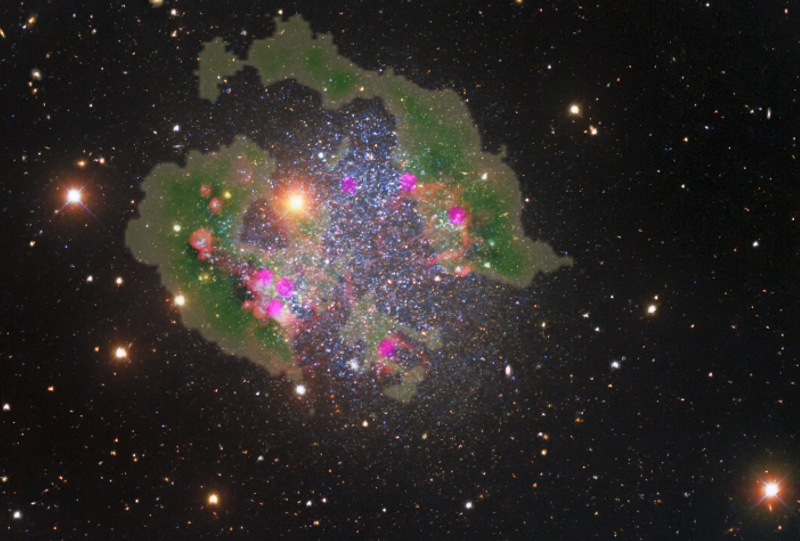}

    \caption{ Visible images of IC 1613 (left) and Sextans A (right) showing the targets used in this study as magenta stars, and the HI 21 cm integrated intensity from LITTLE THINGS as the green transparent color scale. The image of IC 1613 is from OmegaCam on ESO's Very Large Telescope (VLT, credit: ESO) and the image of Sextans A is from the Nicholas U. Mayall 4-meter Telescope at Kitt Peak National Observatory (Credit: KPNO/NOIRLab/NSF/AURA Data obtained and processed by: P. Massey (Lowell Obs.), G. Jacoby, K. Olsen, \& C. Smith (AURA/NSF) Image processing: T.A. Rector (University of Alaska Anchorage/NSF’s NOIRLab), M. Zamani (NSF’s NOIRLab) \& D. de Martin (NSF’s NOIRLab)). }
    \label{fig:plot_show_targets}
\end{figure*}

\indent The target sample is shown in Figure \ref{fig:plot_show_targets} for IC 1613 and Sextans A, along with HI 21 cm intensities from LITTLE THINGS and visible images of those galaxies from OmegaCam on ESO's Very Large Telescope (VLT) and Nicholas U. Mayall 4-meter Telescope at Kitt Peak National Observatory, respectively.

\begin{deluxetable}{lccrc} 
\tablecaption{Targets and observations with COS/HST, including archival datasets. \label{tab:obs}}
\tablehead{\colhead{Target} & \colhead{grating/cenwave}  & \colhead{LP\tablenotemark{a}} & \colhead{T$_{\rm exp}$} & \colhead{PID\tablenotemark{b}} \\   & & & \colhead{s} & }
\startdata
IC1613-61331 & G130M/1291 & LP4 & 9000 & GO-15880 \\ 
IC1613-61331 & G130M/1309 & LP3 & 2900 & GO-15880 \\ 
IC1613-62024 & G130M/1291 & LP4 & 29904 & GO-15880 \\ 
IC1613-62024 & G130M/1309 & LP3 & 4800 & GO-15880 \\ 
IC1613-64066 & G130M/1291 & LP4 & 4545 & GO-15880 \\ 
IC1613-64066 & G130M/1309 & LP3 & 3177 & GO-15880 \\ 
IC1613-67559 & G130M/1291 & LP4 & 11845 & GO-15880 \\ 
IC1613-67559 & G130M/1309 & LP3 & 2419 & GO-15880 \\ 
IC1613-67684 & G130M/1291 & LP4 & 6900 & GO-15880 \\ 
IC1613-67684 & G130M/1309 & LP3 & 2200 & GO-15880 \\ 
IC1613-A13 & G130M/1291 & LP4 & 1680 & GO-15880 \\ 
IC1613-A13 & G130M/1291 & LP1 & 1911 & GO-12867 \\ 
IC1613-A13 & G130M/1327 & LP1 & 2729 & GO-12867 \\ 
IC1613-B11 & G130M/1291 & LP4 & 4210 & GO-15880 \\ 
IC1613-B11 & G130M/1291 & LP1 & 1852 & GO-12867 \\ 
IC1613-B11 & G130M/1327 & LP1 & 2729 & GO-12867 \\ 
IC1613-B2 & G130M/1291 & LP4 & 3123 & GO-15880 \\ 
IC1613-B3 & G130M/1291 & LP4 & 13150 & GO-15880 \\ 
IC1613-B3 & G130M/1309 & LP3 & 6150 & GO-15880 \\ 
IC1613-B7 & G130M/1291 & LP4 & 18650 & GO-15156 \\ 
SEXTANS-A-s050 & G130M/1291 & LP4 & 17150 & GO-15880 \\ 
SEXTANS-A-s050 & G130M/1309 & LP3 & 8140 & GO-15880 \\ 
SEXTANS-A-s014 & G130M/1291 & LP4 & 22240 & GO-15880 \\ 
SEXTANS-A-s014 & G130M/1309 & LP3 & 9840 & GO-15880 \\ 
SEXTANS-A-s022 & G130M/1291 & LP4 & 14239 & GO-15880 \\ 
SEXTANS-A-s022 & G130M/1309 & LP3 & 4836 & GO-15880 \\ 
SEXTANS-A-s038 & G130M/1291 & LP4 & 14582 & GO-15880 \\ 
SEXTANS-A-s038 & G130M/1309 & LP3 & 4836 & GO-15880 \\ 
SEXTANS-A-s029 & G130M/1291 & LP4 & 10726 & GO-15967 \\ 
SEXTANS-A-s037 & G130M/1291 & LP4 & 12846 & GO-16767 \\ 
SEXTANS-A-SA2 & G130M/1291 & LP4 & 10180 & GO-16767 \\ 
SEXTANS-A-s021 & G130M/1291 & LP5 & 10184 & GO-16920 \\ 
\enddata
\tablenotetext{a}{COS lifetime positions used for observations.}
\tablenotetext{b}{Proposal ID corresponding to the observations.}
\end{deluxetable}

\subsection{Observations: COS spectroscopy}
\label{sec:obs}

\indent The COS spectra were obtained with the G130M grating, with some sightlines having the benefit of additional archival COS G160M observations. All observations are summarized in Table \ref{tab:obs}. 

\indent Hydrogen column density measurements are needed to derive abundances and depletion. The \ion{H}{1} column density is typically measured using the Ly-$\alpha$ ($\lambda$1216 \AA) line. At the time of the survey design, it was unclear whether Ly-$\alpha$ could be observed at LP4 given the COS2025 restrictions (Ly-$\alpha$ falls in the gain-sag detector gap of G130M/1291). Therefore, Ly-$\alpha$ was observed with the G130M/1309 at LP3, which was not impacted by gain-sag at the time of the observations.

\indent We used the G130M/1291 setting of COS/FUV at LP4 (FP-POS 3 and 4) to cover interstellar absorption lines from S and Fe: \ion{Fe}{2} $\lambda\lambda\lambda$ 1142, 1143, 1144 \AA~ and \ion{S}{2} $\lambda\lambda\lambda$ 1250, 1253, 1259 \AA. In addition, the 1291 spectra cover Si (\ion{Si}{2} $\lambda\lambda\lambda\lambda$ 1190, 1193, 1260, 1304 \AA), C (\ion{C}{2} $\lambda$1334 \AA) and O (\ion{O}{1} $\lambda$1302 \AA) lines. Those lines are strongly saturated and as such not usable for abundance measurements using standard techniques. The \ion{Mg}{2} lines (\ion{Mg}{2} $\lambda\lambda$ 1239, 1240 \AA) are also covered in the 1291 spectra but not detected in any sightline except for IC 1613-62024.

\indent The observations targeted S/N = 15 with COS G130M/1291 to detect equivalent widths $W_{\lambda} > 50$ m\AA~ at 10$\sigma$, which is expected for the metallicity and N(H) of the sample. We used spectral dithering with two FP-POS positions (3 and 4) to mitigate the effects of the fixed pattern noise on the effective S/N. In the case of the targets with archival COS/FUV observations (see Table \ref{tab:star} inputs marked with $a$), we had to make sure that the S/N of the archival data were sufficient to conduct the hydrogen, S, and Fe column density measurements. As the S/N of the archival observations of the B11 and A13 stars in IC 1613 was not satisfactory, additional exposures with G130M/1291 were taken as part of METAL-Z/GO-15880. 

\subsection{Data reduction}

\indent The COS spectra were retrieved from the MAST archive and processed with version 3.3.10 of the COS calibration pipeline, CalCOS. The different exposures for a given target were observed to be well aligned in wavelength, and no extra wavelength alignment was necessary.

\indent The different spectra obtained for a given target were co-added using the following approach. First, the {\it x1d} spectra were re-sampled using the nearest neighbor interpolation on a common wavelength grid. This method avoids covariance and correlated errors between pixels. The wavelength grid is set to encompass the wavelength coverage of all the input exposures, with the coarsest dispersion amongst the input exposures (note that all spectra used in this work have comparable spectral resolution). For each wavelength sample, the co-added flux and errors are computed by performing the weighted sum of all the input fluxes, where the weight is set to the product of the sensitivity (net counts/flux) and exposure time. This weight provides an unbiased and noiseless estimate of the number of counts expected to fall in a given wavelength bin and is superior to a weighting scheme based on inverse squared errors because COS errors are asymmetric, particularly in the low count regime \citep[COS ISR 2021-03, ][]{cos-johnson2021}. As a result, with an error-based weighting scheme, fluxes in noise troughs would be weighted higher than fluxes in noise peaks, resulting in a negative bias of the co-added fluxes. This bias can exceed 10\% in the low count regime.

\indent In addition to calculating co-added fluxes, the gross counts, background count rates, and flat-field-corrected sensitivities multiplied by exposure times were summed over the wavelength-interpolated {\it x1d} spectra. These values are stored to allow use of a Poisson likelihood function when comparing spectrum models to the data. The motivation for not using the commonly-adopted Gaussian approximation for spectrum uncertainties is the low signal-to-noise ratio of some of the data, which leads to as few as $\approx10$ photons per pixel. The Poisson likelihood function is described in \S \ref{sec:cd-measurement:likelihood}.

\begin{deluxetable}{lccc}
\tablecaption{Central wavelength and oscillator strength of spectral lines used for depletion measurements.  \label{tab:lines}}
\tablehead{\colhead{Element/Ion}  & \colhead{Wavelength} & \colhead{$\log$ $\lambda f_{\lambda}$} \\
&  \colhead{\AA} &\colhead{\AA}   }
\startdata
         \ion{S}{2} &  1250.578 & 0.809 \\
             &  1253.805 & 1.113 \\
             &  1259.518 & 1.295 \\
       \ion{Fe}{2}  &  1142.366 & 0.661 \\
             &  1143.226 & 1.342 \\
             &  1144.938 & 1.978 \\
             &  1608.451 & 1.968 \\
\enddata
\tablerefs{The line strengths come from \citet{kisielius2014} for \ion{S}{2} and \citet{morton2003} for \ion{Fe}{2}.}
\end{deluxetable}

\section{Gas-phase Column Density Measurements}
\label{sec:colden}

\indent In this section, we describe the novel line profile fitting method used to perform ISM abundance measurements from the METAL-Z data. The measurements were performed only for \ion{Fe}{2} and \ion{S}{2} lines (parameters of observed lines are summarized in Table \ref{tab:lines}); other metal lines were either too saturated or undetected (Section \ref{sec:obs}). In addition, we describe the measurements of the \ion{H}{1} column density through line profile fitting of the Ly-$\alpha$ line. 

\indent Alongside the profile fitting, we have implemented a Curve of Growth (CoG) method to measure the column density of S and Fe from METAL-Z data. Due to the resolution of COS and the limited S/N of the data, the CoG resulted in highly scattered measurements with large error bars. We present the results of these measurements in Appendix \ref{sec:app-cog} to illustrate that the Curve of Growth method does not provide sufficiently accurate results from these spectra to pursue our analysis of the variations of depletions and D/G.

\begin{deluxetable}{lcc}
\tablecaption{Gas column density measurements for IC 1613/Sextans A and the Milky Way, as described in Section \ref{sec:hi}. \label{tab:lognh}}
\tablehead{\colhead{Target} & \colhead{$\log$ N(\ion{H}{1})$_{\rm gal}$\tablenotemark{a}}  & \colhead{$\log$ N(\ion{H}{1})$_{\rm MW}$\tablenotemark{b}} \\ & \colhead{cm$^{-2}$} &    \colhead{cm$^{-2}$} } 
\startdata
IC1613-61331 & 20.84 $\pm$ 0.02 &  20.11    $\pm$ 0.12  \\ 
IC1613-62024 & 21.05 $\pm$ 0.03 &  20.39    $\pm$ 0.12  \\ 
IC1613-64066 & 20.88 $\pm$ 0.03 &  20.34    $\pm$ 0.11  \\ 
IC1613-67559 & 20.47 $\pm$ 0.09 &  20.42    $\pm$ 0.14  \\ 
IC1613-67684 & 20.44 $\pm$ 0.03 &  20.23    $\pm$ 0.07  \\ 
IC1613-A13 & 20.26 $\pm$ 0.03 &  20.29    $\pm$ 0.03  \\ 
IC1613-B11 & 20.37 $\pm$ 0.03 &  20.22    $\pm$ 0.06  \\ 
IC1613-B2 & 21.17 $\pm$ 0.03 &  20.35    $\pm$ 0.15  \\ 
IC1613-B3 & 20.64 $\pm$ 0.05 &  20.23    $\pm$ 0.14  \\ 
IC1613-B7 & 20.73 $\pm$ 0.04 &  20.37    $\pm$ 0.06  \\ 
SEXTANS-A-s050 & 20.46 $\pm$ 0.04 &  20.61    $\pm$ 0.03  \\ 
SEXTANS-A-s014 & 20.71 $\pm$ 0.03 &  20.55    $\pm$ 0.05  \\ 
SEXTANS-A-s022 & 21.17 $\pm$ 0.02 &  20.40    $\pm$ 0.11  \\ 
SEXTANS-A-s038 & 21.11 $\pm$ 0.02 &  20.44    $\pm$ 0.10  \\ 
SEXTANS-A-s029 & 20.90 $\pm$ 0.05 &  20.39    $\pm$ 0.20  \\ 
SEXTANS-A-s037 & 21.11 $\pm$ 0.05 &  20.52    $\pm$ 0.22  \\ 
SEXTANS-A-SA2 & 21.07 $\pm$ 0.03 &  20.23    $\pm$ 0.21  \\ 
SEXTANS-A-s021 & 20.38 $\pm$ 0.04 &  20.38    $\pm$ 0.20  \\ 
\enddata
\tablenotetext{a}{$\log$ N(\ion{H}{1}) of IC 1613/Sextans A towards a listed sightline}
\tablenotetext{b}{$\log$ N(\ion{H}{1}) of Milky Way towards a listed sightline}
\end{deluxetable}

\subsection{Measurements of the atomic hydrogen column density}
\label{sec:hi}

\indent The hydrogen column density N(H) is used to normalize metal abundances. The total N(H) includes contributions from hydrogen in atomic (\ion{H}{1}) and molecular (H$_2$) forms. The fraction of the latter in the cold ISM varies with gas column density and metallicity \citep{tumlinson2002}. 

\indent METAL-Z galaxies are sufficiently separated in velocity from the MW ($-$233 km s$^{-1}$ for IC 1613 and 324 km s$^{-1}$ for Sextans A) to use the Ly-$\alpha$ $\lambda$1216 \AA~absorption line for N(\ion{H}{1}) measurements. We measured N(\ion{H}{1}) by modeling the Lorentzian absorption profile of the Ly-$\alpha$ line using two components, one for the MW and one for the galaxy of interest (Sextans A or IC 1613). We used a modified continuum reconstruction method outlined in \citet{diplas1994, romanduval2019}, which we summarize in this Section (method illustrated in Figure \ref{fig:hi}). 

\indent First, we determined the heliocentric velocities of the Milky Way (MW) and the galaxy towards a specific sightline by measuring the velocity offsets of the \ion{S}{2} $\lambda$1250 \AA~ absorption line from its rest-frame wavelength for both the MW and METAL-Z galaxies. These velocity offsets were adopted as central velocities for their respective Ly-$\alpha$ absorption components.
\indent Second, we fitted a linear function to the continuum around the Ly-$\alpha$ line, excluding ISM, photospheric, and stellar wind lines, such as the \ion{N}{5} P Cyg profile around $\lambda$1240 \AA, from the fits. We then normalized the spectra using this linear continuum.
\indent Third, we modeled the Ly-$\alpha$ profile in the normalized spectra using two Lorentzian profiles, one for the MW and one for IC 1613/Sextans A using the velocities derived from \ion{S}{2} in the first step. The profile fits were performed by minimizing the $\chi^2$ between the observed absorption profile and a two-dimensional grid of two-component Lorentzian model profiles sampling the MW and galaxy \ion{H}{1} column density (N(\ion{H}{1})$_{\mathrm{MW}}$ and N(\ion{H}{1})$_{\mathrm{gal}}$). The grid covers a range of column densities $\log$ N(\ion{H}{1}) = 18.5 -- 22.5 cm$^{-2}$ for each of the components (MW and IC 1613/Sextans A). We excluded ISM metal absorption lines, stellar absorption lines, and the very bottom trough of the Ly-$\alpha$ absorption profile, which has zero or negative counts, from the fits by manually selecting spectral windows included in the $\chi^2$ estimation. The fitting resulted in N(\ion{H}{1}) measurements for both the MW and galaxy along each METAL-Z sightline. The measured $\log$ N(\ion{H}{1}) values ranged from $\sim$ 20.5 to $\sim$ 21.2 cm$^{-2}$ and are presented in Table \ref{tab:lognh}.

\indent To estimate the uncertainties of the N(\ion{H}{1}) measurements, we translated the $\chi^2$ to a two-dimensional likelihood $L \sim exp(-\chi^2/2)$. Following \citet{lampton1976} and given our two parameters (N(\ion{H}{1})$_{\mathrm{MW}}$ and N(\ion{H}{1})$_{\mathrm{gal}}$), we adopted $L_{1\sigma}$ $=$ $L_{\rm max} \times 0.32$, corresponding to $\chi^2 = \chi_{\rm min} + 2.3$, as the 1$\sigma$ uncertainty. 

\indent We estimated the molecular gas fraction for our sightlines based on measured N(\ion{H}{1}) and the dependence of the molecular gas fraction on hydrogen column density observed in the LMC and SMC in \citet{welty2012}. In the SMC, which has the closest metallicity to our sample, the molecular gas fraction rapidly decreases below $\log$ N(H) = 21.8 cm$^{-2}$ \citep[see Figure 17 in][]{welty2012}. Below this threshold column density, the molecular fraction is only a few percent or less. Based on the \ion{H}{1} measurements described above, the H$_2$ fraction toward our sightlines, which have even lower metallicity than the SMC, should be at most a few percent and can be neglected. At the same time, the contribution from ionized hydrogen (\ion{H}{2}) at neutral column densities typical of our sample (N(\ion{H}{1} $>$ 19.5 cm$^{-2}$) are $<<$ 1\% \citep{tchernyshyov2015}. Therefore, we adopt the atomic hydrogen column density N(\ion{H}{1}) as a measure of total gas density N(H). 

\subsection{Continuum fitting around metal absorption lines}
\label{sec:cont}
\begin{figure}

    \includegraphics[width=\columnwidth]{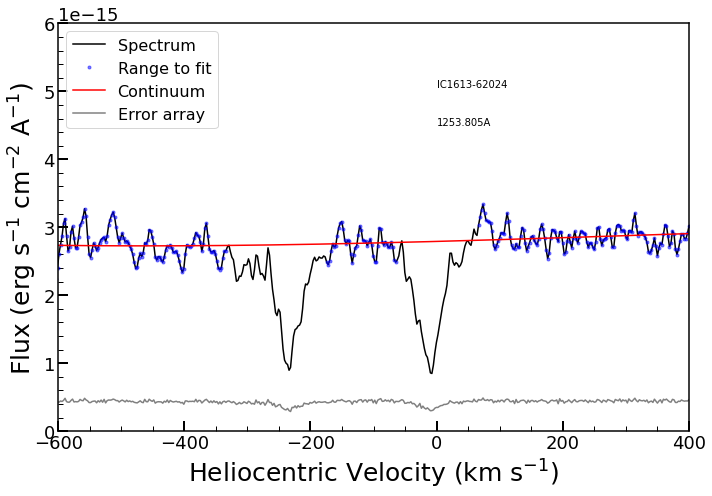}

    \caption{An example of continuum fitting procedure. The spectrum (black) is centered on the line of interest (here \ion{S}{2} $\lambda$1253 \AA) in the Milky Way velocity frame while the \ion{S}{2} from IC 1613 is visible around $-$200 kms$^{-1}$. Blue dots mark regions of the spectrum included in the fit; the red line shows the polynomial fit to the continuum. We selected the fitting windows to exclude Milky Way, ISM, and stellar lines.}
    
    \label{fig:cont}
\end{figure}

\indent We fitted 3$^{\mathrm{rd}}$ order Legendre polynomials \citep[following the fitting approach in ][]{sembach1992, jenkins1996} to the continuum around each measured line, within $\pm$ 1000 $-$ 1500 kms$^{-1}$ from the line center. Figure \ref{fig:cont} shows an example of continuum fit. We followed the error estimation method described in \citet{sembach1992} to calculate the continuum fit errors. In summary, this method considers errors in the flux measurements and the Legendre polynomial fitting and, following the standard $\chi^2$ statistic, provides the error estimation on the continuum fit. The error on the continuum is added in quadrature with other sources of uncertainties (e.g., intensity error) in the measurements.

\subsection{Column density measurements with forward optical depth modeling}
\label{sec:cd-measurement}
\indent We make two non-standard choices in our method for measuring column densities. These choices are motivated by the relatively low signal-to-noise ratio (SNR) of part of the dataset and by the large width of the COS line spread function (LSF), which have a full width half maximum (FWHM) of about 16 km s$^{-1}$, compared to the typical width of a metal line arising in the neutral ISM (FWHM of 3-10 km s$^{-1}$).

\indent To deal with the low SNR, we use a Poisson likelihood function and work with photon counts instead of using a Gaussian likelihood function and working with fluxes. To avoid model selection issues in Voigt profile fitting, we create a mix of that technique with the apparent optical depth (AOD) method \citep{savage1991}.

\subsubsection{The likelihood function}
\label{sec:cd-measurement:likelihood}

\indent The natural noise model for a photon-counting detector is the Poisson distribution. At high count rates, the Normal distribution becomes an acceptable approximation to the Poisson distribution. For some of the Sextans A spectra in the dataset, the typical number of photons at the continuum around the 1140 \AA\ \ion{Fe}{2} lines is 10, too low a number for an uncorrected normal approximation to be accurate.

\indent Our solution to this issue is to use the Poisson distribution instead. Let $f[\lambda]$\footnote{$X[\lambda]$ is the value of a quantity $X$ at a pixel with central wavelength $\lambda$.} be a model of spectrum flux, $T[\lambda]$ be the instrumental sensitivity function multiplied by the integration time, and $b[\lambda]$ be the expected number of background photon counts. The expected number of photon counts $\mu[\lambda]$ is equal to $f[\lambda]\times T[\lambda] + b[\lambda]$. The likelihood function for observed counts $c[\lambda]$ is a Poisson distribution with rate parameter $\mu[\lambda]$.

\subsubsection{Forward optical depth modeling}
\label{sec:cd-measurement:fod}

\indent The advantage of the AOD method is that it is non-parameteric; the advantage of Voigt profile fitting is that it accounts for the LSF through forward modeling. We propose the forward optical depth (FOD) method, in which the optical depth distribution is described using a dense set of basis functions (as in to AOD) instead of using a small number of flexible basis functions (as in Voigt profile fitting). The basis functions we use are Gaussians with fixed $b$ parameters and centroids placed densely over velocity regions that are expected to contain absorption.
The absorption model is specified by choosing the amplitude of each Gaussian.

\indent To perform Bayesian inference using this setup, it is necessary to define a prior over the amplitudes. Allowing negative amplitudes would be unphysical. Constraining amplitudes to be strictly positive would mean that there would be some absorption at every velocity covered by a basis element. We use a solution inspired by the ``spike and slab" family of priors \citep{ishwaran_2005}, which combines a broad distribution away from zero (the slab) with a delta function at zero (the spike).

\indent A scale parameter $s$, which we will see corresponds to the prior in the column density as a function of velocity $N(v)$, follows a uniform prior between $s_{min}$ and $s_{max}$. The pseudo-amplitudes of the basis functions, $a[v]$, follow a Normal prior with mean zero and variance $e^{2s}$:
\begin{align}
s &\sim \text{Uniform}(s_{min}, s_{max})\\
a[v] &\sim \text{Normal}(0, e^{2s}).
\end{align}
\noindent Let $G$ be the matrix or operator that multiplies the Gaussians that make up the basis by the pseudo-amplitudes and sums them, and let $y[v]=Ga[v]$.
The column density as a function of velocity $N[v]$ is $y[v]$ when $y[v]$ is positive and zero otherwise:
\begin{equation}
N[v] = \begin{cases}
    0 & y[v]\leq 0\\
    y[v] & y[v]>0.
\end{cases}
\end{equation}

\indent The resulting prior over positive $N[v]$ is uniform in $\log N[v]$ between $s_{min}$ and $s_{max}$ with tails extending below $s_{min}$ and above $s_{max}$. There is also a spike at $N[v]$=0 which contains half the total probability mass. Results are not sensitive to the exact values of $s_{min}$ and $s_{max}$ so long as $s_{min}$ corresponds to a line depth too small to detect and $s_{max}$ corresponds to at least the largest plausible column density.

\indent It is also necessary to choose the width of the Gaussians and their spacing. We use a $b$ parameter of 2 km s$^{-1}$. This is narrower than most \ion{Fe}{2} and \ion{S}{2} absorption components seen in high-resolution spectra of the neutral ISM \citep{welty1997}. We find that using $b$=1 or 1.5 km s$^{-1}$ does not change the results of numerical experiments (but does require a longer computation time). Component centroids are separated by a factor of FWHM/3 (i.e., 1.57 km s$^{-1}$ for $b$ = 2 km s$^{-1}$) and spectral synthesis is done on a wavelength grid with pixel separation smaller than FWHM/3.

\subsubsection{Complete model}
\label{sec:cd-measurement:combining}

\indent The absorption model uses the FOD parameterization described above with Gaussian basis elements covering foreground and target galaxy velocity ranges. For IC 1613, the foreground range is from $-$75 to $+$50 km s$^{-1}$ and the target galaxy range is from $-$300 to $-$200 km s$^{-1}$. For Sextans A, the foreground range is from $-$75 to $+$200 km s$^{-1}$ and the target galaxy range is from $+$220 to $+$420 km s$^{-1}$. The foreground range for Sextans A is wider because of the presence of strong high-velocity cloud absorption in the spectrum. 

\indent To marginalize over continuum placement uncertainties, the model includes the coefficients of a degree two polynomial perturbation about the continuum derived in \S\ref{sec:cont}.
\indent The absorption and continuum models are multiplied to produce a complete flux model and the complete flux model is used to compute the likelihood function as is described in \S\ref{sec:cd-measurement:likelihood}.

\indent We implement the model in the \texttt{numpyro} probabilistic programming language. The posterior probability density function over the model parameters is explored through Markov Chain Monte Carlo (MCMC) sampling using the \texttt{numpyro} implementation of the No U Turns Sampler \citep{hoffman2014gelman}. For each problem, we generate three independent MCMC chains running each for 200 (subsequently discarded) burn-in steps and 500 kept steps. We assess convergence using the rank-normalized $\hat{R}$ statistic \citep{vehtari2021ranknorm}.

\begin{figure}
    \centering
    \includegraphics[width=\linewidth]{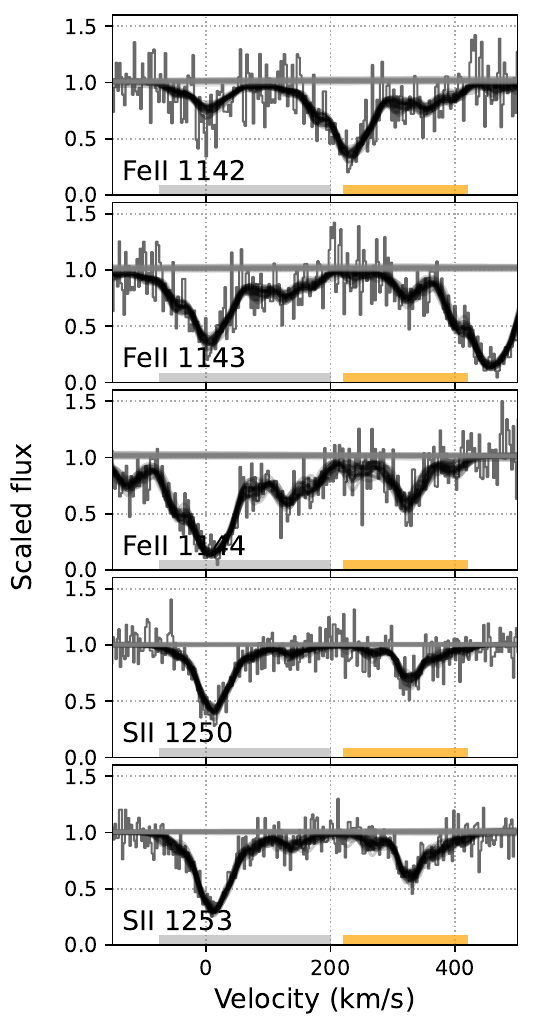}
    \caption{Forward optical depth method fits to \ion{Fe}{2} and \ion{S}{2} absorption toward the star Sextans A LGN s022. The velocity ranges of foreground Milky Way and Sextans A-associated gas are indicated by gray and orange horizontal bars. The observed spectrum is shown using a stepped gray line, continuum uncertainty is shown using smooth gray lines, and the absorption model is shown using black lines.}
    \label{fig:demo-sextans-a}
\end{figure}

\begin{figure}
    \centering
    \includegraphics[width=\linewidth]{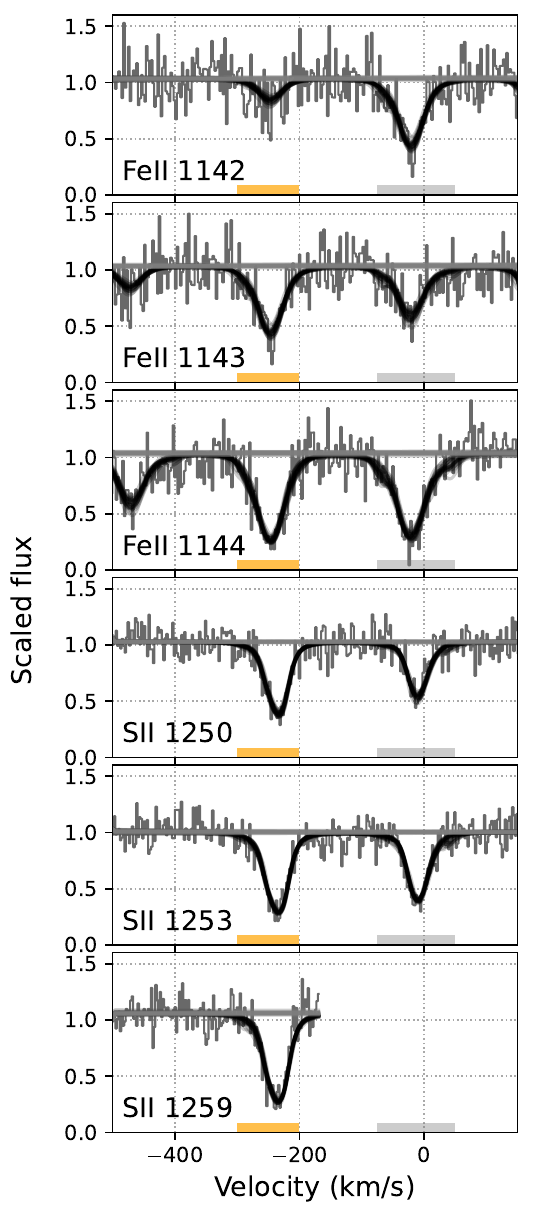}
    \caption{Forward optical depth model fits to \ion{Fe}{2} and \ion{S}{2} absorption toward the star IC1613-B2. The velocity ranges of foreground and IC1613-associated gas are indicated by gray and orange horizontal bars. The observed spectrum is shown using a stepped gray line, continuum uncertainty is shown using smooth gray lines, and the absorption model is shown using black lines.}
    \label{fig:demo-ic1613}
\end{figure}

\indent Figures \ref{fig:demo-sextans-a} and \ref{fig:demo-ic1613} show examples of fits to absorption toward stars in Sextans A and IC1613. We summarize the estimated total column densities $\log_{10} N$ in the target galaxies by taking means and standard deviations of the posterior probability distributions over $\log_{10}N$. These values are reported in \autoref{tab:fes-ncol-d}.

\indent To test the reliability of the column density estimates, we create artificial spectra with known input column densities and use different methods to try to recover the inputs. These numerical experiments are described in Appendix \ref{appendix:method-tests}.

\begin{deluxetable*}{lccccc}
\tablecaption{Column density,  Fe and S depletions for METAL-Z sightlines.\label{tab:fes-ncol-d}}
\tablehead{\colhead{Target} & \colhead{$\log$ N(H)} & \colhead{$\log$ N(\ion{S}{2})} &\colhead{$\log$ N(\ion{Fe}{2})} & \colhead{$\delta(\rm S)$} &\colhead{$\delta(\rm Fe)$}  \\
& \colhead{cm$^{-2}$} &\colhead{cm$^{-2}$} &  \colhead{cm$^{-2}$} & & }
\startdata
IC1613-61331 & 20.84 $\pm$ 0.03& 15.35  $\pm$ 0.05 & 14.59$\pm$ 0.06 & 0.04 $\pm$ 0.06 &-1.12 $\pm$ 0.07 \\ 
IC1613-62024 & 21.05 $\pm$ 0.03& 15.54  $\pm$ 0.07 & 14.84$\pm$ 0.07 & 0.02 $\pm$ 0.08 &-1.08 $\pm$ 0.08 \\ 
IC1613-64066 & 20.88 $\pm$ 0.04& 15.43  $\pm$ 0.06 & 14.80$\pm$ 0.07 & 0.08 $\pm$ 0.07 &-0.95 $\pm$ 0.08 \\ 
IC1613-67559 & 20.47 $\pm$ 0.14& 15.16  $\pm$ 0.07 & 14.61$\pm$ 0.06 & 0.22 $\pm$ 0.15 &-0.73 $\pm$ 0.15 \\ 
IC1613-67684 & 20.44 $\pm$ 0.04& 14.97  $\pm$ 0.07 & 14.54$\pm$ 0.07 & 0.06 $\pm$ 0.08 &-0.77 $\pm$ 0.08 \\ 
IC1613-A13 & 20.26 $\pm$ 0.03& 15.11  $\pm$ 0.08 & 14.69$\pm$ 0.09 & 0.38 $\pm$ 0.09 &-0.44 $\pm$ 0.09 \\ 
IC1613-B11 & 20.37 $\pm$ 0.05& 15.19  $\pm$ 0.07 & 14.41$\pm$ 0.08 & 0.35 $\pm$ 0.09 &-0.83 $\pm$ 0.09 \\ 
IC1613-B2 & 21.17 $\pm$ 0.03& 15.65  $\pm$ 0.07 & 14.99$\pm$ 0.06 & 0.01 $\pm$ 0.08 &-1.05 $\pm$ 0.07 \\ 
IC1613-B3 & 20.64 $\pm$ 0.06& 15.28  $\pm$ 0.08 & 14.49$\pm$ 0.08 & 0.17 $\pm$ 0.10 &-1.02 $\pm$ 0.10 \\ 
IC1613-B7 & 20.73 $\pm$ 0.03& 15.24  $\pm$ 0.05 & 14.71$\pm$ 0.04 & 0.04 $\pm$ 0.06 &-0.89 $\pm$ 0.05 \\ 
SEXTANS-A-s050 & 20.46 $\pm$ 0.04& 15.17  $\pm$ 0.05 & 14.74$\pm$ 0.07 & 0.54 $\pm$ 0.06 &-0.27 $\pm$ 0.08 \\ 
SEXTANS-A-s014 & 20.71 $\pm$ 0.03& 15.06  $\pm$ 0.07 & 14.70$\pm$ 0.09 & 0.18 $\pm$ 0.08 &-0.56 $\pm$ 0.09 \\ 
SEXTANS-A-s022 & 21.17 $\pm$ 0.02& 15.35  $\pm$ 0.07 & 14.73$\pm$ 0.09 & 0.01 $\pm$ 0.08 &-0.99 $\pm$ 0.09 \\ 
SEXTANS-A-s038 & 21.11 $\pm$ 0.02& 15.27  $\pm$ 0.05 & 14.86$\pm$ 0.07 & -0.01 $\pm$ 0.06 &-0.80 $\pm$ 0.07 \\ 
SEXTANS-A-s029 & 20.90 $\pm$ 0.07& 15.13  $\pm$ 0.08 & 14.69$\pm$ 0.12 & 0.06 $\pm$ 0.11 &-0.76 $\pm$ 0.14 \\ 
SEXTANS-A-s037 & 21.11 $\pm$ 0.07& 15.29  $\pm$ 0.09 & 15.08$\pm$ 0.12 & 0.01 $\pm$ 0.11 &-0.58 $\pm$ 0.14 \\ 
SEXTANS-A-SA2 & 21.07 $\pm$ 0.04& 15.39  $\pm$ 0.07 & 14.98$\pm$ 0.10 & 0.15 $\pm$ 0.08 &-0.64 $\pm$ 0.11 \\ 
SEXTANS-A-s021 & 20.38 $\pm$ 0.04& 15.29  $\pm$ 0.10 & 14.72$\pm$ 0.11 & 0.74 $\pm$ 0.11 &-0.21 $\pm$ 0.11 \\ 
\enddata
\end{deluxetable*}

\subsection{Gas-phase abundances and depletions of Fe and S}\label{sec:dep_meas}

\indent We calculated gas-phase abundances of Fe and S by taking the ratio of the measured column densities to the total hydrogen column densities derived in Section \ref{sec:hi}. We assume that an element's total (gas + dust) abundance is equal to that element's abundance in the photospheres of young stars recently formed in the ISM. By comparing measured abundances from the gas phase with the abundances in stellar photospheres, we can derive the fraction of metal locked in the dust (see Equation \ref{eq:dep})

\indent The stellar abundances used in this work, alongside their sources in the literature, are summarized in Table \ref{tab:stellar_ab}. Both METAL-Z galaxies have measured stellar Fe abundances. We note that the stellar abundance of Fe in IC 1613 reported in the literature varies between [Fe/H] = -0.3 and 0.8 \citep[see][for full discussion]{garcia2014}. Here, we adopted the value of [Fe/H] = -0.69 \citep{bouret2015}, consistently found in massive stars.

\indent IC 1613 and Sextans A lack direct measurements of S stellar abundances. Therefore we used other $\alpha$ elements as the reference, assuming their abundance relative to solar is comparable to S. We used O for IC 1613 \citep{bresolin2007}, and Mg in Sextans A \citep[][as it does not have a measured O stellar abundance either]{kaufer2004}, and assumed [S/H] $\simeq$ [O/H] $\simeq$ [Mg/H]. 

\begin{deluxetable}{l|cc}
\tablecaption{Stellar abundances of METAL-Z galaxies. \label{tab:stellar_ab}}
\tablehead{\colhead{Galaxy} & \colhead{[Fe/H]} & \colhead{[$\alpha$/H] }}
\startdata
       IC 1613 & -0.67 $\pm$ 0.09 \tablenotemark{a} & -0.79 $\pm$ 0.08\tablenotemark{b} \\
       Sextans A & -0.99 $\pm$ 0.04\tablenotemark{c} & -1.09$^{+0.02}_{-0.19}$\tablenotemark{c}\\
\enddata
\tablecomments{Due to lack of S abundance measurements we adopted $\alpha$ = O for IC 1613 and $\alpha$ = Mg for Sextans A.}
\tablerefs{(a) \citet{tautvaisiene2007}, (b) \citet{bresolin2007}, (c) \citet{kaufer2004} } 
\end{deluxetable}

\section{The dependence of depletions on gas density and metallicity}
\label{sec:depl}

\indent In this section, we examine the dependence of metal depletion on gas density and metallicity, since these two parameters are expected to drive the dust growth timescale in the ISM \citep{zhukovska2016, asano2013}. We present the relation between the depletion of Fe and S in Section \ref{sec:dep} and their dependence on the hydrogen column density and metallicity in Section \ref{sec:nhdep}.

\subsection{Relation between depletions of different elements}
\label{sec:dep}

\begin{figure}
\centering
	\includegraphics[width=\columnwidth]{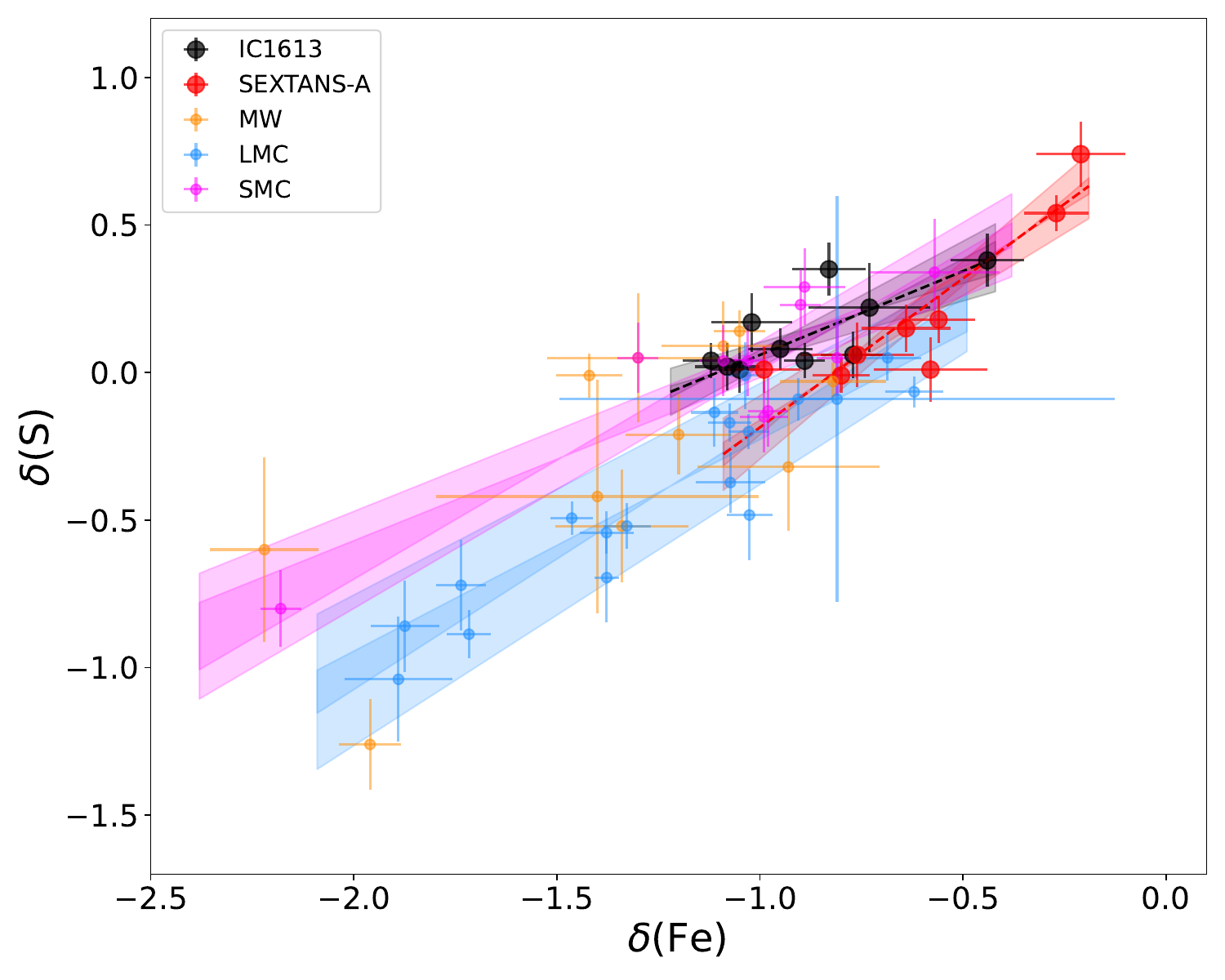}

    \caption{Elemental depletions of sulfur versus iron. METAL-Z measurements of depletion of these two elements are shown in black (IC 1613) and red (Sextans A). Depletions of S and Fe derived in the MW \citep{jenkins2009}, LMC \citep{romanduval2021}, and SMC \citep{jenkins2017} are also shown in orange, blue, and magenta, respectively. For each set of measurements, a linear fit and associated uncertainties are shown in transparency. The relation between $\delta$(S) and $\delta$(Fe) in IC 1613 and Sextans A follows the trends observed in the LMC and SMC, with IC 1613 being closer to the SMC and Sextans A to the LMC. We note that $\delta(S) > 0$ indicates no sulfur depletion at metallicities below 20 \% solar. The all-positive depletions of sulfur can result from uncertain stellar abundances in IC 1613 and Sextans A and/or a contribution from ionized gas to \ion{S}{2}. }
    \label{fig:dep}
\end{figure}

\indent \citet{jenkins2017, romanduval2021, romanduval2022a} have shown a strong correlation between the depletion of various elements and that of Fe in the SMC and LMC, respectively, and that this correlation weakly depends on metallicity, at least in the Milky Way, LMC, and SMC. There are some exceptions: \citet{romanduval2022a} showed that Ti and Mg deplete less relative to Fe in the Magellanic Clouds compared to the Milky Way, while \citet{jenkins2017} found that Mn depletes more rapidly relative to Fe in the SMC than in the MW.\\

\indent Fe column densities are easier to measure than other elements, thanks to their multiple transitions with varying oscillator strengths throughout the UV range. An invariance of the relation between depletions of different elements with metallicity would therefore prove very useful in estimating depletions for dust-building elements from Fe depletions alone in low metallicity, more distant systems where other UV lines may not be detected.

\indent We test the invariance of the relation between Fe and S depletions with metallicity down to $Z = 0.1 Z_{\odot}$ using the METAL-Z observations. In Figure \ref{fig:dep}, we plot the Fe and S depletion measurements for IC 1613 and Sextans A, as well as the relation between $\delta(\rm Fe)$ and $\delta(\rm S)$ derived in the MW, LMC, and SMC. The measurements obtained in IC 1613 and Sextans A are consistent with the trends established in the MW, LMC, and SMC, with all trends within 2$\sigma$ of each other (as shown by the overlap between the transparent bands representing the linear relations and their 1$\sigma$ uncertainties in Figure \ref{fig:dep}). The linear relation between $\delta$(Fe) and $\delta$(S) in IC 1613 is slightly closer to that observed in the SMC, while that same relation observed in Sextans A is in marginally better agreement with measurements in the LMC. This result suggests that there is no significant evolution of this trend with metallicity, within the uncertainties. This is a good indication that the relation between the depletions of different elements at first order may not significantly vary with metallicity. This will allow us later in the paper to assume the MW relation between the depletion of Fe and other elements to estimate D/G in IC 1613 and Sextans A based on the measured Fe depletions in those galaxies (see Section \ref{sec:dg}). We note, however, that the METAL-Z program only probes S and Fe, and that neither METAL-Z nor previous studies in the LMC and SMC included the major source of metal and dust mass: carbon and oxygen.

\indent Depletions toward sightlines in IC 1613 and Sextans A vary between $-$1.1 to $-$0.2 for $\delta(\rm Fe)$ and 0 to 1 for $\delta(\rm S)$. The $\delta(S) > 0$ measurements would indicate no sulfur depletion, but there remains the possibility of slightly depleted slightlines if the stellar abundances of sulfur are under-estimated. Indeed, stellar abundances of sulfur are not directly measured. Instead, we assumed [S/H]  = [Mg/H] in Sextans A and [S/H] = [O/H] in IC 1613 (see Section \ref{sec:dep_meas} and Table \ref{tab:stellar_ab}).

\begin{figure}
\centering
	\includegraphics[width=\columnwidth]{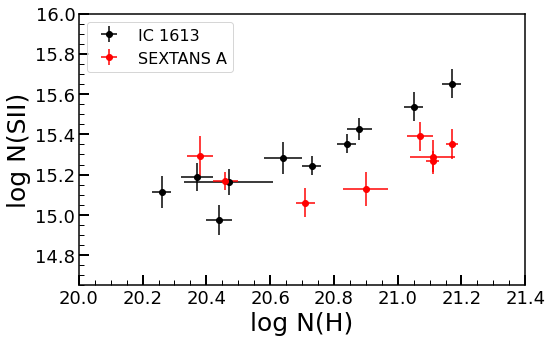}

    \caption{Column density measurements of \ion{S}{2} and H in IC 1613 (black) and Sextans A (red). The expected linear relation between the column density of sulfur and hydrogen flattens at low gas densities of $\log$ N(H) $< 20.6$ cm$^{-2}$. Unexpectedly high N(\ion{S}{2}) happening mostly at lower N(H) can be explained by an ionized layer of gas that is sub-dominant at high gas densities but apparent at lower values of N(H). }
    \label{fig:nsii}
\end{figure}

\indent Another possible explanation for the positive depletions of S in IC 1613 and Sextans A is the contribution of ionized gas to \ion{S}{2}. Since \ion{S}{2} has a significantly higher ionization potential than hydrogen (23 eV), \ion{S}{2} in ionized gas surrounding the target O or early B star may contribute significantly to the \ion{S}{2} column density. In Figure \ref{fig:nsii}, we show the relation between column densities of \ion{S}{2} and H. At $\log$ N(H) $<$ 20.6 cm$^{-2}$, the otherwise linear relation flattens, resulting in higher measurements of $\log$ N(\ion{S}{2}) than expected, signifying a contamination from ionized gas. Since abundances are only normalized to H, this would result in overestimated gas-phase abundances. For sightlines that are only very mildly depleted in S, as in the case of the low metallicities observed here, the contamination of N(\ion{S}{2}) by ionized gas would, therefore, result in positive depletions $\delta(\rm S)> 0$ \citep{jenkins2009, jenkins2017}. 

\indent Positive S depletions are seen up to gas column densities $\log$ N(\ion{H}{1}) $\sim$ 21 cm$^{-2}$ in IC 1613 and Sextans A. For this reason, measurements of sulfur depletions cannot be used to draw robust conclusions. Because of the positive S depletions $\delta(S) > 0 $, we do not attempt to fit the $\delta(\rm Fe)$--$\delta(\rm S)$ relation for IC 1613 and Sextans A. 

\subsection{The variations of Fe and S depletion with hydrogen column density and metallicity}\label{sec:nhdep}

\indent The timescale for the accretion of gas-phase metals onto dust grains is inversely proportional to the metallicity and the gas density \citep{asano2013, zhukovska2016}. Therefore, we expect to find a metal depletion and dust abundance dependence on these parameters. Depletion studies in the Milky Way, LMC and SMC have shown that the fraction of metals in the gas (depletion) decreases with increasing N(H) \citep{wakker2000, jenkins2009, tchernyshyov2015, romanduval2021, romanduval2022a} in systems down to metallicity of 20 \% solar. This indicates that the fraction of metals locked in dust, and subsequently the abundance of dust, increases with increasing gas density owing to the accretion of gas-phase metals onto dust grains as the ISM becomes denser. We would expect similar trends in lower metallicity galaxies such as those observed as part of METAL-Z. 

\indent For a face-on galaxy, variations in hydrogen column density are caused by density fluctuations in the medium and changes in the scale length perpendicular to the galaxy disk. \citet{romanduval2022a} argued that the magnitude of path length variations in the LMC is negligible. Therefore, variations in N(H) between lines-of-sight probe variations in the mean gas density along those sightlines. We assume the same is true for IC 1613 and Sextans A, with inclinations of $38^{\circ}$ and $33^{\circ}$, respectively \citep{hunter2012}, and that N(H) represents the average behavior of the gas density structure in the ISM of these objects.

\indent  For the LMC, SMC and MW, \citet{romanduval2021, romanduval2022a} introduced a linear relation between metal depletion and $\log$ N(H):

\begin{equation}
   \delta(X) = B_{\rm{H}}(X) + A_{\rm{H}}(X)(\log N(\rm{H}) - \log N_{{\rm{H}_0}}(X))
 \end{equation}

\noindent where $A_{\rm{H}}$ is the slope of the linear relation, $B_{\rm{H}}$ is the zero-point, and $\log N_{\rm{H}_0}(\rm{X})$ is a weighted mean hydrogen column density computed for each sample galaxy and element X. $\log N_{\rm{H}_0}(\rm{X})$ is introduced to remove the covariance between the intercept and the slope of the linear function and is defined as:

\begin{equation}
    \log N_{\rm{H}_0}(\rm{X}) = \frac{\sum_{\rm l.o.s} \frac{\log N(\rm{H})}{\sigma(\delta({\rm{X}))^2}}}{\sum_{\rm l.o.s} \frac{1}{\sigma(\delta(\rm{X}))^2}}
\end{equation}

\noindent where the sum is performed over all lines of sight in a given galaxy, and $\sigma$($\delta$(X)) is the uncertainty on the depletion measurement of X along each sightline. \\
\indent The zero-points of the relation between hydrogen column density and depletions, i.e., $B_{\rm{H}}$ as well as the intercept of the linear relation given by $B_{\rm{H}}$ $-$ $A_{\rm{H}}$$\log N_{\rm{H}_0}(\rm{X})$, are observed to increase with decreasing metallicity in the MW, LMC, and SMC \citep{romanduval2022a}, indicating that, for a given gas density, metals become less depleted from the gas-phase as the total metallicity decreases. As a result, \citet{romanduval2022a} showed that the gas-phase metallicity of the SMC is similar to the gas-phase metallicity of the MW, because, while the metallicity of the SMC is five times lower than the MW, the fraction of metals in the gas is also five times higher than in the MW. \\
\indent To test how metals deplete in the lower metallicity galaxies IC 1613 and Sextans A, we fitted a similar linear function to the relation between Fe depletions and $\log$ N(H) measured in those galaxies. We used a Bayesian linear regression method to find the best-fit parameters $A_{\mathrm{H}}$ and $B_{\mathrm{H}}$ and estimate their uncertainties. To create the model, we constructed a grid of coefficient values for the slope, $A_{\rm{H}}$ (ranging from -1 to 1), and the zero-point $B_H$ (ranging from -2 to 2). For $A_{\rm{H}}$ and $B_{\rm{H}}$ parameters, we used flat priors limited to the grid range. 

\begin{figure*}
    \centering
 	\includegraphics[width=2\columnwidth]{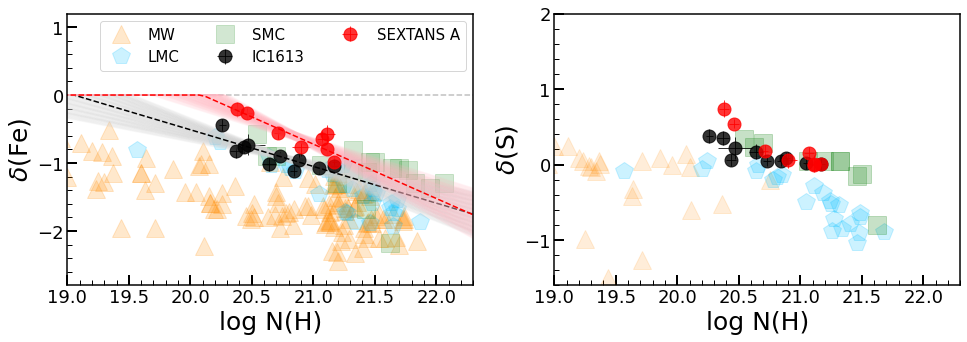}

    \caption{Relation between depletion of Fe (left), S (right) and hydrogen column density N(H). Galaxies in the METAL-Z sample are marked in red (Sextans A) and black (IC 1613). We compare our measurements with the literature: Milky Way \citep[orange triangles,][]{jenkins2009}, LMC \citep[blue pentagons,][]{romanduval2021}, and SMC \citep[green squares,][]{jenkins2017}. In IC 1613 and Sextans A, iron depletion decreases with increasing column density, similar to trends in the MW, LMC, and SMC. We fit the linear relation between Fe depletions and $\log$ N(H) (dotted line) and show the uncertainties as shaded areas. Fe depletions also show a dependence on total metallicity: Sextans A, the lowest metallicity galaxy in the sample, shows less depletion than higher metallicity galaxies in the same N(H) range. The positive depletions of sulfur are consistent with no depletion and show no variation with gas column density.}
 \label{fig:nhdep-fit}
 \end{figure*}

\indent We adopted the parameter pair with the highest probability (the smallest $\chi^2$ as the best-fitting model. We used the \citet{lampton1976} method to establish the errors and adapted a $\chi^2 = \chi_{min} + 2.3$ as the 1$\sigma$ uncertainty. The derived fit parameters for METAL-Z galaxies are listed in Table \ref{tab:bh}. In Figure \ref{fig:nhdep-fit}, we show the fitted relation between $\log$ N(H) and depletion of Fe and compare them to the MW, LMC, and SMC measurements. Wherever the relation predicts $\delta(X) > 0$, we plot $\delta(X) = 0$ instead. As $\delta(S) > 0$ for both galaxies, we do not establish the relation with gas density for sulfur; we plot the depletion measurements in Figure \ref{fig:nhdep-fit}.

\indent From previous results in the MW, LMC, and SMC \citep{romanduval2022a}, we expected to see two effects: 1) a decrease in iron depletion with increasing hydrogen column density (i.e., an increase in the fraction of metals locked in dust with increasing gas density) and 2) an increase in depletion value with decreasing galaxy metallicity (i.e., a decrease in the fraction of metals locked in dust with decreasing metallicity). The negative slopes of the relation between $\delta(Fe)$ and $\log$ N(H) show that, in the low metallicity environments of IC 1613 and Sextans A, iron is indeed more depleted in the higher density ISM. The slopes $A_{\mathrm{H}}$(Fe) in IC 1613 and Sextans A are $-$0.54$\pm$0.19 and $-$0.80$\pm$0.23, respectively, which is in line with the slopes reported by \citet{romanduval2022a} in the MW ($-$0.41$\pm$0.03), LMC ($-$0.71$\pm$0.03), and SMC ($-$0.59$\pm$0.04). \\
\indent Simultaneously, the zero-point $B_\mathrm{H}$ of the relation between $\log$ N(H) and $\delta$(Fe) increases with decreasing metallicity, with $B_{\mathrm{H}}$(Fe) $=$ $-$0.91$\pm$0.04 in IC 1613 and $B_{\mathrm{H}}$(Fe) $=$ $-$0.61$\pm$0.06 in Sextans A, compared to $B_{\mathrm{H}}$(Fe) $=$ $-$1.62$\pm$0.02 in the MW, $-$1.39$\pm$0.01 in the LMC, and $-$1.18$\pm$0.01 in the SMC \citep{romanduval2022a}. Admittedly, $B_{\mathrm{H}}$(Fe) corresponds to the depletion level at $\log N_{\rm{H}_0}(\rm{Fe})$, which the weighted mean hydrogen column density of each sample and varies from galaxy to galaxy. Therefore, the evolution of the $B_{\mathrm{H}}$(Fe) parameter with metallicity is not an exact representation of the changes in the depletion zero-levels with metallicity. Therefore, we also report in Table \ref{tab:bh} the Fe depletions at the lower and higher ends of the $\log$ N(H) range probed by the sample, $\log$ N(H) = 20.3 cm${-3}$ and $\log$ N(H) = 21 cm$^{-2}$. At $\log$ N(H) = 21 cm$^{-2}$, the Fe depletion level in the MW is -1.58 \citep{romanduval2022a}, and thus the Fe depletion value increases by 0.53 dex (factor 3.4) from the MW to IC 1613 and by 0.86 dex (factor 7.2) from the MW to Sextans A. This indicates that the zero-point level of depletions continues to increase (i.e., the fraction of metals locked in dust continues to decrease) as metallicity decreases below that of the SMC (20\% solar).

\begin{deluxetable}{c c  c }
\tablecaption{Fitted parameters $A_{\rm{H}}$, $B_{\rm{H}}$ and $\log N_{\rm{H}_0}$ of the relation between Fe depletion $\delta(\rm Fe)$ and hydrogen column density $\log$ $N$(H) for IC 1613 and Sextans A. \label{tab:bh}}
\tablehead{ \colhead{} & \colhead{IC 1613} & \colhead{Sextans A} }
\startdata
$A_{\rm{H}}$(Fe) & -0.54 $\pm$ 0.19 &  -0.80 $\pm$ 0.23  \\
$B_{\rm{H}}$(Fe) & -0.91 $\pm$ 0.04  & -0.61 $\pm$ 0.06 \\
$\log N_{\rm{H}_0}(\rm{Fe})$ & 20.74 & 20.86 \\
$\delta$(Fe) at $\log$ N = 20.3 cm$^{-2}$ & -0.67 $\pm$ 0.09 & -0.16 $\pm$ 0.14 \\
$\delta$(Fe) at $\log$ N = 21 cm$^{-2}$ & -1.05 $\pm$ 0.06 & -0.72 $\pm$ 0.07\\
\enddata
\end{deluxetable}

\subsection{Evolution of the iron fraction in dust with metallicty}
\label{sec:fe}

\indent Having measured the depletions of Fe in IC 1613 and Sextans A, we compute the fraction of iron in dust and the ``iron dust-to-gas mass ratio", D$_{\rm Fe}$/G, given below: 

\begin{equation}
D_{\rm Fe} / G = \frac{1}{1.36}(1 – 10^{\delta(\rm Fe)}) \bigg(\frac{N(\rm Fe)}{N(\rm H)}\bigg)_{\rm tot} W(\rm Fe), 
\end{equation}
where $W(\rm Fe)$ is the atomic weight of iron. The $1 – 10^{\delta(\rm Fe)}$ alone gives the fraction of Fe atoms locked in the dust form. $D_{\rm Fe} / G$ corresponds to the abundance of iron dust relative to gas.

\begin{figure}
    \centering
 	\includegraphics[width=\columnwidth]{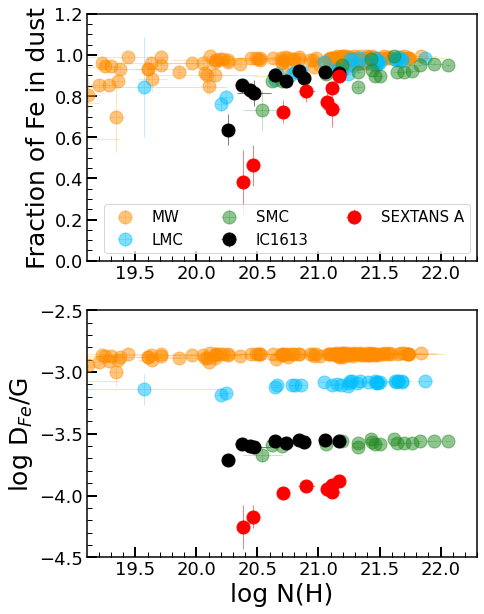}
    \caption{Fraction of Fe in dust versus N(H) column density (top), and Fe dust-to-gas ratio (bottom). METAL Z galaxies are marked in black (IC 1613) and red (Sextans A). Literature data from \citet{romanduval2022a} is shown in orange for the Milky Way, cyan for the LMC, and green for the SMC.}
 \label{fig:mfe}
 \end{figure}

\indent In the top panel of Figure \ref{fig:mfe}, we show the fraction of Fe locked in the dust at different gas densities. Alongside IC 1613 and Sextans A, we plot the fraction of Fe in dust for the MW, LMC, and SMC taken from \citet{jenkins2009}, \citet{romanduval2021} and \citet{jenkins2017}. The MW, LMC, and SMC show comparable fractions of Fe in dust ($\sim$1) for $\log$ N(H) $>$ 21.5 cm$^{-2}$. At lower column densities, the fraction of Fe in dust drops faster for lower metallicity systems. This is expected, as the dust accretion timescale is inversely proportional to density and metallicity. IC 1613, with an iron content that is similar to the SMC, aligns with the Magellanic Clouds with 70---80\% of iron in dust at $\log$ N(H) = 20.5 cm$^{-2}$. In the lowest metallicity system, Sextans A, the fraction of Fe in the dust is about 0.1--0.4 dex lower than in IC 1613 and the Magellanic Clouds.

\indent The bottom panel of Figure \ref{fig:mfe} shows $D_{\rm Fe} / G$ for the MW, LMC, SMC, IC 1613 and Sextans A, as a function of $\log$ N(H). While $D_{\rm Fe} / G$ in IC 1613 follows the trend observed in the SMC very closely, Sextans A has a D$_{\rm Fe}$/G that is 0.5---70.5 lower than the SMC and IC 1613. Since D$_{\rm Fe}$/G depends on the iron abundance and the fraction of iron in dust, some of these variations are purely due to the varying metal content of these galaxies (metallicity). For example, for Sextans A, the lower metallicity contributes a factor of 10 in the lower $D_{\rm Fe}/G$ compared to the MW. However, the $D_{\rm Fe}/G$ in Sextans A is 20 times lower than the MW because the fraction of Fe in dust also decreases with metallicity, as shown in Figure \ref{fig:mfe}.


\section{Estimation of D/G and D/M from iron depletion}
\label{sec:dg}


\indent In Section \ref{sec:dep}, we showed that the relation between the depletions of iron and sulfur in IC 1613 and Sextans A is consistent with that of the MW, LMC and SMC (Section \ref{sec:dep}). The MW and Magellanic Clouds, the depletions of many measured elements (Si, Mg, Ni, Cr, Zn, S, Ti) correlate tightly with the depletion of Fe \citep{romanduval2022a}. It is therefore not unreasonable to assume that the relation between the depletion of Fe and other elements in IC 1613 and Sextans A also closely follows the MW, LMC, and SMC relation. Under this assumption, Fe can be used as a tracer of the collective depletion level of various elements, from which D/G and D/M can be estimated. 

\indent In previous sections, we have also shown that 1) the depletion of Fe (i.e., the fraction of Fe in the gas phase) in IC 1613 and Sextans A decreases with increasing gas density with a slope consistent with that of the MW, LMC and SMC; and 2) that the zero-point of Fe depletions increases with decreasing metallicity from the MW to the LMC, SMC, IC 1613 and Sextans A. Since different elements deplete from the gas collectively, as shown by the tight correlations between the depletions of different elements observed in the MW, LMC, SMC \citep{romanduval2022a}, and between Fe and S in IC 1613 and Sextans A (see Section \ref{sec:dep}), it is safe to infer that the variations in Fe depletions observed in IC 1613 and Sextans A indicate that the dust-to-metal ratio and therefore also the dust-to-gas ratio increases with increasing gas density, at a rate that is similar between galaxies spanning 10\% to 100\% solar metallicity (the slopes of Fe depletions versus $\log$ N(H) are similar in all five galaxies). The variations of the Fe depletion zero-point implies that the dust-to-metal ratio decreases with decreasing metallicity. This in turn, implies that the dust-to-gas ratio, given by D/G = D/M $\times$ Z, should decrease faster than metallicity.  \\

\indent The measurements of S and Fe depletions presented in this paper are not sufficient for the derivation of the dust-to-gas ratio in IC 1613 and Sextans A. But by using Fe as a proxy for the overall depletion level of other elements and by assuming that the MW relation between $\delta$(Fe) and $\delta$(X) is invariant with metallicity, one can estimate D/G and D/M from the iron depletions measured in IC 1613 and Sextans A, and investigate how D/G and D/M vary with gas density and metallicity down to $\sim$10\% solar metallicity.\\

\indent We note that C and O comprise most of the metal and dust mass. However, C and O depletions have only been measured in the MW, and even in the MW, the trend of C depletions with $F_*$ (gas density) is quite uncertain due to the sparsity and large errors of the measurements \citep{jenkins2009}. Therefore, the invariance of the relation between C or O depletions and Fe depletions with metallicity has not been tested, not even in the LMC and SMC \citep{romanduval2022a}. While assuming the invariance of the $\delta$(Fe)---$\delta$(C or O) relation with metallicity allows us to estimate D/G and D/M and explore their variations with metallicity down to a yet unexplored metallicity regime, it does introduce a significant systematic uncertainty on the derived D/G and D/M values. Further constraining the relations between depletions in low metallicity environments requires the sensitivity of next-generation UV telescopes, such as the Habitable Worlds Observatory.

\begin{figure}

	\includegraphics[width=\columnwidth]{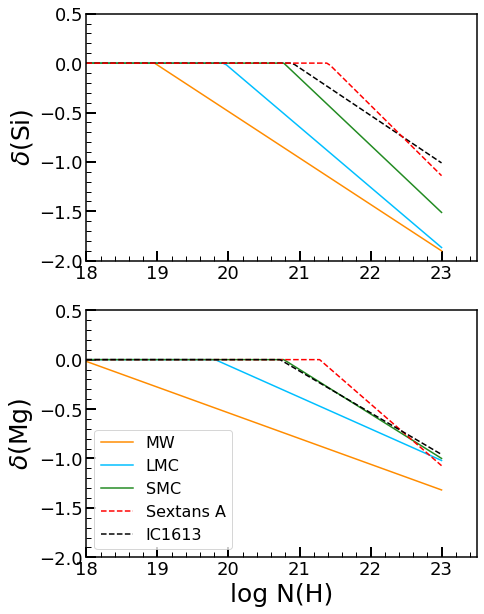}
    \caption{Depletion of Si (top) and Mg (bottom) as a function of $\log$ N(H) in the MW (orange), LMC (blue), SMC (green), IC 1613 (black), and Sextans A (red). In the MW and Magellanic Clouds, the relations were taken directly from the fits given in \citep{romanduval2022a}. We derived those relations in IC 1613 and Sextans A based on 1) the measured Fe depletions and 2) the MW relation between Fe depletion and depletion of other elements (since Si and Mg were not measured as part of METAL-Z observations). Such relations were derived for all main constituents of dust in order to estimate the D/G ratio for METAL-Z galaxies.}
    \label{fig:si}
\end{figure}

\begin{figure}
    \centering
 	\includegraphics[width=\columnwidth]{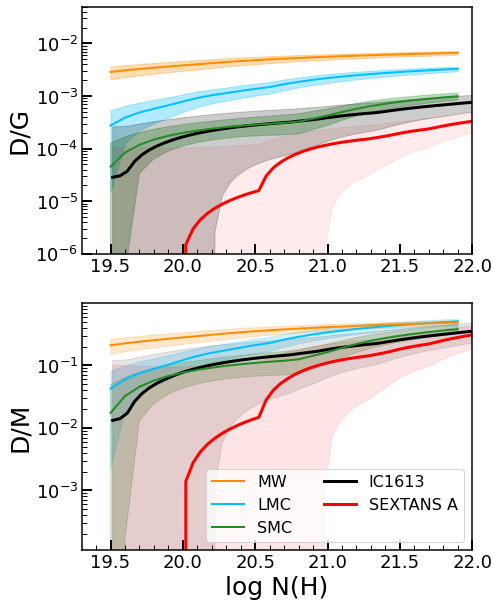}
    \caption{Dust-to-gas ratio (top) and dust-to-metal ratio (bottom) as a function of $\log$ N(H) in Sextans A (red), IC 1613 (black), the SMC (green), LMC (blue) and MW (orange). The transparent bands mark the 1$\sigma$ uncertanity. The metallicity drops from solar at MW on the top to 10\% solar with Sextans A at the bottom. D/M decreases with decreasing density and metallicity. As a result, D/G decreases faster than metallicity. IC 1613 and SMC, which have similar metallicities, have comparable D/G ratios. }
 \label{fig:dg-comp}
 \end{figure} 

\subsection{Estimation of the dust-to-gas and dust-to-metal ratios}

\begin{deluxetable}{lcccc}
\tablecaption{Abundances of elements included in D/G calculations: solar, and scaled for IC 1613 and Sextans A. \label{tab:ab_sol}}
\tablehead{\colhead{Element} & \colhead{W$_{\rm X}$} & \multicolumn{3}{c}{12 + $\log$(X/H)${\rm tot}$} \\
&&\colhead{MW\tablenotemark{a}} & \colhead{IC 1613\tablenotemark{b}} & \colhead{SEXTANS A\tablenotemark{b}} }
\startdata
\multicolumn{5}{c}{$\alpha$ elements\tablenotemark{c}} \\
\hline
C & 12.01 & 8.46 & 7.67 & 7.37 \\ 
O & 16.00 & 8.76 & 7.97 & 7.67 \\ 
Mg & 24.30 & 7.62 & 6.83 & 6.53 \\ 
Si & 28.10 & 7.61 & 6.82 & 6.52 \\ 
S & 32.06 & 7.26 & 6.47 & 6.17 \\ 
\hline
\multicolumn{5}{c}{Fe peak elements\tablenotemark{d}}\\
\hline
Ti & 47.87 & 5.00 & 4.33 & 4.01 \\ 
Cr & 52.00 & 5.72 & 5.05 & 4.73 \\ 
Fe & 55.85 & 7.54 & 6.87 & 6.55 \\ 
Ni & 58.70 & 6.29 & 5.62 & 5.30 \\ 
Cu & 63.55 & 4.34 & 3.67 & 3.35 \\ 
Zn & 65.40 & 4.70 & 4.03 & 3.71 \\ 
\enddata
\tablerefs{(a) \citet{morton2003}} 
\tablenotetext{b}{See Table \ref{tab:stellar_ab} for references to stellar abundances.}
\tablenotetext{c}{Alpha elements abundances were scaled from O abundance in IC 1613 and Mg in Sextans A}
\tablenotetext{d}{Iron peak abundances were scaled from Fe abundance in both IC 1613 and Sextans A}
\end{deluxetable}

\indent From this point of our analysis, we make the fundamental assumption that the relation between the depletion of Fe and other elements does not vary with metallicity in order to infer D/G and D/M in IC 1613 and Sextans A, using Equation 8 of \citet{romanduval2022a}:

\begin{equation}
\label{eq:dg}
    D/G = \frac{1}{1.36} \sum_{X} \big(1 - 10^{\delta(X)} \big) \bigg( \frac{N(X)}{N_H} \bigg)_{\rm tot} W(X).
\end{equation}

\noindent where $N$(X)/$N$(H$_{\rm tot}$) is the total abundance of an element X in the galaxy and $W$(X) is the atomic weight of element X. We include different dust-building elements in the computation of D/G: C, O, Mg, Si, S, Ti, Cr, Ni, Cu, Zn and Fe. The dust-to-metal ratio D/M gives the fraction of the mass of metals locked in dust. It can be derived from D/G = D/M $\times$ Z, or directly from Equation 9 of \citet{romanduval2022a}: 

\begin{equation}
\label{eq:dm}
D/M = \frac{ \sum_{X} \big(1 - 10^{\delta(X)} \big) \bigg( \frac{N(X)}{N_H} \bigg)_{\rm tot} W(X)}{\sum_{X} \bigg( \frac{N(X)}{N_H} \bigg)_{\rm tot} W(X)} .
\end{equation}

\indent We use the photospheric abundances of Fe and S in B supergiants listed in Table \ref{tab:stellar_ab}, and assume [X/H] = [Fe/H] for iron-peak elements (Fe, Zn, Ni, Cr), and [X/H] = [Mg/H] (Sextans A) or [O/H] (IC 1613) for $\alpha$ elements (C, O, Si ,Mg, Ti). Such derived abundances, together with reference solar abundances, are listed in Table \ref{tab:ab_sol}.

\indent To compute the depletions of elements other than Fe, we assume that the relation between $\delta$(Fe) and $\delta$(X) in IC 1613 and Sextans A is the same as for the MW \citep{jenkins2009}, where depletion for all those elements are measured. This assumption is based on the facts that 1) the relation between Fe and S depletion for METAL-Z galaxies is consistent with that derived for the MW and Magellanic Clouds (see Section \ref{sec:dep}), and 2) that \citet{romanduval2022a} showed that the relation between the depletion of different elements and that of Fe does not vary significantly with metallicity (except for Mg, Ti and Mn, see Section \ref{sec:dep}). Therefore, as a 0th-order approximation, we assume the same is true for METAL-Z galaxies. Thus, we compute $F_*$ from the fitted relation of $\delta$(Fe) vs $\log$ N(H) (described in Section \ref{sec:nhdep}) in IC 1613 and Sextans A using the $A_{\rm{Fe}}$, $B_{\rm{Fe}}$ and $z_{\rm{Fe}}$ given in \citet{jenkins2009}. The $F_{*}$ parameter, introduced by \citet{jenkins2009} to describe the collective behavior of the depletion in MW, provides a good reference frame for comparing depletion of different elements. The $F_{*}$ relates to depletion through a linear relation.

\begin{equation}
\delta(X) = A_x \times (F_{*}  - z_x) + B_x 
\end{equation} 

\noindent where the $A_x$ (slope), $B_x$ (zero-point), and $z_x$ (zero-point reference displacement) coefficients are given in \citet{jenkins2009}. Then, we calculated $\delta(X)$ from $F_{*}$ for other elements (Si, Mg, S, C, O, Zn, Ti, Cr, Cu, and Ni) using the relevant $A_x$, $B_x$ and $z_x$ coefficients derived in \citet{jenkins2009} for the MW.

\indent We obtain relations between $\delta$(X) and $\log$ N(H) for all elements accounted for in the D/G. We do not allow positive depletion values; extrapolated relations between depletion and gas density are set to zero for $\delta(X) > 0$. 

\indent We use a Monte Carlo method to propagate uncertainties from the fitted relation between $\log$ N(H) and $\delta$(Fe) to the extrapolated relations between $\delta$(X) and $\log$ N(H). We sampled from the Gaussian distributions of the coefficients from the relations between $\delta(Fe)$ -- $F_{*}$ and $\delta(X)$ -- $F_{*}$. Assuming that the intercept and slope of the relationship are not covariant, we generated all possible relations that fit the data points and selected the one at $1\sigma$ deviation from the linear relationship used as the final uncertainty.

\indent Figure \ref{fig:si} shows the relations between depletion of Si and Mg scaled from $\delta(\rm Fe)$ and $\log$ N(H) in the MW, LMC, SMC (solid lines) and IC 1613 and Sextans A (dashed lines). For Si, IC 1613 shows a similar relation to the LMC, while Sextans A has a similar slope but a different zero point (as expected, since metals become less depleted with decreasing metallicity). For Mg, the extrapolated relation for METAL-Z galaxies has different slopes from the MW, LMC, and SMC. This difference arises from the slight slope variations between the LMC and MW relations between $\delta$(Mg) and $\log$ N(H). Si is not as affected because the relation between $\delta$(Si) and $\delta$(Fe) has an almost identical slope in the MW and LMC. This example shows how subtle differences between elements can be amplified under our assumptions. 

\indent With the $\delta(X)$ -- $\log$ N(H) relations in hand, we can sum the contributions of different elements to estimate D/G and D/M using Equations \ref{eq:dg} and \ref{eq:dm}. 

\indent We remind the reader of two caveats. First, depletions for O and C have never been observed outside the MW. Therefore, the variations of the contributions of these two elements to the dust mass and composition with metallicity are uncertain. This limitation constitutes a source of significant systematic uncertainty in our D/G estimates, particularly because C and O constitute the main reservoir of metal mass in the ISM. Second, the relative invariance of the relation between the depletion of different elements with metallicity has never been tested below SMC metallicity (except for Fe and S, see Section \ref{sec:dep}).
 
\begin{deluxetable*}{l| cc | cc}
\tablecaption{D/G and D/M measurements for different gas column densities in IC 1613 and Sextans A. \label{tab:dg}}
\tablehead{\multicolumn{3}{c}{IC 1613} & \multicolumn{2}{c}{Sextans A}  \\ \hline \colhead{$\log$ N(H)}  & \colhead{D/G} &\colhead{D/M}  & \colhead{D/G} & \colhead{D/M}  }
\startdata
20.3 cm$^{-2}$ \tablenotemark{a} & (2.26 +/- 2.23)$\times10^{-4}$ &0.10 +/- 0.10 & (0.12 +/- 1.02)$\times10^{-4}$ & 0.01 +/- 0.09 \\ 
21 cm$^{-2}$ & (3.81 +/- 2.50)$\times10^{-4}$ &0.18 +/- 0.12 & (1.15 +/- 1.14)$\times10^{-4}$ & 0.11 +/- 0.11 \\ 
22 cm$^{-2}$ & (7.56 +/- 2.69)$\times10^{-4}$ &0.35 +/- 0.12 & (3.28 +/- 1.36)$\times10^{-4}$ & 0.30 +/- 0.13 \\ 
\enddata
\tablenotetext{a}{D/G and D/M at the lowest measured gas column density, $\log$ N(H) = 20.26 cm$^{-2}$ for IC 1613 and $\log$ N(H) = 20.38 cm$^{-2}$ for Sextans A.}
\end{deluxetable*}

\subsection{The variations of D/G and D/M with metallicity and gas density}

\indent We show the evolution of D/G and D/M with gas column density ($\log$ N(H)) in Figure \ref{fig:dg-comp}. We compare the relation between D/G (resp. D/M) and gas column density in the MW, LMC, SMC, IC 1613, and Sextans A in the top (resp. bottom) panel of Figure \ref{fig:dg-comp}. \\

\indent D/M decreases with decreasing gas density and metallicity. In addition, the variation of D/M with gas density is substantially steeper at low metallicity than at solar metallicity. For example, D/M varies from 0.52 down to 0.3, or a factor of 1.7, in the MW between $\log$ N(H) $=$ 22 cm$^{-2}$ and $\log$ N(H) = 20 cm$^{-2}$ \citep{romanduval2022a}. For IC 1613 and Sextans A, where we do not have measurements down to $\log$ N(H) = 20 cm$^{-2}$, we quote the D/G and D/M at the lowest measured column density $\log$ N(H) = 20.3 cm$^{-2}$. In Sextans A at 10\% solar metallicity, D/M decreases from 0.30 at $\log$ N(H) = 22 cm$^{-2}$ to 0.01 at the lowest $\log$ N(H) probed by the sample, or 20.3 cm$^{-2}$ (factor 30 decrease). In IC 1613, which has $\sim$15\% solar metallicity, D/M decreases from 0.35 to 0.10, or a factor 3, over $\log$ N(H) = 22 to 20.3 cm$^{-2}$ (see Table \ref{tab:dg}). \\

\indent From the variations reported in Figure \ref{fig:dg-comp} and in Table \ref{tab:dg}, as well as those reported in Table 4 of \citet{romanduval2022a}, it is also evident that the zero-point of D/M for a given column density also decreases with decreasing metallicity. At the highest column densities ($\log$ N(H) = 22 cm$^{-2}$), D/M in Sextans A is 1.7 times lower than in the MW, while in the diffuse ISM ($\log$ N(H) $=$ 20.3 cm$^{-2}$), D/M in Sextans A is 30 times lower than in the MW. \\

\indent The steeper variations of D/M with gas density at lower metallicity stems from the combination of two effects. First, the fraction of an element in dust is given by (1 - 10$^{\delta(\rm{X})}$), and the function $y(x)$ $=$ (1-10$^x$) takes a steep downturn when x approaches zero from negative values. As a result, the fraction of X in dust varies from 0 to 50\% between depletion values of 0 and -0.3. Second, as the metallicity decreases, the zero-point of the depletions becomes less negative (i.e., metals are less depleted). As a result, the gas column density at which elements become un-depleted ($\delta(X)$ = 0) moves to higher values as metallicity decreases (see the location of the kinks in Figure \ref{fig:nhdep-fit} moving to the right as the galaxy metallicity decreases). The combination of these two effects leads to the fraction of various elements in dust, and therefore D/M, dropping steeply over an increasingly wide range of gas column densities as the metallicity decreases. The direct consequence of this effect is that, at low metallicity, the variations of D/M, and subsequently D/G, over the range of gas column densities between $\log$N(H) = 20 cm$^{-2}$ and 22 cm$^{-2}$ can be significantly larger than at higher metallicity, as illustrated in Figure \ref{fig:dg-comp}. \\

\indent Because D/G is the product of metallicity and D/M, the variations of D/M with gas density and metallicity observed in the bottom panel of Figure \ref{fig:dg-comp} are reflected in the variations of D/G seen in the top panel, with the additional imprint of the large metallicity changes between galaxies in the sample (i.e., from Sextans A at $\sim$10\% solar metallicity to the MW). Variations in D/G with gas density are larger at low metallicity than at solar metallicity (similar to variations in the D/M). And because D/M decreases with decreasing metallicity, the D/G decreases faster than metallicity. At the extreme end of $\log$ N(H) = 20.3 cm$^{-2}$, D/G in Sextans A (0.12$\times$10$^{-4}$, Table \ref{tab:dg}) is over 300 times lower than in the MW \citep[3.83$\times$10$^{-3}$, see][]{romanduval2022a}. At the higher column density end, this difference is reduced to a factor of 20 (3.28$\times$10$^{-4}$ in Sextans A versus 6.74$\times$10$^{-3}$ in the MW). 

\indent The increase in D/M and correspondingly D/G with gas density for a fixed metallicity results from the evolution of the steady-state, equilibrium abundance between one the one hand, dust formation in stellar sources (AGB stars, SNe) and dust growth in the ISM through the accretion of gas-phase metals onto dust grains, and on the other hand, dust destruction by SN shocks and dust dilution by pristine inflows. The timescale for dust growth is inversely proportional to gas density \citep{asano2013, feldmann2015, zhukovska2016}. In turn, the timescale for dust destruction by SN shocks increases as the gas density increases, because SN shocks cannot propagate at high speeds in a dense medium, and because dust destruction only occurs at speeds $>$ 50 km s$^{-1}$ \citep{jones1996}. As a result, the D/M and D/G, which result from the steady-state balance between dust growth and destruction, increase with increasing gas density.\\
\indent In addition, because D/M is lower in low metallicity galaxies than the MW, with an even more pronounced difference in the diffuse ISM, there are more metals in the gas-phase available to accrete onto dust grains in those environments. This larger reservoir of potential dust components in the gas results in a steeper and wider variation of the D/M and D/G from the diffuse to dense ISM, where a significant fraction of the gas-phase metals have accreted onto dust grains.

\begin{figure*}
	\includegraphics[width=2\columnwidth]{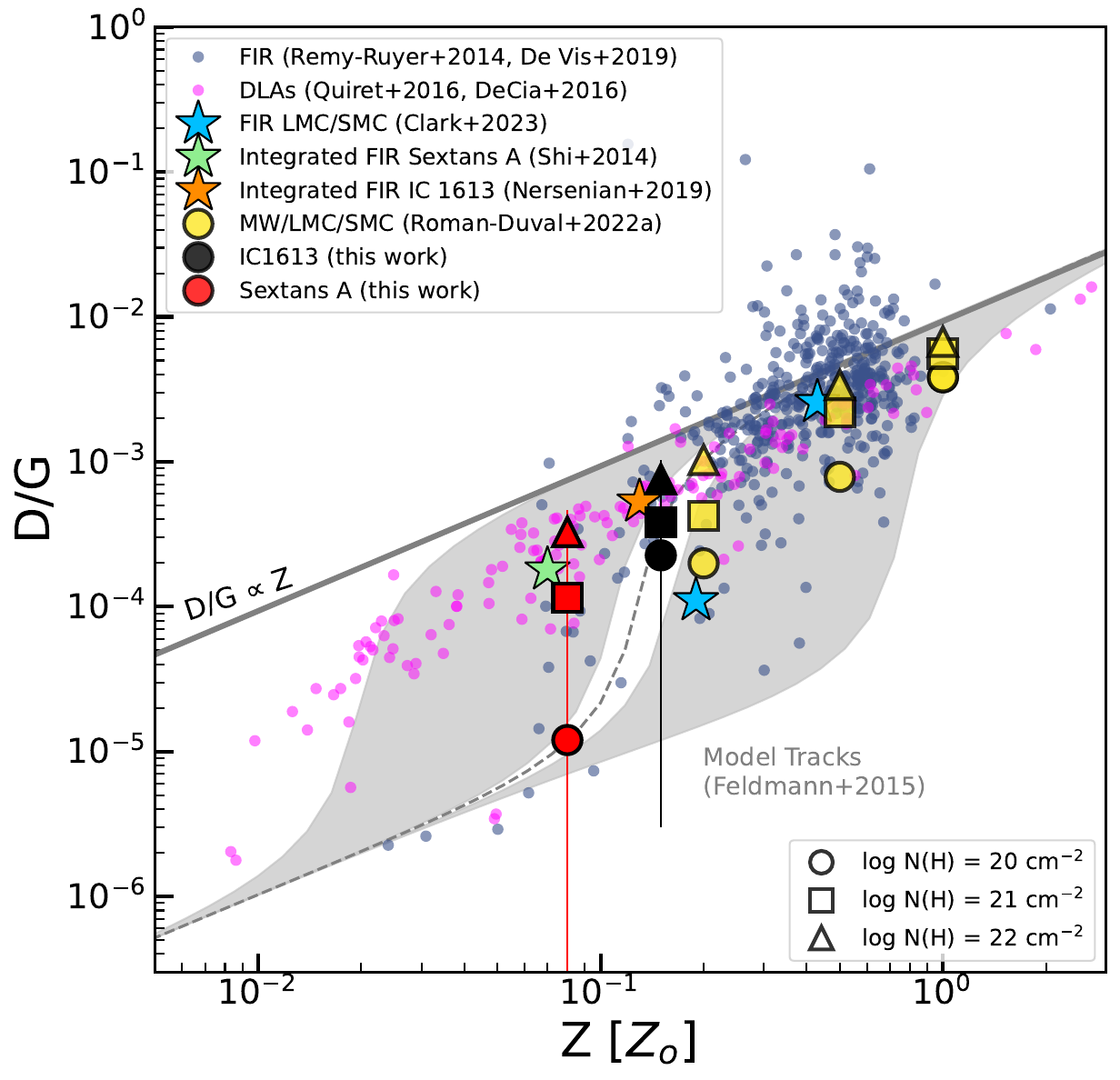}
    \caption{Dust-to-gas mass ratio as a function of (total) metallicity for different samples and types of observations. The dark blue points correspond to FIR emission measurements in nearby galaxies from \citet{remyruyer2014, devis2019}, LMC/SMC \citep{clark2023}, IC 1613 \citep{nersenian2019}, and Sextans A \citep{shi2014}. The mean hydrogen column densities over the apertures used for the FIR-based D/G measurements in IC 1613 and Sextans A are 4.04$\times$10$^{20}$cm$^{-2}$ and 1.4$\times$10$^{21}$cm$^{-2}$, respectively. In magenta are D/G estimates in DLAs \citet{decia2016, quiret2016} from \citet{romanduval2022b}. Dust depletion measurements from this work (IC 1613 in black, Sextans A in red) and from the literature (MW, LMC, SMC in yellow \citet{romanduval2022a}) are shown as symbols for three hydrogen column density values: $\log$ $N$(H) = 20 cm$^{-2}$ (circle), 21 cm$^{-2}$ (square), and 22 cm$^{-2}$ (triangle). Lastly, the gray shaded area shows the chemical evolution model from \citet{feldmann2015} for a range of $\gamma$ parameters ($2\times 10^3 -10^6$). The dashed black line represents the fiducial $\gamma$ = 3$\times$10$^4$ model, in best agreement with the trend obtained from FIR measurements. A solid grey line marks the linear relation between D/G and metallicity. }
    \label{fig:dg}
\end{figure*}

\subsection{Variations of D/G with metallicity: Comparison of different observational constraints}

\indent In Figure \ref{fig:dg}, we present a comparison of the evolution of D/G with metallicity between chemical evolution models \citep{feldmann2015} and different observational estimates of D/G: Depletions derived from UV spectroscopy in nearby galaxies \citep[this work for IC 1613 and Sextans A][and references therein for the MW, LMC, SMC]{romanduval2022a}; FIR + 21 cm + CO (1-0) emission maps of nearby galaxies \citep{remyruyer2014, devis2019}; depletions derived from rest-frame UV spectroscopy and calibrations of abundance ratios in DLAs \citep{decia2016, quiret2016, Peroux2020, romanduval2022b}. \\

\indent D/G estimates from depletions in the MW, LMC, and SMC \citep{romanduval2022a} and this work (IC 1613 and Sextans A) are plotted for different $\log$ N(H) values ($\log$ N(H) = 20, 21, 22 cm$^{-2}$ in Figure \ref{fig:dg}. \\

\indent Theoretical models from \citet[][gray]{feldmann2015} are also shown in Figure \ref{fig:dg}, and predict a turnover and steep decrease in the D/G ratio below a critical metallicity of Z $<$ 0.1-0.2 $Z_{\odot}$. In the model, the critical metallicity is determined by the ratio $\gamma$ of the timescale for the depletion of molecular gas by star-formation, or about 2-4 Gyr \citep{bigiel2008}, to the timescale for dust growth in the ISM at solar metallicity and densities of 100 cm$^{-3}$, or about 1.5$\times$10$^5$ yr \citep{hirashita2000, feldmann2015}. For a fiducial parameter $\gamma$ = 3$\times$10$^4$, the critical metallicity is about 10\% solar. This critical metallicity marks a transition point where dust growth in the ISM becomes insufficient to counterbalance the effects of dust destruction in SN shocks and dilution through gas inflows. As a result, the D/G below the critical metallicity is driven by stellar inputs of dust (evolved stars, SNe), resulting in a low D/M and D/G. Combining D/G measurements across a wide range of metallicities allows us to test the theoretically predicted D/G ratio dependence on metallicity.

\indent The most extensive sample of D/G measurements comes from FIR dust emission \citep[][dark blue points]{remyruyer2014, devis2019}. We also show recent FIR measurements of D/G in the LMC and SMC \citep{clark2023} at $\log$ N(H) = 21 cm$^{-2}$ (blue stars), as well as IC 1613 \citep{nersenian2019} and Sextans A \citep{shi2014}. The mean hydrogen column densities over the apertures used for the FIR-based D/G measurements in IC 1613 and Sextans A are 4.04$\times$10$^{20}$cm$^{-2}$ and 1.4$\times$10$^{21}$cm$^{-2}$, respectively. The dust surface density is derived from FIR emission, while the atomic gas and molecular gas surface densities are traced by 21 cm and CO rotational emission, respectively. FIR measurements for Z $>$ 0.1 $Z_{\odot}$ show a significant scatter around the linear relation between D/G and metallicity. At lower metallicities, FIR-based D/G measurements mostly follow the model-predicted steep turnover in D/G for $\gamma$ $\sim$ $3\times10^4$. However, the resolved FIR-measurements in IC 1613 and Sextans A tend to lie on the upper envelope of the scatter in this trend. 

\indent In DLAs, only gas-phase abundances can be measured, and stellar abundances are not known. In this case, depletions are therefore derived from the [Zn/Fe] abundance ratio tied to a calibration of $\delta$(Zn) versus [Zn/Fe] in the MW, LMC, or SMC \citep{decia2016, romanduval2022b}. Depletion-based estimates of D/G in DLAs \citep{quiret2016, decia2016, romanduval2022b} follow a slightly sub-linear relation with metallicity, even below 10\% solar. As a result, the trend of D/G versus metallicity seen in DLAs lies significantly higher than the trend derived from FIR + 21 cm + CO measurements in nearby galaxies from the \citet{remyruyer2014} and \citet{nersenian2019} samples. However, the resolved FIR measurements in IC 1613 and Sextans A are in good agreement with the trend obtained in DLAs. In addition, the D/G estimates in DLAs are higher than the predictions from the \citet{feldmann2015} chemical evolution model with a fiducial $\gamma$ parameter of $3\times10^4$. The trend of D/G versus metallicity in DLAs suggests very high values of $\gamma$, of order 10$^6$ (left edge of the gray tracks in Figure \ref{fig:dg}). This implies a very fast dust growth timescale compared to the star-formation timescale.

\indent At high metallicity (above 10\% solar, which includes IC 1613), the model and observed trends of D/G versus metallicity generally agree between all tracers and systems (FIR in nearby galaxies, UV spectroscopic depletions in nearby galaxies, and rest-frame UV spectroscopy in DLAs). We do note the factor 3-4 discrepancy between the resolved FIR-based \citep{clark2023} and depletion-based \citep{romanduval2022a} D/G in the SMC however, which might be due to a combination of factor examined later in this Section. \\

\indent At metallicities below 10\% solar (i.e., in Sextans A), the trend of D/G versus metallicity obtained from depletions lies in excellent agreement with the trend seen in DLAs for column densities $\log$ N(H) = 21-22 cm$^{-2}$. The FIR-based resolved D/G measurement in Sextans A, which was derived in an aperture where the gas column density is about 1.4$\times$10$^{21}$ cm$^{-2}$, is also in excellent agreement with the depletion-based measurement for those gas column densities. However, at hydrogen column densities of $\log$ N(H) $\sim$ 20 cm$^{-2}$, the depletion-based D/G estimate in Sextans A lies almost perfectly on the trend of D/G versus metallicity obtained from large sample of FIR measurements \citep{remyruyer2014, devis2019}, a significant fraction of which are unresolved.

\indent In a nutshell, there is a general tension between two types of constraints for the trend of D/G versus metallicity: from large samples of FIR measurements in nearby galaxies on the one hand, which suggest a steep drop in D/M and D/G at a critical metallicity of about 20\% solar; and from rest-frame UV spectroscopy, both in local and high-redshift systems on the other hand, which suggest a slightly sub-linear relation between D/G and metallicity, but does not exhibit any such steep drop in D/G below 10\% solar metallicity. There are, however, exceptions to this discrepancy: the resolved FIR measurements of D/G in Sextans A at 7\% solar metallicity are in lie on top of the trend derived from DLAs, and are in excellent agreement with the depletion-based D/G estimates for $\log$ N(H) $>$ 21 cm$^{-2}$; Conversely, the depletion-based D/G estimates for $\log$ N(H) $\sim$ 20 cm$^{-2}$ in Sextans A lie on top of the trend of D/G versus metallicity obtained from large samples of FIR measurements.\\

\indent Chemical evolution models can be tuned to match both trends by adjusting the ratio of the timescale for the consumption of molecular gas by star-formation to the timescale for dust growth in the ISM. The $\gamma$ value in best agreement with the trend observed in the FIR is $\gamma$ $\sim$ $3\times10^4$, while the trend seen from depletions is consistent with $\gamma$ $=$ 10$^6$, suggesting a much faster dust growth timescale compared to the gas consumption by star-formation. In this case, dust could keep growing in low-metallicity environments.

\indent A few culprits could explain the tension between FIR-based and UV-spectroscopy-based estimates of the trend of D/G versus metallicity, and the occasional discrepancy between D/G estimates from depletions and the FIR for a given galaxy (e.g., the SMC). As pointed out in \citet{romanduval2022b}, the conversion of FIR emission to dust mass relies on an assumed FIR opacity, which is only constrained observationally in the MW and has been shown from models and laboratory studies to depend on the composition, size, and fractal aspect of dust grains \citep[e.g.][]{demyk2017}. However, this FIR opacity is unconstrained observationally and may well vary significantly with metallicity and density, since there is strong evidence that the composition, grain size, and optical properties of dust change significantly within and between galaxies. Examples of such evidence include studies based on the FIR \citep[M31, M33, M101, M74, M83, and other nearby galaxies see][]{fritz2012, smith2012, relano2018, chiang2018, clark2019, lamperti2019}, UV dust extinction curves in the Magellanic Clouds \citep{gordon2003}, and depletion studies in the Magellanic Clouds \citep{romanduval2022a}. \\

\indent Furthermore, many 21 cm and FIR observations involved in the determination of the D/G in the \citet{remyruyer2014} and \citet{devis2019} samples are either unresolved or do not have aperture matching between the dust mass and gas mass estimates. It is, therefore, entirely possible that the D/G estimates obtained with this method may be underestimated due to the dilution of the dust mass in a larger, diffuse gas volume of low average column density. In this scenario, the steep variations of D/G with gas density at low metallicity, such as the one seen from depletions in Sextans A, could be playing a role in the apparent discrepancy between the trend of D/G versus metallicity derived from FIR and from depletions in nearby galaxies and DLAs. In particular, the unresolved measurements in galaxies where the dust disk could likely occupy a smaller volume than the gas disk could well be affected by a dilution of the dust component in the larger diffuse gas component, resulting in an artificially low D/G estimate (corresponding to a low average gas column density).\\
\indent The D/G estimate from UV-spectroscopy is not immune to systematics either. As mentioned earlier in this paper, the large contribution of C and O to the dust budget is highly uncertain owing to the sparsity and uncertainties of C measurements in the MW and the lack of C and O depletion measurements altogether outside of the MW. It is entirely possible that variations in nucleosynthetic histories, particularly with metallicity, could lead to varying abundance ratios and chemical affinities for dust grains. This would result in departures in the relation between Fe depletions and depletions of other elements from the MW relations. This could lead to large (but not quantifiable) systematic uncertainties in D/G estimates based on depletions. \\

\indent Nevertheless, the METAL-Z large Hubble program shows the potential for depletion measurements in low metallicity systems to help resolve the tension between D/G estimates from the FIR and from spectroscopy of DLAs. Depletion measurements in nearby low metallicity galaxies where both interstellar and stellar abundances can be measured can be especially impactful for these calibrations as DLAs are often used to trace the metal enrichment at higher redshift, where low metallicity systems are more common. Additional measurements in Sextans A and other low metallicity dwarfs in the Local Group alongside robust stellar abundances will further constrain relations between elemental depletion and dust-to-gas ratio, deepening our understanding of dust at metallicity below 20 \% solar. 



\section{Conclusions}
\label{sec:concl}
\indent We presented the results of the METAL-Z Hubble program targeting 18 sightlines toward massive stars in two local low metallicity dwarf galaxies, IC 1613 (10\% solar oxygen metallicity, 20\% solar iron metallicity) and Sextans A (10\% solar metallicity). We used COS FUV spectroscopy toward those massive stars to measure gas-phase column densities of hydrogen, iron, and sulfur and, in turn, calculate the gas-phase abundances and elemental depletion of Fe and S.

\indent The relation between the depletions of Fe and S in IC 1613 and Sextans A is consistent within errors with the relations between these elements established in the MW, LMC, and SMC. 

\indent Similarly to findings in the MW, LMC, and SMC, the depletions of Fe in IC 1613 and Sextans A show a declining trend with increasing gas column density, indicating that the fraction of Fe locked in dust increases when the gas density of the ISM increases due to dust growth in the ISM through the accretion of gas-phase metals onto dust grains. We also found that the base level of Fe depletion increases with decreasing metallicity. At $\log$ N(H) = 21 cm$^{-2}$, the Fe depletion value increases by 0.55 dex from the MW to IC 1613 and by 0.85 dex from the MW to Sextans A. This corresponds to a reduction of the fraction of Fe in dust from 0.97 to 0.81 between the MW and Sextans A at $\log$ N(H) = 21 cm$^{-2}$, and implies that metals become less depleted from the gas as the metallicity of a galaxy decreases. At lower column densities ($\log$ N(H) = 20.3 cm$^{-2}$), the reduction in the fraction of Fe in dust from the MW to Sextans A is even more dramatic, going from 0.97 to 0.32 (the fraction of Fe in dust decreases from 0.97 in MW to 0.74 in IC 1613 at this column density).


\indent To estimate the depletions of other, unobserved metals, we used the relation between the depletion of Fe and the depletion of other elements measured in the MW, where abundances and depletions for a complete set of elements can be obtained. This assumption is based on the findings that 1) the relation between the depletions between various elements appears relatively invariant with metallicity in the MW, LMC and SMC, and 2) the relation between the depletions of Fe and S in IC 1613 and Sextans A is consistent with that found in the MW, LMC, and SMC. The caveat to this assumption is two-fold. First, the depletion of C and O, which are major mass reservoirs of metals for dust, have never been measured outside the MW. Given that the stellar C/O ratio in low metallicity galaxies tend to be lower than in the MW, the chemical affinities of dust grains could change in low metallicity systems, which would result in a different relationship between the C or O depletion and the depletion of Fe at low metallicity. Second, the metallicity invariance of the relation between the depletions of different element has not been tested at metallicities lower than the SMC (20\% solar), other than for Fe and S (this work). Nevertheless, this assumption allows us to estimate the dust-to-gas and dust-to-metal ratios from the depletions of Fe measured in IC 1613 and Sextans A.




\indent Based on the fundamental assumption that the depletion of Fe and other elements in Sextans A and IC 1613 follow the same relations as in the MW, we estimate the depletions of all constituents of dust as a function of $\log$ N(H) and derive the dust-to-gas and dust-to-metal ratios in IC 1613 and Sextans A, as a function of $\log$ N(H). As indicated by the trends obtained from Fe alone, we find that D/M increases with increasing column density, by factors of 3.5 and 35 in IC 1613 and Sextans A, respectively, between $\log$ N(H) = 20.3 and 22 cm$^{-2}$. These variations imply that dust growth does occur at low metallicity, resulting in significant changes in the dust abundance with the ISM density.

\indent In addition, D/M decreases with decreasing metallicity for a given gas column density, from $0.41 \pm 0.5$ in the MW at $\log$ N(H) = 21 cm$^{-2}$ to $0.28 \pm 0.12$ in IC 1613 and to $0.19 \pm 0.13$ in Sextans A at the same column density. The variation of D/M with metallicity is more pronounced at lower column density, with a factor 40 variation from the MW to Sextans A at $\log$ N(H) = 20 cm$^{-2}$. As a result of the varying D/M, D/G decreases slightly faster than metallicity (but not as fast as observed in the FIR). 

\indent We compare the evolution of D/G with metallicity derived from depletion-based estimates in this work (Sextans A, IC 1613) and previous work (SMC, LMC, MW) with D/G estimates obtained from two different techniques and samples: 1) a combination of FIR emission to trace dust and 21 cm + CO rotational emission to trace atomic and molecular gas; and 2) spectroscopic measurements of the [Zn/Fe] abundance ratio in DLAs, from which D/G can be estimated using calibrations derived in the MW \citep{decia2016}. At metallicities below 20\% solar, the depletion-based measurements in Sextans A and IC 1613 are in good agreement with the trend obtained in DLAs, which is only slightly sub-linear with metallicity. Conversely, the trend obtained from FIR observations sees a steep decline of D/M and D/G below 20\% solar metallicity, which is not observed from depletions in nearby galaxies and DLAs. This solidifies the tension between the trends of D/G vs. metallicity derived from FIR in nearby galaxies on the one hand and rest-frame UV spectroscopy in nearby galaxies and DLAs on the other hand. 

\indent Constraining the dust abundance and the dust-to-gas ratio in low metallicity environments is crucial for understanding the ISM of galaxies and chemical enrichment across cosmic time. With this work, we showed that depletion measurements based on UV spectroscopy are a powerful tool for estimating the abundance of dust at low metallicity. In the future, we will expand the studies to more galaxies, constraining the dust abundances in a broad range of metallicity and gas density environments.

\begin{acknowledgments}

A.H., E.B.J. and K.T. acknowledge support from grant No. HST-GO-15880. This work is based on observations with the NASA/ESA Hubble Space Telescope obtained at the Space Telescope Science Institute, which is operated by the Associations of Universities for Research in Astronomy, Incorporated, under NASA contract NAS5-26555. These observations are associated with program 15880. Support for Program number 15880 was provided by NASA through a grant from the Space Telescope Science Institute, which is operated by the Association of Universities for Research in Astronomy, Incorporated, under NASA contract NAS5-26555. M.G. acknowledges support by grants PID2019-105552RB-C41, PID2022-137779OB-C41 and PID2022-140483NB-C22, and MDM-2017-0737 Unidad de Excelencia ’Maria de Maeztu’ - Centro de Astrobiologia (CSIC-INTA), funded by MCIN/AEI/10.13039/501100011033. The HST data presented in this article were obtained from the Mikulski Archive for Space Telescopes (MAST) at the Space Telescope Science Institute. The specific observations analyzed can be accessed via \dataset[DOI: 10.17909/1hzc-8k61]{https://doi.org/10.17909/1hzc-8k61}.

\end{acknowledgments}

\appendix
\twocolumngrid
\section{Metal column densities measured with Curve of Growth}
\label{sec:app-cog}
\indent In this appendix, we provide an overview of our metal line column density measurements achieved by utilizing the Curve of Growth (CoG) method. The Curve of Growth method is a classical and widely used approach for determining column densities of absorption lines. Nevertheless, our analysis reveals that when applied to the low S/N COS spectra of the METAL-Z sample, this method proves insufficient in yielding conclusive measurements.

\subsection{Equivalent width measurements}
\label{sec:eqw}

\indent The Curve of the Growth method uses equivalent widths of the lines to calculate the column density of the ion. To measure equivalent widths, we first had to determine the integration limits for each line. We assumed that all ISM lines have the same velocity dispersion towards a given sightline. We then used the \ion{S}{2} $\lambda$1253 \AA~ (IC 1613) or 21 cm HI profile towards a particular sightline \citep[Sextans A, from LITTLE-THINGS survey][]{ott2012} to determine the velocity integration limits and then apply those limits to all lines. We chose \ion{S}{2} $\lambda$1253 \AA~ as it is a medium-strength isolated line, for which our approach worked best. However, for Sextans A, where all metal lines are weaker (due to the lower abundance), we found that using the strong 21 cm \ion{H}{1} line provides the most optimal results.

\indent We designed a method to determine integration limits in an objective, automated, and reproducible process. First, we calculated the derivative of the \ion{S}{2} $\lambda$1253 \AA~ (or 21 cm \ion{H}{1}) line equivalent width as a function of the distance from the line center. The resulting derivative curve grows from the center of the line outwards until it starts to plateau, indicating that the flux has reached the continuum level. We fit a Gaussian function to the derivative curve for each sightline and use 4$\sigma$ to mark the place where the line recovers to the continuum. We found that the 3$\sigma$ cut-off underestimates the contribution from line wings, while 5$\sigma$ often incorporates wings of neighboring lines. Therefore we settled with 4$\sigma$ as the most robust representation of the limit between the line and continuum dominant contribution to the equivalent width measurement. 

\begin{figure*}

	\includegraphics[width=\columnwidth]{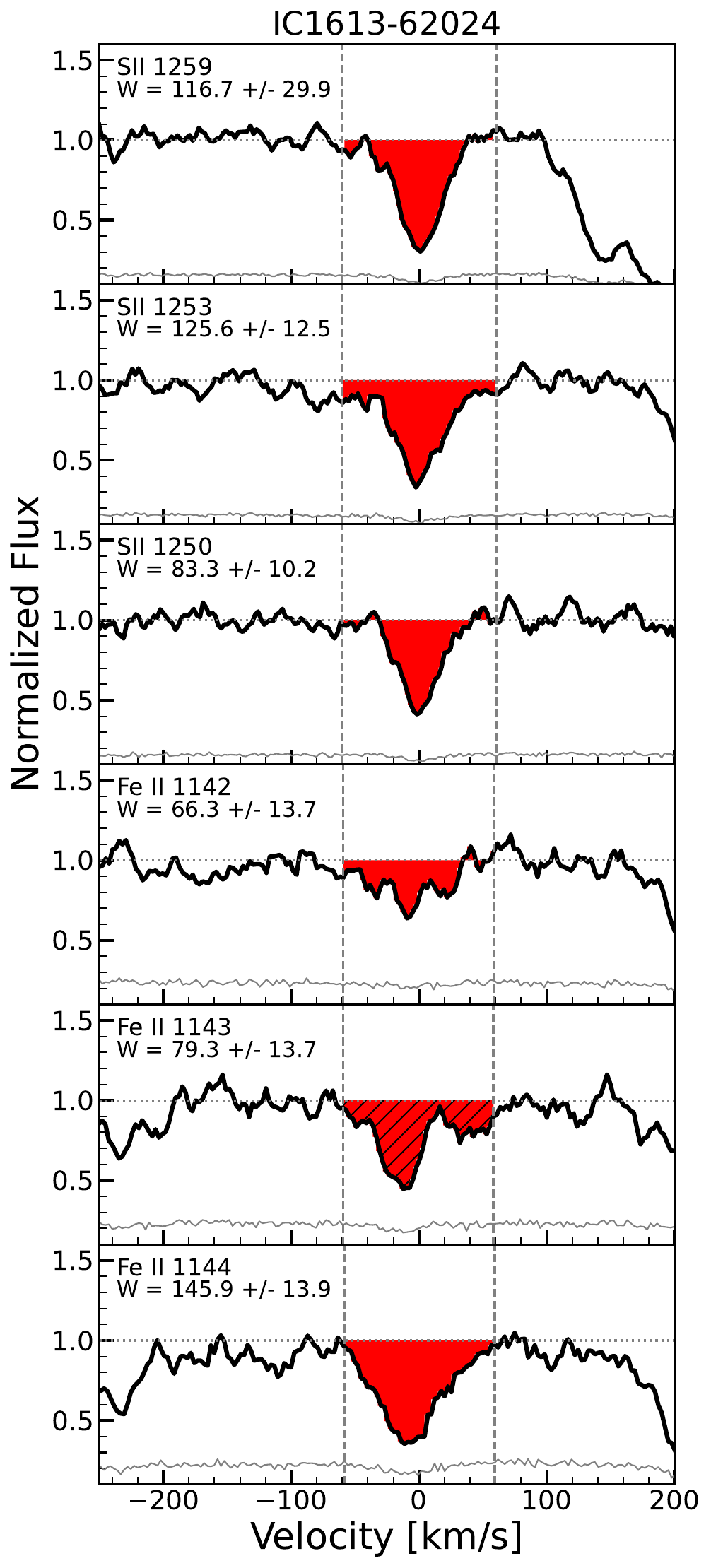}
	\includegraphics[width=\columnwidth]{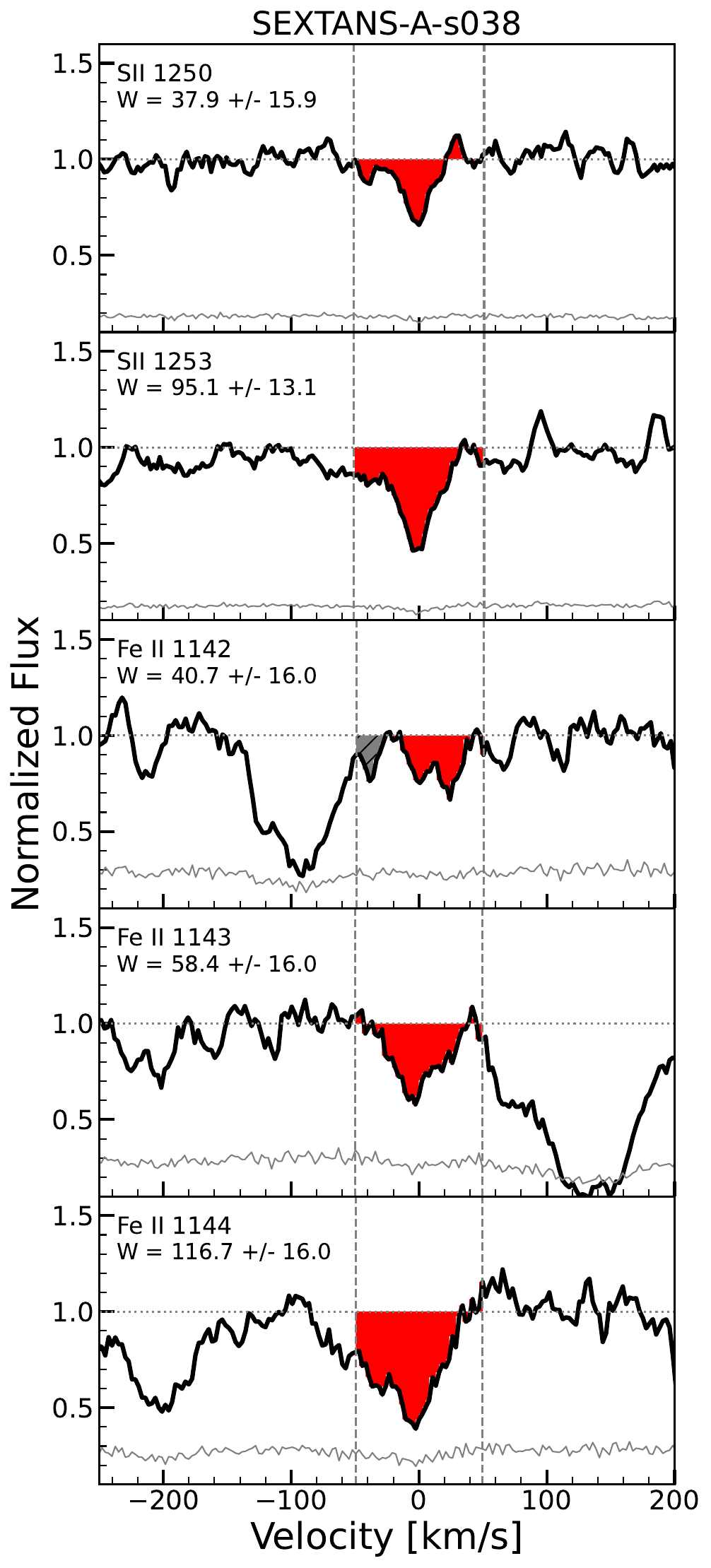}

    \caption{Examples of line measurements for selected sightlines in IC 1613 and Sextans A. We show all lines used for abundance measurements. In black, we plot smoothed, continuum--subtracted spectra centered at the line in the velocity frame of the particular galaxy. Dashed vertical lines mark the equivalent width integration limits, while the red fill shows the whole integration range. In IC 1613, \ion{Fe}{2} $\lambda$1143 \AA~ line is completely blended with MW \ion{Fe}{2} $\lambda$1142 \AA~ line and is hatched (see Section \ref{sec:eqw} for full discussion). For Sextans A, we do not show \ion{S}{2} $\lambda$1259 \AA\ as it is contaminated by an MW line. In grey, we mark contaminated line regions excluded from the measurements.}
    \label{fig:linestack}   
\end{figure*}

\indent Once integration limits are determined for all sightlines, we can proceed with equivalent width measurements for S and Fe lines. We measured three \ion{Fe}{2} lines: \ion{Fe}{2} $\lambda\lambda\lambda$1142, 1143, 1144 \AA, (see Figure \ref{fig:linestack} and Table \ref{tab:lines}). For four sightlines covered by the G160M, $\lambda$1608 \AA~ was also available, and we included it in the measurements. The short wavelength part of the spectrum proved challenging due to the dropping sensitivity of COS. Therefore for ten sightlines, the weak \ion{Fe}{2} $\lambda$1142 \AA~ line remains undetected, and we report upper limits on the equivalent width. In Sextans A, the \ion{Fe}{2} $\lambda$1142 \AA~ line is slightly blended with the Milky Way's \ion{Fe}{2} $\lambda$1143 \AA\ line. For equivalent width measurements of this line, we had to manually mask the contaminated part of the line within automatically selected velocity limits (the masked region is marked in Figure \ref{fig:linestack}). 

\indent Due to the velocity difference between IC 1613 and the MW, the \ion{Fe}{2} $\lambda$1143 \AA~ line originating from IC 1613 completely blends with the Milky Way's \ion{Fe}{2} $\lambda$1142 \AA\ line. As we cannot disentangle them from profile shapes, we used other \ion{Fe}{2} lines originating from the MW to estimate the equivalent width of the blended MW $\lambda$1142 \AA~ line. Specifically, we measured the equivalent width of the $\lambda$1143 and $\lambda$1144 \AA~ MW lines and fitted a Curve of Growth (details of the method in Section \ref{sec:cog}). We found that the blended \ion{Fe}{2} $\lambda$1142 \AA~ and $\lambda$1143 \AA~ lines in both IC 1613 and MW are optically thin - they lay on the linear part of the Curve of Growth. In this regime, the equivalent width of blended lines equals the sum of their individual equivalent widths. Therefore, we estimated the equivalent width of the IC 1613's $\lambda$1143 \AA\ line by subtracting the MW's $\lambda$1142 \AA~ equivalent width from the one measured over the blended line. We present the measurements for each step of this procedure (equivalent widths of MW lines, blend, results of Curve of Growth and derived \ion{Fe}{2} $\lambda$1143 \AA~for IC 1613) in Table \ref{tab:blend}.  

\indent For the second measured element, S, we cover three \ion{S}{2} lines: $\lambda\lambda\lambda$1250, 1253, 1259 \AA. In Sextans A, the latter is blended with the stronger MW \ion{Si}{2} $\lambda$1259 \AA~ absorption line. Therefore for that galaxy we only report measurements for the \ion{S}{2} $\lambda$1250 \AA~ and $\lambda$1253 \AA. 

\indent We summarize the measurements of equivalent width for all lines in the online materials, for which an excerpt is shown in Table \ref{tab:eqw}.

\indent COS Line Spread Functions (LSF) have broad wings that can contain up to 30\% of the observed flux \citep{coslp}. To test whether the limiting velocity selection influences our equivalent width measurement, we have integrated the LSF outside the line integration limits. We found that the fraction of missing flux ranges between 1-10\% for most sightlines, well within the flux measurement uncertainty. The fractions of flux missing due to the LSF wings for each sightline can be found in Table \ref{tab:lognh}, but it is not incorporated into final measurements as it is smaller than measurement errors. 

\begin{deluxetable}{llcrcc}
\tablecaption{Equivalent width measurements of \ion{Fe}{2} and \ion{S}{2} lines in METAL-Z targets. \label{tab:eqw}}
\tablehead{\colhead{Sightline} & \colhead{Ion} & \colhead{$\lambda$} & \colhead{$W_{\lambda}$} & \colhead{$v_{\rm limit}$\tablenotemark{a}} & \colhead{LSF\tablenotemark{b}} \\ & & \colhead{\AA} & \colhead{m\AA} & \colhead{kms$^{-1}$} & \colhead{\%}}
\startdata
IC1613-61331 & \ion{S}{2} & 1250.578 & 66.21 $\pm$ 8.95 & 52  & 4 \\ 
IC1613-61331 & \ion{S}{2} & 1253.805 & 157.81 $\pm$ 13.01 & 52 & 4 \\ 
IC1613-61331 &\ion{S}{2} & 1259.518 & 108.70 $\pm$ 27.18 & 52 & 4\\ 
IC1613-61331 & \ion{Fe}{2} & 1142.366 & 8.40 $\pm$ 11.30 & 52 & 5\\ 
IC1613-61331 & \ion{Fe}{2} & 1143.226 & 67.10 $\pm$ 11.30 & 52 & 5 \\ 
IC1613-61331 & \ion{Fe}{2} & 1144.938 & 108.82 $\pm$ 11.45 & 52 & 5 \\ 
IC1613-62024 & \ion{S}{2} & 1250.578 & 83.30 $\pm$ 10.22 & 60 & 3 \\ 
IC1613-62024 & \ion{S}{2} & 1253.805 & 125.57 $\pm$ 12.50 & 60 & 3 \\ 
IC1613-62024 & \ion{S}{2} & 1259.518 & 116.72 $\pm$ 29.91 & 60 & 3 \\ 
IC1613-62024 & \ion{Fe}{2} & 1142.366 & 66.32 $\pm$ 13.73 & 60 & 3 \\ 
\enddata
\tablecomments{The full, machine-readable version of the table is available online as supplemental material}
\tablenotetext{a}{The velocity limits used for the equivalent width calculations}
\tablenotetext{b}{The fraction of missing flux due to the LSF wings outside of the equivalent width integration limits}
\end{deluxetable}

\begin{deluxetable*}{lccccc}
\tablecaption{\ion{Fe}{2} equivalent width derivation from blended IC1613/1143 + MW/1142 lines.  \label{tab:blend} }
\tablehead{\colhead{Target} &\colhead{MW $\lambda 1144$\tablenotemark{a}} & \colhead{MW $\lambda 1143$\tablenotemark{b}} &  \colhead{IC1613/$\lambda1143$ + MW/$\lambda1142$\tablenotemark{c}} & \colhead{ MW $\lambda 1142$\tablenotemark{d}} &
\colhead{IC1613 $\lambda 1143$\tablenotemark{e}} 
\\& \colhead{m\AA} & \colhead{m\AA} & \colhead{m\AA} & \colhead{m\AA} & \colhead{m\AA}}
\startdata
IC1613-61331 & 180.47 $\pm$ 19.22 & 82.68 $\pm$ 14.71 & 68.79 $\pm$ 9.51 & 21.48 $\pm$ 11.66 & 47.31 $\pm$ 15.05\\
IC1613-62024 & 184.24 $\pm$ 18.82 & 80.53 $\pm$ 15.10 & 113.66 $\pm$ 14.33 & 20.64 $\pm$ 11.76 & 93.02 $\pm$ 18.54\\
IC1613-64066 & 163.44 $\pm$ 19.64 & 66.98 $\pm$ 15.85 & 89.38 $\pm$ 12.85 & 16.55 $\pm$ 11.66 & 72.82 $\pm$ 17.35 \\
IC1613-67559 & 163.48 $\pm$ 18.55 & 79.92 $\pm$ 13.63 & 59.84 $\pm$ 11.28 & 21.70 $\pm$ 11.35 & 38.14 $\pm$ 16.00 \\
IC1613-67684 & 161.79 $\pm$ 15.77 & 78.65 $\pm$ 14.14 & 39.11 $\pm$ 8.55 & 21.24 $\pm$ 11.34 & 17.88 $\pm$ 14.20 \\
IC1613-A13 & 170.19 $\pm$ 19.28 & 78.78 $\pm$ 16.82 & 75.36 $\pm$ 12.36 & 20.49 $\pm$ 11.53 & 54.87 $\pm$ 16.91\\
IC1613-B11 & 178.22 $\pm$ 18.73 & 119.34 $\pm$ 16.32 & 42.31 $\pm$ 12.91 & 43.42 $\pm$ 11.10 & $>$ 17.03 \\
IC1613-B2 & 164.71 $\pm$ 19.45 & 107.58 $\pm$ 20.78 & 126.40 $\pm$ 13.87 & 37.60 $\pm$ 10.95 & 88.80 $\pm$ 17.68 \\
IC1613-B3 & 143.99 $\pm$ 16.64 & 42.72 $\pm$ 12.53 & 9.55 $\pm$ 12.26 & 9.39 $\pm$ 12.33 & $>$ 17.39 \\
IC1613-B7 & 180.42 $\pm$ 23.93 & 75.16 $\pm$ 16.26 & 68.83 $\pm$ 11.84 & 18.55 $\pm$ 11.82 & 50.28 $\pm$ 16.73 \\ 
\enddata
\tablenotetext{a}{Equivalent width of \ion{Fe}{2} $\lambda 1144$ from Milky Way}
\tablenotetext{b}{Equivalent width of \ion{Fe}{2} $\lambda 1143$ from Milky Way}
\tablenotetext{c}{Equivalent width of blended \ion{Fe}{2} IC 1613 $\lambda1143$ + Milky Way $\lambda1142$}
\tablenotetext{d}{Equivalent width of \ion{Fe}{2} $\lambda 1142$ from Milky Way,  derived through CoG}
\tablenotetext{e}{The deblended equivalent width of IC 1613 \ion{Fe}{2} $\lambda 1143$}
\end{deluxetable*}

\begin{figure}
	\includegraphics[width=\columnwidth]{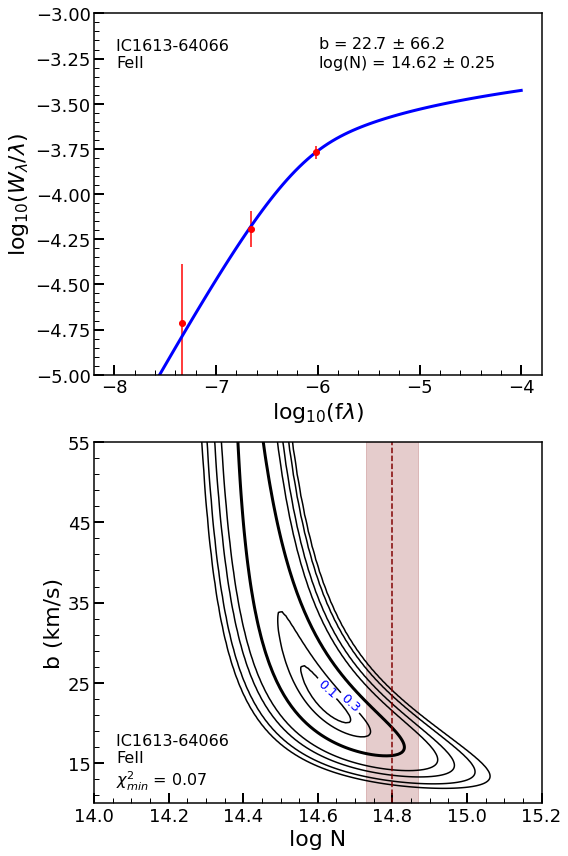}
    \caption{Curve of Growth model best-fit parameters and probability contours. The top panel shows the best-fitting model (blue curve) with measurements (red points) and corresponding parameters listed. The bottom panel displays the probability contours for the model's $N$ and $b$ parameters, with the 1$\sigma$ contour highlighted in bold. Results from Voigt profile fitting are shown as the red band.}
    \label{fig:cog-ex}
\end{figure}

\subsection{Column density measurements with the Curve of Growth method}
\label{sec:cog}

\indent In the CoG approach, for each element, we find the best fitting model for the relation between observed equivalent width and oscillator strength $f_{\lambda}$ of all measured lines. The model is given by Equations 2--4 of \citet{jenkins1996} and links the line's equivalent width with the gas physical parameters: column density $N$ and velocity dispersion $b$.

\indent The central optical depth is defined as
\begin{equation}
    \tau_0 = \frac{\pi^{1/2}e^{2}f\lambda}{m_ec} \bigg(\frac{N}{b}\bigg)
\end{equation}

The equivalent width is defined as

\begin{equation}
    \frac{W_{\lambda}}{\lambda} = \frac{2bF(\tau_0)}{c} ,
\end{equation}

where

\begin{equation}
    F(\tau_0) =  \int_{0}^{\infty}[1-\rm exp(-\tau_0 e^{-x^2})] \,dx 
\end{equation}

\indent We created a grid of models with a range of (N, b) parameters $\log$(N) = 10--20 cm$^{-2}$, with a spacing of 0.01 cm$^{-2}$  and $b$ = 5--100 km s$^{-1}$ with a spacing of 0.1 kms$^{-1}$. We then searched the grid for the best-fit model by minimizing the $\chi^2$ between the model and the measured equivalent widths for the different transitions. We calculate the $\chi^2$ as

\begin{equation}
   \chi^{2}_{ij}  = \Sigma_{k=0}^{N} \frac{(w_{k} - m_{jk})^2}{s_{k}^2},   
\end{equation}

where $w_{k}$ and $s_{k}$ are the equivalent width measurements and associated errors for all lines for a given element, m$_{jk}$ is the model value of the equivalent width for a given $N$ and $b$.

\indent The model is best constrained for three or more measurements of a particular ion. In several cases, we only had two measurements (e.g. one of the \ion{S}{2} lines in Sextans A were fully blended with a MW interloper), resulting in unconstrained $b$. For these cases, we assumed $b = 20^{+\infty}_{-20}$ kms$^{-1}$ (Table \ref{tab:col-dep}). Column density is less affected, and even with only two measurements, we could reasonably constrain $N$.

\indent The CoG method is illustrated in Figure \ref{fig:cog-ex}. In the top panel, we present the measurements of the equivalent widths of three \ion{Fe}{2} lines as a function of oscillator strength. The best-fit CoG model obtained by $\chi^2$ minimalization is also shown. 

\indent To estimate the uncertainties on the best-fit parameters $\log$(N) and $b$, we adopt the approach from \citet{lampton1976, romanduval2021}. On the lower panel of Figure \ref{fig:cog-ex}, we see the representation of probability contours in the (N, $b$) parameters space. Following the \citet{lampton1976} approach for two parameters, the contour of $\chi^2$ corresponding to $\chi^2_{\rm min}$ + 2.3 provides the 1$\sigma$ uncertainties. The best-fit parameters (N, $b$) have the lowest value of $\chi^2$, and we adopt them as the final measurement (Table \ref{tab:col-dep}).

\begin{figure}

	\includegraphics[width=\columnwidth]{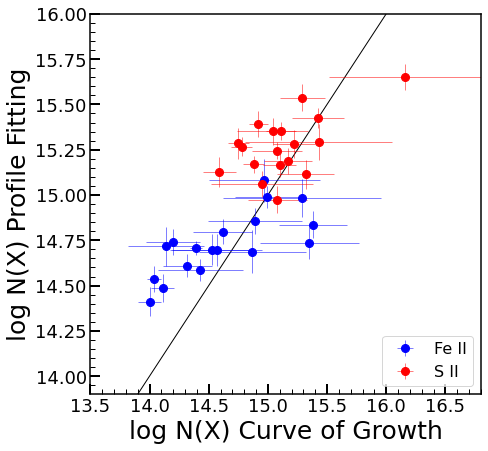}
    \caption{Comparison of column density measurements between Curve of Growth and Profile Fitting methods for \ion{Fe}{2} (blue) and \ion{S}{2} (red). A 1:1 line is marked in black; dashed lines mark the 1$\sigma$ dispersion. Most measurements are similar within uncertainties, and the outliers have significant errors. Two measurement methods provide comparable results, we chose the Curve of Growth results for further analysis.}
    \label{fig:cog-pf}
\end{figure}

\begin{deluxetable*}{lccccccc}
\tablecaption{Column density and depletion measurements derived using Curve of Growth method. \label{tab:col-dep}}
\tablehead{\colhead{Target} & \colhead{$\log$ N(\ion{H}{1})\tablenotemark{a}} & \colhead{$\log$ N(\ion{S}{2})} & \colhead{\textit{b}} & \colhead{$\log$ N(\ion{Fe}{2})} & \colhead{\textit{b}}  & \colhead{$\delta(S)$\tablenotemark{b}}  & \colhead{$\delta(\rm Fe)$\tablenotemark{b}} \\
& \colhead{cm$^{-2}$} &\colhead{cm$^{-2}$}  & \colhead{km/s} & \colhead{cm$^{-2}$} &\colhead{km/s} & & }
 \startdata
IC1613-61331 & 20.84 $\pm$ 0.03 & 15.11$^{+ 0.14 }_{- 0.13 }$ & 21.0 $^{+ \infty}_{- 8 }$ & 14.43 $^{+ 0.28 }_{- 0.36 }$ &12.0 $^{+ \infty }_{- 4 }$ & -1.28 $^{+0.36 }_{- 0.28 }$ & -0.13 $^{+ 0.13 }_{- 0.14 }$ \\ 
IC1613-62024 & 21.05 $\pm$ 0.03 & 15.29$^{+ 0.45 }_{- 0.19 }$ & 12.0 $^{+ 8}_{- 4 }$ & 15.38 $^{+ 0.41 }_{- 0.29 }$ &9.0 $^{+ 2 }_{- 2 }$ & -0.54 $^{+0.29 }_{- 0.41 }$ & -0.16 $^{+ 0.19 }_{- 0.45 }$ \\ 
IC1613-64066 & 20.88 $\pm$ 0.04 & 15.42$^{+ 0.92 }_{- 0.22 }$ & 12.0 $^{+ 8}_{- 5 }$ & 14.62 $^{+ 0.20 }_{- 0.25 }$ &23.0 $^{+ 8 }_{- 5 }$ & -1.13 $^{+0.25 }_{- 0.20 }$ & 0.14 $^{+ 0.22 }_{- 0.92 }$ \\ 
IC1613-67559 & 20.47 $\pm$ 0.14 & 15.10$^{+ 0.17 }_{- 0.14 }$ & 16.0 $^{+ 26}_{- 6 }$ & 14.32 $^{+ 0.30 }_{- 0.21 }$ &22.0 $^{+ \infty }_{- 10 }$ & -1.02 $^{+0.25 }_{- 0.33 }$ & 0.23 $^{+ 0.20 }_{- 0.22 }$ \\ 
IC1613-67684 & 20.44 $\pm$ 0.04 & 15.08$^{+ 0.19 }_{- 0.25 }$ & 5.0 $^{+ 3}_{- 5 }$ & 14.04 $^{+ 0.35 }_{- 0.06 }$ &20.0 $^{+ \infty }_{- 20 }$ & -1.27 $^{+0.07 }_{- 0.35 }$ & 0.24 $^{+ 0.25 }_{- 0.19 }$ \\ 
IC1613-A13 & 20.26 $\pm$ 0.03 & 15.32$^{+ 1.36 }_{- 0.24 }$ & 9.0 $^{+ 7}_{- 4 }$ & 14.57 $^{+ 0.39 }_{- 0.38 }$ &9.0 $^{+ 12 }_{- 3 }$ & -0.56 $^{+0.38 }_{- 0.39 }$ & 0.66 $^{+ 0.24 }_{- 1.36 }$ \\ 
IC1613-B11 & 20.37 $\pm$ 0.05 & 15.17$^{+ 0.36 }_{- 0.20 }$ & 10.0 $^{+ 9}_{- 4 }$ & 14.00 $^{+ 0.47 }_{- 0.10 }$ &20.0 $^{+ \infty }_{- 20 }$ & -1.24 $^{+0.11 }_{- 0.47 }$ & 0.40 $^{+ 0.21 }_{- 0.36 }$ \\ 
IC1613-B2 & 21.17 $\pm$ 0.03 & 16.16$^{+ 3.74 }_{- 0.64 }$ & 9.0 $^{+ 5}_{- 4 }$ & 14.99 $^{+ 0.32 }_{- 0.27 }$ &12.0 $^{+ 4 }_{- 3 }$ & -1.05 $^{+0.27 }_{- 0.32 }$ & 0.59 $^{+ 0.64 }_{- 3.75 }$ \\ 
IC1613-B3 & 20.64 $\pm$ 0.06 & 15.22$^{+ 0.27 }_{- 0.18 }$ & 14.0 $^{+ 19}_{- 5 }$ & 14.11 $^{+ 0.22 }_{- 0.10 }$ &20.0 $^{+ \infty }_{- 20 }$ & -1.40 $^{+0.12 }_{- 0.23 }$ & 0.18 $^{+ 0.19 }_{- 0.28 }$ \\ 
IC1613-B7 & 20.73 $\pm$ 0.03 & 15.08$^{+ 0.36 }_{- 0.21 }$ & 17.0 $^{+ \infty}_{- 9 }$ & 14.39 $^{+ 0.22 }_{- 0.05 }$ &20.0 $^{+ \infty }_{- 20 }$ & -1.21 $^{+0.06 }_{- 0.22 }$ & -0.05 $^{+ 0.21 }_{- 0.36 }$ \\ 
SEXTANS-A-s050 & 20.46 $\pm$ 0.04 & 14.88$^{+ 0.12 }_{- 0.09 }$ & 20.0 $^{+ \infty}_{- 20 }$ & 14.20 $^{+ 1.33 }_{- 0.23 }$ &22.0 $^{+ \infty }_{- 17 }$ & -0.81 $^{+0.23 }_{- 1.33 }$ & 0.43 $^{+ 0.10 }_{- 0.13 }$ \\ 
SEXTANS-A-s014 & 20.71 $\pm$ 0.03 & 14.95$^{+ 0.35 }_{- 0.43 }$ & 6.0 $^{+ \infty}_{- 6 }$ & 14.53 $^{+ 0.44 }_{- 0.36 }$ &8.0 $^{+ 11 }_{- 3 }$ & -0.73 $^{+0.36 }_{- 0.44 }$ & 0.25 $^{+ 0.43 }_{- 0.35 }$ \\ 
SEXTANS-A-s022 & 21.17 $\pm$ 0.03 & 15.04$^{+ 1.16 }_{- 0.30 }$ & 11.0 $^{+ \infty}_{- 6 }$ & 15.35 $^{+ 0.46 }_{- 0.42 }$ &5.0 $^{+ 1 }_{- 5 }$ & -0.37 $^{+0.42 }_{- 0.46 }$ & -0.12 $^{+ 0.30 }_{- 1.16 }$ \\ 
SEXTANS-A-s038 & 21.11 $\pm$ 0.02 & 14.78$^{+ 0.39 }_{- 0.09 }$ & 20.0 $^{+ \infty}_{- 20 }$ & 14.89 $^{+ 0.57 }_{- 0.40 }$ &8.0 $^{+ 4 }_{- 3 }$ & -0.77 $^{+0.40 }_{- 0.57 }$ & -0.32 $^{+ 0.09 }_{- 0.39 }$ \\ 
SEXTANS-A-s029 & 20.90 $\pm$ 0.07 & 14.59$^{+ 0.17 }_{- 0.14 }$ & 20.0 $^{+ \infty}_{- 20 }$ & 14.87 $^{+ 0.42 }_{- 0.45 }$ &5.0 $^{+ 2 }_{- 5 }$ & -0.58 $^{+0.46 }_{- 0.43 }$ & -0.30 $^{+ 0.16 }_{- 0.18 }$ \\ 
SEXTANS-A-s037 & 21.11 $\pm$ 0.07 & 14.75$^{+ 0.17 }_{- 0.09 }$ & 20.0 $^{+ \infty}_{- 20 }$ & 14.97 $^{+ 0.43 }_{- 0.47 }$ &5.0 $^{+ 2 }_{- 5 }$ & -0.69 $^{+0.48 }_{- 0.44 }$ & -0.35 $^{+ 0.11 }_{- 0.18 }$ \\ 
SEXTANS-A-SA2 & 21.07 $\pm$ 0.04 & 14.92$^{+ 0.36 }_{- 0.08 }$ & 20.0 $^{+ \infty}_{- 20 }$ & 15.29 $^{+ 0.58 }_{- 0.67 }$ &5.0 $^{+ 3 }_{- 5 }$ & -0.33 $^{+0.67 }_{- 0.58 }$ & -0.14 $^{+ 0.09 }_{- 0.36 }$ \\ 
SEXTANS-A-s021 & 20.38 $\pm$ 0.04 & 15.43$^{+ 0.43 }_{- 0.62 }$ & 5.0 $^{+ 14}_{- 5 }$ & 14.14 $^{+ 0.55 }_{- 0.32 }$ &10.0 $^{+ \infty }_{- 5 }$ & -0.79 $^{+0.32 }_{- 0.55 }$ & 1.06 $^{+ 0.62 }_{- 0.45 }$ \\ 
\enddata
\tablenotetext{a}{\ion{H}{1} column density from Ly-$\alpha$ profile fitting}
\tablenotetext{b}{Elemental depletion}
\end{deluxetable*}

\indent In Figure \ref{fig:cog-pf}, we present a comparison between the column density measurements obtained through the Curve of Growth method and the profile fitting method for two ions, \ion{S}{2} (in red) and \ion{Fe}{2} (in blue). To assess the agreement between the two methods, we include a 1:1 line in the plot and mark a 1$\sigma$ dispersion with a dashed line. Most measurements agree with errors, with Fe column density measurements showing the largest discrepancies. The CoG method shows larger error bars, especially for Fe measurement, which are crucial for the analysis presented in this paper. While CoG provides results comparable to the profile fitting method, the constraints are insufficient for robust conclusions. 

\section{Testing the Forward Optical Depth Method} 
\label{appendix:method-tests}

In this Appendix, we use ``artificial absorption`` tests to quantify the performance of the proposed forward optical depth (FOD) method.
We take ground truth component structures, synthesize corresponding artificial spectra with noise properties representative of those in the sample, and use FOD to estimate total column densities.
For comparison, we also estimate total column densities using Voigt profile fitting and the apparent optical depth (AOD) method. 
The first component structure is an \ion{Fe}{2} absorption system taken from an analysis of a high resolution (FWHM=4 km s$^{-1}$) spectrum by \citet{welty1997}.
This structure serves as a proof of concept, demonstrating that FOD can work on a realistic absorption system.
The second component structure is a single Voigt profile with varying total column density
We use this structure to characterize the performance of FOD at different absorption line depths.

Each set of tests is done at continuum noise levels chosen to represent the span of the real spectra analyzed the main part of this work.
We parameterize the noise level in terms of the \emph{continuum photon rate}, or \emph{CPR}.
This is a natural parameterization to use since the likelihood function used in the analysis operates on photon counts rather than on fluxes (see \S\ref{sec:cd-measurement:likelihood}).
The CPRs are 10 and 30, for \ion{Fe}{2}, and 30 and 80, for \ion{S}{2}.
Note that these are CPRs for the intrinsic HST-COS pixel scale, not for a resolution element.

When creating a synthetic spectrum, we set the continuum and instrumental sensitivity (relative to a normalized spectrum) to one.
The expected number of photons $\mu$ at a pixel is then the CPR multiplied by the absorption spectrum at that pixel. 
$\mu$ is then used as the rate parameter for the Poisson distribution from which we draw an ``observed'' number of photons.
We estimate a Gaussian flux uncertainty from the photon counts using the \texttt{poisson\_conf\_interval} function in \texttt{astropy} with the ``frequentist-confidence'' interval calculation method.
This is how the photon count uncertainty is currently estimated in \texttt{calcos}.

The FOD implementation is similar to the one described in \S\ref{sec:cd-measurement:combining}, the only change being that the velocity range is -100 to +100 km s$^{-1}$. 
Voigt profile fitting is implemented as a \texttt{numpyro} model with uniform component parameter priors between 12 and 18 for $\log_{10} (N/\text{cm}^{-2}$, 2 and 60 for $b$, and -50 and 100 for the centroid.
The model likelihood and continuum handling is the same as for FOD.
AOD is done using a utility in the \texttt{linetools} package over the range -100 to 100 km s$^{-1}$.

We quantify the correctness of solutions using two metrics: the root mean square error (RMSE), which reflects the accuracy of the point estimates, and the root mean square standardized error (RMSSE), which reflects the accuracy of the uncertainty estimates.
The expression for the RMSE is 
\begin{equation}
    \text{RMSE} = \left(\frac{1}{K} \sum_{k=1}^K \left(\log_{10}N_k - \log_{10}N_{inp} \right)^2 \right)^{1/2},
\end{equation}
where $k$ is an index over $K$ replications of a particular test problem, $\log_{10}N_k$ is the mean of the posterior probability distribution over $\log_{10}$ total column density for problem $k$, and $\log_{10}N_{inp, k}$ is the input $\log_{10}$ total column density for problem $k$.
The expression for the RMSSE is
\begin{equation}
    \text{RMSSE} = \left(\frac{1}{K} \sum_{k=1}^K \left(\frac{\log_{10}N_k - \log_{10}N_{inp, k}}{\sigma_k} \right)^2 \right)^{1/2},
\end{equation}
where $\sigma_k$ is the estimated standard deviation of $\log_{10}N_k$. 
For well-calibrated $\sigma_k$ and Gaussian-like posterior probability distributions, the RMSSE should be close to one. 
Underestimated and overestimated $\sigma_k$ will give RMSSEs above and below one, respectively.

The first ground truth structure we test is Milky Way foreground \ion{Fe}{2} towards the star Sk 108 in the Small Magellanic Cloud \citep{welty1997}.
The absorption system consists of nine components with velocity centroids spanning 64 km s$^{-1}$, individual component $b$-parameters between 2.5 and 4 km s$^{-1}$, and a total column density of $6\times 10^{14}$ cm$^{-2}$.
We generate 30 artificial spectra with this structure for CPRs of 10 and 30 and estimate total column densities using FOD, single Voigt profile fitting, and AOD done using the 1142, 1143, and 1144Å \ion{Fe}{2} lines.
AOD has a high RMSE with any of these lines, being biased due to saturation for the 1143 and 1144Å lines and being dominated by noise for the weaker 1142Å line.
FOD and single Voigt profile fitting have similar RMSEs of about 0.1 dex, but FOD has an RMSSE of 1.0 while Voigt profile fitting has an RMSSE of 1.4. 

Adding a second Voigt profile component does not improve the RMSSE and is not favored by frequentist or Bayesian model comparison methods.
For the frequentist method, we find maximum likelihood solutions with one and two Voigt profile components and use the likelihood ratio test assuming a chi-squared distribution with three degrees of freedom for the test statistic sampling distribution. 
For the Bayesian method, we use product space MCMC \citep{carlin1995chib} to calculate the Bayes factor for a two component model relative to a single component model.
We apply these methods to the CPR=30 spectra and find that none of the 30 artificial spectra show strong evidence (likelihood ratio test statistic in excess of 95\%, Bayes factor greater than 10) in favor of a second component.

The second ground truth structure is a single Voigt profile at column densities spanning the range seen in the observations.
The profile has a broadening parameter of 10 km s$^{-1}$, a centroid of 0 km s$^{-1}$, and input column densities ranging from $10^{14.12}$ to $10^{15.1}$ cm$^{-2}$, for \ion{Fe}{2}, and from $10^{14.8}$ to $10^{15.8}$ cm$^{-2}$, for \ion{S}{2}.
The synthetic spectra cover the 1142\AA, 1143\AA, and 1144\AA lines, for \ion{Fe}{2}, and the 1250\AA and 1253\AA lines for \ion{S}{2}. 
This set of lines for \ion{S}{2} reflects the coverage that is available for the Sextans A sightlines, where the stronger 1257\AA line is redshifted into  much stronger Milky Way \ion{Si}{2} absorption.

We generate three artificial observations for each combination of input total column density and CPR and measure column densities using FOD, MCMC Voigt profile fitting, and AOD. 
For AOD, we consider line-by-line measurements as well as a ``line oracle'' case, where the adopted measurement is the one that is closest to the truth. 
This is the \emph{oracle} version because it is as though an oracle told the investigator which line was least affected by noise and saturation.

\begin{figure*}
    \centering
    \includegraphics[width=\linewidth]{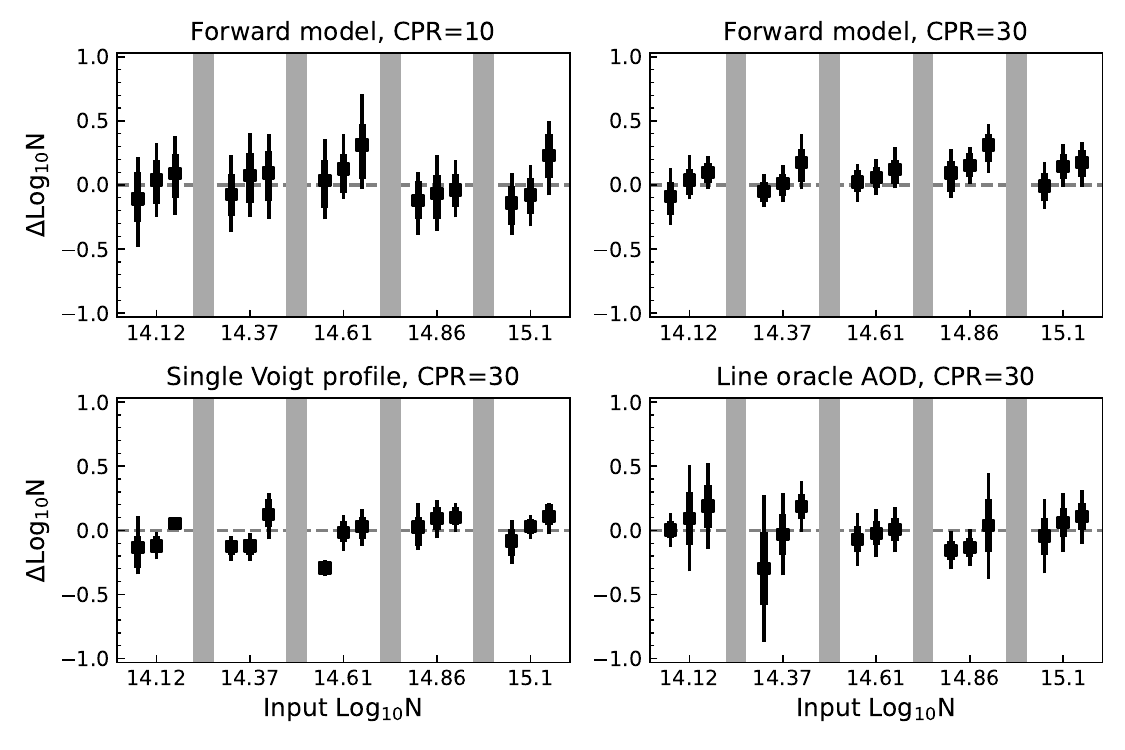}
    \caption{Results of estimating the column density of a injected single-component \ion{Fe}{2} absorber using different methods. Panels corresponds to different solution methods or continuum photon count rates. Each panel shows $\log$ column density residuals (recovered value minus true value) as a function of true, input, column densities. For each input column density, we show the results of analyzing three spectrum realizations (i.e., three draws of a spectrum with the same input absorption profile but with different photon noise). Sets of results are separated by vertical gray bars. See \S\ref{appendix:method-tests} for a complete description of these tests.}
    \label{fig:appendix:feii-test}
\end{figure*}

\autoref{fig:appendix:feii-test} show residuals as a function of true column density for the \ion{Fe}{2} tests.
All of the single-line AOD results have high RMSEs for some part of the true column density range.
FOD, fitting a single Voigt profile, and line oracle AOD provide point estimates with RMSEs of 0.1 to 0.2 (depending on CPR).
FOD and line oracle AOD also provide accurate estimates of the uncertainty on the point estimates, with RMSSEs of 1.0, while the single Voigt profile fits have underestimated uncertainties, with an RMSSE of 2.0.

\begin{figure*}
    \centering
    \includegraphics[width=\linewidth]{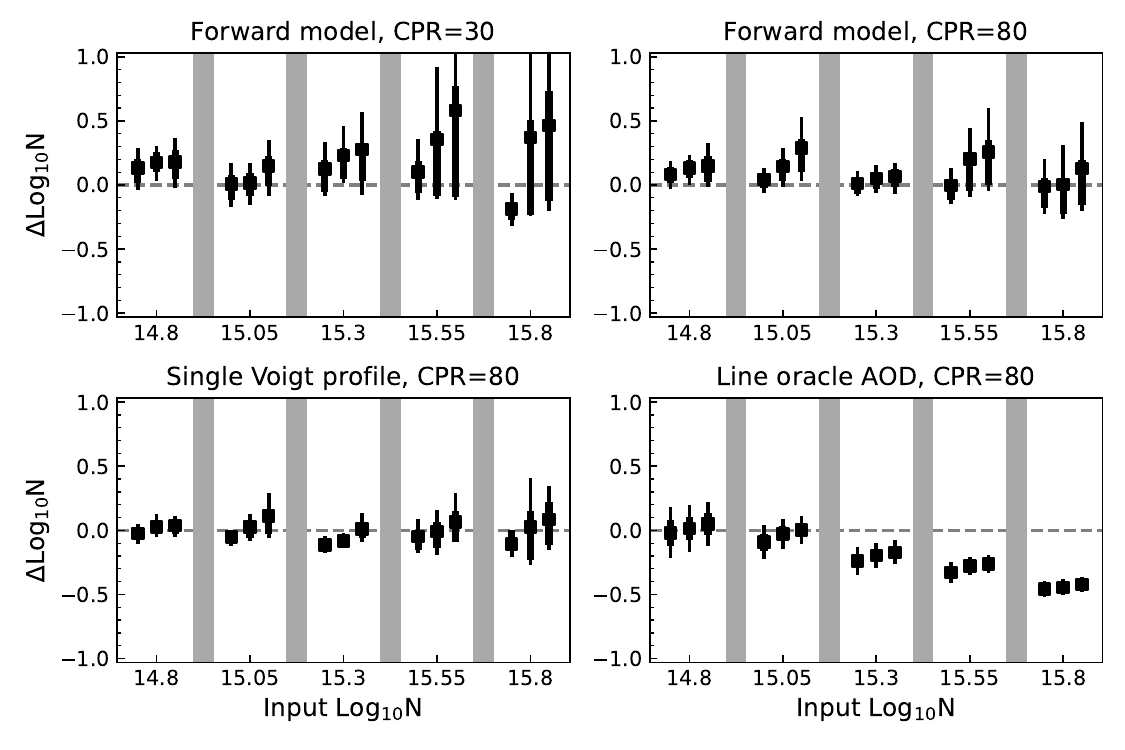}
    \caption{Results of estimating the column density of a injected single-component \ion{S}{2} absorber using different methods. See \autoref{fig:appendix:feii-test} for a description of figure elements.}
    \label{fig:appendix:sii-test}
\end{figure*}

\autoref{fig:appendix:sii-test} shows residuals for the \ion{S}{2} tests.
At the highest true column densities, both \ion{S}{2} lines are saturated.
As a result, even line oracle AOD produces results that are biased low.
The posterior probability distributions from FOD and Voigt profile fitting are skewed to high values, with the FOD probability distributions being wider.
This greater width is reflected in the RMSSEs: 1.1 for FOD and 1.4 for Voigt profile fitting.

FOD and Voigt profile point estimates for the total column have similar accuracy across both sets of tests, but FOD also produces well-calibrated uncertainties while Voigt profile fitting underestimates them.
Line oracle AOD suggests a qualitative explanation for the performance of FOD.
Forward modeling of the optical depth distribution accounting for the instrumental LSF allows the method to ``detect'' if any of the available absorption lines are in the informative regime.
If there are, FOD performs similarly to line oracle AOD.
If not, the flexibility of the model allows the uncertainties to (correctly) blow up.

\bibliography{biblio.bib}{}

\begin{thebibliography}{}
\expandafter\ifx\csname natexlab\endcsname\relax\def\natexlab#1{#1}\fi
\providecommand{\url}[1]{\href{#1}{#1}}
\providecommand{\dodoi}[1]{doi:~\href{http://doi.org/#1}{\nolinkurl{#1}}}
\providecommand{\doeprint}[1]{\href{http://ascl.net/#1}{\nolinkurl{http://ascl.net/#1}}}
\providecommand{\doarXiv}[1]{\href{https://arxiv.org/abs/#1}{\nolinkurl{https://arxiv.org/abs/#1}}}

\bibitem[{{Asano} {et~al.}(2013){Asano}, {Takeuchi}, {Hirashita}, \&
  {Nozawa}}]{asano2013}
{Asano}, R.~S., {Takeuchi}, T.~T., {Hirashita}, H., \& {Nozawa}, T. 2013,
  \mnras, 432, 637, \dodoi{10.1093/mnras/stt506}

\bibitem[{{Bigiel} {et~al.}(2008){Bigiel}, {Leroy}, {Walter}, {Brinks}, {de
  Blok}, {Madore}, \& {Thornley}}]{bigiel2008}
{Bigiel}, F., {Leroy}, A., {Walter}, F., {et~al.} 2008, \aj, 136, 2846,
  \dodoi{10.1088/0004-6256/136/6/2846}

\bibitem[{{Bolatto} {et~al.}(2011){Bolatto}, {Leroy}, {Jameson}, {Ostriker},
  {Gordon}, {Lawton}, {Stanimirovi{\'c}}, {Israel}, {Madden}, {Hony},
  {Sandstrom}, {Bot}, {Rubio}, {Winkler}, {Roman-Duval}, {van Loon},
  {Oliveira}, \& {Indebetouw}}]{bolatto2011}
{Bolatto}, A.~D., {Leroy}, A.~K., {Jameson}, K., {et~al.} 2011, \apj, 741, 12,
  \dodoi{10.1088/0004-637X/741/1/12}

\bibitem[{{Bouret} {et~al.}(2015){Bouret}, {Lanz}, {Hillier}, {Martins},
  {Marcolino}, \& {Depagne}}]{bouret2015}
{Bouret}, J.~C., {Lanz}, T., {Hillier}, D.~J., {et~al.} 2015, \mnras, 449,
  1545, \dodoi{10.1093/mnras/stv379}

\bibitem[{{Bresolin} {et~al.}(2007){Bresolin}, {Urbaneja}, {Gieren},
  {Pietrzy{\'n}ski}, \& {Kudritzki}}]{bresolin2007}
{Bresolin}, F., {Urbaneja}, M.~A., {Gieren}, W., {Pietrzy{\'n}ski}, G., \&
  {Kudritzki}, R.-P. 2007, \apj, 671, 2028, \dodoi{10.1086/522571}

\bibitem[{{Camacho} {et~al.}(2016){Camacho}, {Garcia}, {Herrero}, \&
  {Sim{\'o}n-D{\'\i}az}}]{camacho2016}
{Camacho}, I., {Garcia}, M., {Herrero}, A., \& {Sim{\'o}n-D{\'\i}az}, S. 2016,
  \aap, 585, A82, \dodoi{10.1051/0004-6361/201425533}

\bibitem[{Carlin \& Chib(2018)}]{carlin1995chib}
Carlin, B.~P., \& Chib, S. 2018, Journal of the Royal Statistical Society:
  Series B (Methodological), 57, 473,
  \dodoi{10.1111/j.2517-6161.1995.tb02042.x}

\bibitem[{{Chiang} {et~al.}(2018){Chiang}, {Sandstrom}, {Chastenet}, {Johnson},
  {Leroy}, \& {Utomo}}]{chiang2018}
{Chiang}, I.-D., {Sandstrom}, K.~M., {Chastenet}, J., {et~al.} 2018, \apj, 865,
  117, \dodoi{10.3847/1538-4357/aadc5f}

\bibitem[{{Choudhuri} \& {Roy}(2019)}]{choudhuri2019}
{Choudhuri}, S., \& {Roy}, N. 2019, \mnras, 483, 3437,
  \dodoi{10.1093/mnras/sty3342}

\bibitem[{{Clark} {et~al.}(2023){Clark}, {Roman-Duval}, {Gordon}, {Bot},
  {Smith}, \& {Hagen}}]{clark2023}
{Clark}, C. J.~R., {Roman-Duval}, J.~C., {Gordon}, K.~D., {et~al.} 2023, arXiv
  e-prints, arXiv:2302.07378, \dodoi{10.48550/arXiv.2302.07378}

\bibitem[{{Clark} {et~al.}(2019){Clark}, {De Vis}, {Baes}, {Bianchi},
  {Casasola}, {Cassar{\`a}}, {Davies}, {Dobbels}, {Lianou}, {De Looze},
  {Evans}, {Galametz}, {Galliano}, {Jones}, {Madden}, {Mosenkov}, {Verstocken},
  {Viaene}, {Xilouris}, \& {Ysard}}]{clark2019}
{Clark}, C.~J.~R., {De Vis}, P., {Baes}, M., {et~al.} 2019, \mnras, 489, 5256,
  \dodoi{10.1093/mnras/stz2257}

\bibitem[{{De Cia} {et~al.}(2016){De Cia}, {Ledoux}, {Mattsson}, {Petitjean},
  {Srianand}, {Gavignaud}, \& {Jenkins}}]{decia2016}
{De Cia}, A., {Ledoux}, C., {Mattsson}, L., {et~al.} 2016, \aap, 596, A97,
  \dodoi{10.1051/0004-6361/201527895}

\bibitem[{{De Vis} {et~al.}(2019){De Vis}, {Jones}, {Viaene}, {Casasola},
  {Clark}, {Baes}, {Bianchi}, {Cassara}, {Davies}, {De Looze}, {Galametz},
  {Galliano}, {Lianou}, {Madden}, {Manilla-Robles}, {Mosenkov}, {Nersesian},
  {Roychowdhury}, {Xilouris}, \& {Ysard}}]{devis2019}
{De Vis}, P., {Jones}, A., {Viaene}, S., {et~al.} 2019, \aap, 623, A5,
  \dodoi{10.1051/0004-6361/201834444}

\bibitem[{{Demyk} {et~al.}(2017{\natexlab{a}}){Demyk}, {Meny}, {Lu},
  {Papatheodorou}, {Toplis}, {Leroux}, {Depecker}, {Brubach}, {Roy}, {Nayral},
  {Ojo}, {Delpech}, {Paradis}, \& {Gromov}}]{demyk2017a}
{Demyk}, K., {Meny}, C., {Lu}, X.~H., {et~al.} 2017{\natexlab{a}}, \aap, 600,
  A123, \dodoi{10.1051/0004-6361/201629711}

\bibitem[{{Demyk} {et~al.}(2017{\natexlab{b}}){Demyk}, {Meny}, {Leroux},
  {Depecker}, {Brubach}, {Roy}, {Nayral}, {Ojo}, \& {Delpech}}]{demyk2017b}
{Demyk}, K., {Meny}, C., {Leroux}, H., {et~al.} 2017{\natexlab{b}}, \aap, 606,
  A50, \dodoi{10.1051/0004-6361/201730944}

\bibitem[{{Demyk} {et~al.}(2017{\natexlab{c}}){Demyk}, {Meny}, {Leroux},
  {Depecker}, {Brubach}, {Roy}, {Nayral}, {Ojo}, \& {Delpech}}]{demyk2017}
---. 2017{\natexlab{c}}, \aap, 606, A50, \dodoi{10.1051/0004-6361/201730944}

\bibitem[{{Diplas} \& {Savage}(1994)}]{diplas1994}
{Diplas}, A., \& {Savage}, B.~D. 1994, \apjs, 93, 211, \dodoi{10.1086/192052}

\bibitem[{{Eales} {et~al.}(2012){Eales}, {Smith}, {Auld}, {Baes}, {Bendo},
  {Bianchi}, {Boselli}, {Ciesla}, {Clements}, {Cooray}, {Cortese}, {Davies},
  {De Looze}, {Galametz}, {Gear}, {Gentile}, {Gomez}, {Fritz}, {Hughes},
  {Madden}, {Magrini}, {Pohlen}, {Spinoglio}, {Verstappen}, {Vlahakis}, \&
  {Wilson}}]{eales2012}
{Eales}, S., {Smith}, M. W.~L., {Auld}, R., {et~al.} 2012, \apj, 761, 168,
  \dodoi{10.1088/0004-637X/761/2/168}

\bibitem[{{Feldmann}(2015)}]{feldmann2015}
{Feldmann}, R. 2015, \mnras, 449, 3274, \dodoi{10.1093/mnras/stv552}

\bibitem[{{Fritz} {et~al.}(2012){Fritz}, {Gentile}, {Smith}, {Gear}, {Braun},
  {Roman-Duval}, {Bendo}, {Baes}, {Eales}, {Verstappen}, {Blommaert},
  {Boquien}, {Boselli}, {Clements}, {Cooray}, {Cortese}, {De Looze}, {Ford},
  {Galliano}, {Gomez}, {Gordon}, {Lebouteiller}, {O'Halloran}, {Kirk},
  {Madden}, {Page}, {Remy}, {Roussel}, {Spinoglio}, {Thilker}, {Vaccari},
  {Wilson}, \& {Waelkens}}]{fritz2012}
{Fritz}, J., {Gentile}, G., {Smith}, M.~W.~L., {et~al.} 2012, \aap, 546, A34,
  \dodoi{10.1051/0004-6361/201118619}

\bibitem[{{Galliano} {et~al.}(2018){Galliano}, {Galametz}, \&
  {Jones}}]{galliano2018}
{Galliano}, F., {Galametz}, M., \& {Jones}, A.~P. 2018, \araa, 56, 673,
  \dodoi{10.1146/annurev-astro-081817-051900}

\bibitem[{{Garcia} \& {Herrero}(2013)}]{garcia-herrero2013}
{Garcia}, M., \& {Herrero}, A. 2013, \aap, 551, A74,
  \dodoi{10.1051/0004-6361/201219977}

\bibitem[{{Garcia} {et~al.}(2014){Garcia}, {Herrero}, {Najarro}, {Lennon}, \&
  {Alejandro Urbaneja}}]{garcia2014}
{Garcia}, M., {Herrero}, A., {Najarro}, F., {Lennon}, D.~J., \& {Alejandro
  Urbaneja}, M. 2014, \apj, 788, 64, \dodoi{10.1088/0004-637X/788/1/64}

\bibitem[{{Garcia} {et~al.}(2009){Garcia}, {Herrero}, {Vicente}, {Castro},
  {Corral}, {Rosenberg}, \& {Monelli}}]{garcia2009}
{Garcia}, M., {Herrero}, A., {Vicente}, B., {et~al.} 2009, \aap, 502, 1015,
  \dodoi{10.1051/0004-6361/200911791}

\bibitem[{{Gordon} {et~al.}(2003){Gordon}, {Clayton}, {Misselt}, {Landolt}, \&
  {Wolff}}]{gordon2003}
{Gordon}, K.~D., {Clayton}, G.~C., {Misselt}, K.~A., {Landolt}, A.~U., \&
  {Wolff}, M.~J. 2003, \apj, 594, 279, \dodoi{10.1086/376774}

\bibitem[{{Hildebrand}(1983)}]{hildebrand1983}
{Hildebrand}, R.~H. 1983, \qjras, 24, 267

\bibitem[{{Hirashita}(2000)}]{hirashita2000}
{Hirashita}, H. 2000, \pasj, 52, 585, \dodoi{10.1093/pasj/52.4.585}

\bibitem[{Hoffman \& Gelman(2014)}]{hoffman2014gelman}
Hoffman, M.~D., \& Gelman, A. 2014, Journal of Machine Learning Research, 15,
  1593.
\newblock \url{http://jmlr.org/papers/v15/hoffman14a.html}

\bibitem[{{Hunter} {et~al.}(2012){Hunter}, {Ficut-Vicas}, {Ashley}, {Brinks},
  {Cigan}, {Elmegreen}, {Heesen}, {Herrmann}, {Johnson}, {Oh}, {Rupen},
  {Schruba}, {Simpson}, {Walter}, {Westpfahl}, {Young}, \&
  {Zhang}}]{hunter2012}
{Hunter}, D.~A., {Ficut-Vicas}, D., {Ashley}, T., {et~al.} 2012, \aj, 144, 134,
  \dodoi{10.1088/0004-6256/144/5/134}

\bibitem[{Ishwaran \& Rao(2005)}]{ishwaran_2005}
Ishwaran, H., \& Rao, J.~S. 2005, The Annals of Statistics, 33, 730 ,
  \dodoi{10.1214/009053604000001147}

\bibitem[{{Jenkins}(1996)}]{jenkins1996}
{Jenkins}, E.~B. 1996, \apj, 471, 292, \dodoi{10.1086/177969}

\bibitem[{{Jenkins}(2009)}]{jenkins2009}
---. 2009, \apj, 700, 1299, \dodoi{10.1088/0004-637X/700/2/1299}

\bibitem[{{Jenkins} \& {Wallerstein}(2017)}]{jenkins2017}
{Jenkins}, E.~B., \& {Wallerstein}, G. 2017, \apj, 838, 85,
  \dodoi{10.3847/1538-4357/aa64d4}

\bibitem[{{Johnson} {et~al.}(2021){Johnson}, {Plesha}, {Jedrzejewski},
  {Frazer}, \& {Dashtamirova}}]{cos-johnson2021}
{Johnson}, C.~I., {Plesha}, R., {Jedrzejewski}, R., {Frazer}, E., \&
  {Dashtamirova}, D. 2021, {Updated Flux Error Calculations for CalCOS},
  Instrument Science Report COS 2021-03

\bibitem[{{Jones} {et~al.}(1996){Jones}, {Tielens}, \&
  {Hollenbach}}]{jones1996}
{Jones}, A.~P., {Tielens}, A.~G.~G.~M., \& {Hollenbach}, D.~J. 1996, \apj, 469,
  740, \dodoi{10.1086/177823}

\bibitem[{{Kaufer} {et~al.}(2004){Kaufer}, {Venn}, {Tolstoy}, {Pinte}, \&
  {Kudritzki}}]{kaufer2004}
{Kaufer}, A., {Venn}, K.~A., {Tolstoy}, E., {Pinte}, C., \& {Kudritzki}, R.-P.
  2004, \aj, 127, 2723, \dodoi{10.1086/383209}

\bibitem[{{Kisielius} {et~al.}(2014){Kisielius}, {Kulkarni}, {Ferland},
  {Bogdanovich}, \& {Lykins}}]{kisielius2014}
{Kisielius}, R., {Kulkarni}, V.~P., {Ferland}, G.~J., {Bogdanovich}, P., \&
  {Lykins}, M.~L. 2014, \apj, 780, 76, \dodoi{10.1088/0004-637X/780/1/76}

\bibitem[{{Kniazev} {et~al.}(2005){Kniazev}, {Grebel}, {Pustilnik}, {Pramskij},
  \& {Zucker}}]{kniazev2005}
{Kniazev}, A.~Y., {Grebel}, E.~K., {Pustilnik}, S.~A., {Pramskij}, A.~G., \&
  {Zucker}, D.~B. 2005, \aj, 130, 1558, \dodoi{10.1086/432931}

\bibitem[{{Lamperti} {et~al.}(2019){Lamperti}, {Saintonge}, {De Looze},
  {Accurso}, {Clark}, {Smith}, {Wilson}, {Brinks}, {Brown}, {Bureau},
  {Clements}, {Eales}, {Glass}, {Hwang}, {Lee}, {Lin}, {Michalowski},
  {Sargent}, {Williams}, {Xiao}, \& {Yang}}]{lamperti2019}
{Lamperti}, I., {Saintonge}, A., {De Looze}, I., {et~al.} 2019, \mnras, 489,
  4389, \dodoi{10.1093/mnras/stz2311}

\bibitem[{{Lampton} {et~al.}(1976){Lampton}, {Margon}, \&
  {Bowyer}}]{lampton1976}
{Lampton}, M., {Margon}, B., \& {Bowyer}, S. 1976, \apj, 208, 177,
  \dodoi{10.1086/154592}

\bibitem[{{Lorenzo} {et~al.}(2022){Lorenzo}, {Garcia}, {Najarro}, {Herrero},
  {Cervi{\~n}o}, \& {Castro}}]{Lorenzo2022}
{Lorenzo}, M., {Garcia}, M., {Najarro}, F., {et~al.} 2022, \mnras, 516, 4164,
  \dodoi{10.1093/mnras/stac2050}

\bibitem[{{Morton}(2003)}]{morton2003}
{Morton}, D.~C. 2003, \apjs, 149, 205, \dodoi{10.1086/377639}

\bibitem[{{Nersesian} {et~al.}(2019){Nersesian}, {Xilouris}, {Bianchi},
  {Galliano}, {Jones}, {Baes}, {Casasola}, {Cassar{\`a}}, {Clark}, {Davies},
  {Decleir}, {Dobbels}, {De Looze}, {De Vis}, {Fritz}, {Galametz}, {Madden},
  {Mosenkov}, {Tr{\v{c}}ka}, {Verstocken}, {Viaene}, \&
  {Lianou}}]{nersenian2019}
{Nersesian}, A., {Xilouris}, E.~M., {Bianchi}, S., {et~al.} 2019, \aap, 624,
  A80, \dodoi{10.1051/0004-6361/201935118}

\bibitem[{{Ott} {et~al.}(2012){Ott}, {Stilp}, {Warren}, {Skillman},
  {Dalcanton}, {Walter}, {de Blok}, {Koribalski}, \& {West}}]{ott2012}
{Ott}, J., {Stilp}, A.~M., {Warren}, S.~R., {et~al.} 2012, \aj, 144, 123,
  \dodoi{10.1088/0004-6256/144/4/123}

\bibitem[{{P{\'e}roux} \& {Howk}(2020)}]{Peroux2020}
{P{\'e}roux}, C., \& {Howk}, J.~C. 2020, \araa, 58, 363,
  \dodoi{10.1146/annurev-astro-021820-120014}

\bibitem[{{Popping} \& {P{\'e}roux}(2022)}]{popping2022}
{Popping}, G., \& {P{\'e}roux}, C. 2022, \mnras, 513, 1531,
  \dodoi{10.1093/mnras/stac695}

\bibitem[{{Quiret} {et~al.}(2016){Quiret}, {P{\'e}roux}, {Zafar}, {Kulkarni},
  {Jenkins}, {Milliard}, {Rahmani}, {Popping}, {Rao}, {Turnshek}, \&
  {Monier}}]{quiret2016}
{Quiret}, S., {P{\'e}roux}, C., {Zafar}, T., {et~al.} 2016, \mnras, 458, 4074,
  \dodoi{10.1093/mnras/stw524}

\bibitem[{{Rafelski} {et~al.}(2012){Rafelski}, {Wolfe}, {Prochaska},
  {Neeleman}, \& {Mendez}}]{rafelski202}
{Rafelski}, M., {Wolfe}, A.~M., {Prochaska}, J.~X., {Neeleman}, M., \&
  {Mendez}, A.~J. 2012, \apj, 755, 89, \dodoi{10.1088/0004-637X/755/2/89}

\bibitem[{{Rela{\~n}o} {et~al.}(2018){Rela{\~n}o}, {De Looze}, {Kennicutt},
  {Lisenfeld}, {Dariush}, {Verley}, {Braine}, {Tabatabaei}, {Kramer},
  {Boquien}, {Xilouris}, \& {Gratier}}]{relano2018}
{Rela{\~n}o}, M., {De Looze}, I., {Kennicutt}, R.~C., {et~al.} 2018, \aap, 613,
  A43, \dodoi{10.1051/0004-6361/201732347}

\bibitem[{{R{\'e}my-Ruyer} {et~al.}(2014){R{\'e}my-Ruyer}, {Madden},
  {Galliano}, {Galametz}, {Takeuchi}, {Asano}, {Zhukovska}, {Lebouteiller},
  {Cormier}, {Jones}, {Bocchio}, {Baes}, {Bendo}, {Boquien}, {Boselli},
  {DeLooze}, {Doublier-Pritchard}, {Hughes}, {Karczewski}, \&
  {Spinoglio}}]{remyruyer2014}
{R{\'e}my-Ruyer}, A., {Madden}, S.~C., {Galliano}, F., {et~al.} 2014, \aap,
  563, A31, \dodoi{10.1051/0004-6361/201322803}

\bibitem[{{Roman-Duval} {et~al.}(2013){Roman-Duval}, {Elliott}, {Aloisi},
  {Delker}, {Oliveira}, {Osten}, {Osterman}, {Penton}, \&
  {Sonnentrucker}}]{coslp}
{Roman-Duval}, J., {Elliott}, E., {Aloisi}, A., {et~al.} 2013, {COS/FUV Spatial
  and Spectral Resolution at the new Lifetime Position}, Instrument Science
  Report COS 2013-07, 33 pages

\bibitem[{{Roman-Duval} {et~al.}(2014){Roman-Duval}, {Gordon}, {Meixner},
  {Bot}, {Bolatto}, {Hughes}, {Wong}, {Babler}, {Bernard}, {Clayton}, {Fukui},
  {Galametz}, {Galliano}, {Glover}, {Hony}, {Israel}, {Jameson},
  {Lebouteiller}, {Lee}, {Li}, {Madden}, {Misselt}, {Montiel}, {Okumura},
  {Onishi}, {Panuzzo}, {Reach}, {Remy-Ruyer}, {Robitaille}, {Rubio}, {Sauvage},
  {Seale}, {Sewilo}, {Staveley-Smith}, \& {Zhukovska}}]{romanduval2014}
{Roman-Duval}, J., {Gordon}, K.~D., {Meixner}, M., {et~al.} 2014, \apj, 797,
  86, \dodoi{10.1088/0004-637X/797/2/86}

\bibitem[{{Roman-Duval} {et~al.}(2019){Roman-Duval}, {Jenkins}, {Williams},
  {Tchernyshyov}, {Gordon}, {Meixner}, {Hagen}, {Peek}, {Sandstrom}, {Werk}, \&
  {Yanchulova Merica-Jones}}]{romanduval2019}
{Roman-Duval}, J., {Jenkins}, E.~B., {Williams}, B., {et~al.} 2019, \apj, 871,
  151, \dodoi{10.3847/1538-4357/aaf8bb}

\bibitem[{{Roman-Duval} {et~al.}(2021){Roman-Duval}, {Jenkins}, {Tchernyshyov},
  {Williams}, {Clark}, {Gordon}, {Meixner}, {Hagen}, {Peek}, {Sandstrom},
  {Werk}, \& {Yanchulova Merica-Jones}}]{romanduval2021}
{Roman-Duval}, J., {Jenkins}, E.~B., {Tchernyshyov}, K., {et~al.} 2021, \apj,
  910, 95, \dodoi{10.3847/1538-4357/abdeb6}

\bibitem[{{Roman-Duval} {et~al.}(2022{\natexlab{a}}){Roman-Duval}, {Jenkins},
  {Tchernyshyov}, {Clark}, {De Cia}, {Gordon}, {Hamanowicz}, {Lebouteiller},
  {Rafelski}, {Sandstrom}, {Werk}, \& {Merica-Jones}}]{romanduval2022a}
---. 2022{\natexlab{a}}, \apj, 928, 90, \dodoi{10.3847/1538-4357/ac5248}

\bibitem[{{Roman-Duval} {et~al.}(2022{\natexlab{b}}){Roman-Duval}, {Jenkins},
  {Tchernyshyov}, {Clark}, {De Cia}, {Gordon}, {Hamanowicz}, {Lebouteiller},
  {Rafelski}, {Sandstrom}, {Werk}, \& {Yanchulova
  Merica-Jones}}]{romanduval2022b}
---. 2022{\natexlab{b}}, arXiv e-prints, arXiv:2206.03639.
\newblock \doarXiv{2206.03639}

\bibitem[{{Rowlands} {et~al.}(2012){Rowlands}, {Dunne}, {Maddox}, {Bourne},
  {Gomez}, {Kaviraj}, {Bamford}, {Brough}, {Charlot}, {da Cunha}, {Driver},
  {Eales}, {Hopkins}, {Kelvin}, {Nichol}, {Sansom}, {Sharp}, {Smith}, {Temi},
  {van der Werf}, {Baes}, {Cava}, {Cooray}, {Croom}, {Dariush}, {de Zotti},
  {Dye}, {Fritz}, {Hopwood}, {Ibar}, {Ivison}, {Liske}, {Loveday}, {Madore},
  {Norberg}, {Popescu}, {Rigby}, {Robotham}, {Rodighiero}, {Seibert}, \&
  {Tuffs}}]{rowlands2012}
{Rowlands}, K., {Dunne}, L., {Maddox}, S., {et~al.} 2012, \mnras, 419, 2545,
  \dodoi{10.1111/j.1365-2966.2011.19905.x}

\bibitem[{{Rowlands} {et~al.}(2014){Rowlands}, {Dunne}, {Dye},
  {Arag{\'o}n-Salamanca}, {Maddox}, {da Cunha}, {Smith}, {Bourne}, {Eales},
  {Gomez}, {Smail}, {Alpaslan}, {Clark}, {Driver}, {Ibar}, {Ivison},
  {Robotham}, {Smith}, \& {Valiante}}]{rowlands2014}
{Rowlands}, K., {Dunne}, L., {Dye}, S., {et~al.} 2014, \mnras, 441, 1017,
  \dodoi{10.1093/mnras/stu510}

\bibitem[{{Savage} \& {Sembach}(1991)}]{savage1991}
{Savage}, B.~D., \& {Sembach}, K.~R. 1991, \apj, 379, 245,
  \dodoi{10.1086/170498}

\bibitem[{{Schruba} {et~al.}(2012){Schruba}, {Leroy}, {Walter}, {Bigiel},
  {Brinks}, {de Blok}, {Kramer}, {Rosolowsky}, {Sandstrom}, {Schuster},
  {Usero}, {Weiss}, \& {Wiesemeyer}}]{schruba2012}
{Schruba}, A., {Leroy}, A.~K., {Walter}, F., {et~al.} 2012, \aj, 143, 138,
  \dodoi{10.1088/0004-6256/143/6/138}

\bibitem[{{Sembach} \& {Savage}(1992)}]{sembach1992}
{Sembach}, K.~R., \& {Savage}, B.~D. 1992, \apjs, 83, 147,
  \dodoi{10.1086/191734}

\bibitem[{{Shi} {et~al.}(2014){Shi}, {Armus}, {Helou}, {Stierwalt}, {Gao},
  {Wang}, {Zhang}, \& {Gu}}]{shi2014}
{Shi}, Y., {Armus}, L., {Helou}, G., {et~al.} 2014, \nat, 514, 335,
  \dodoi{10.1038/nature13820}

\bibitem[{{Skillman} {et~al.}(1989){Skillman}, {Kennicutt}, \&
  {Hodge}}]{skillman1989}
{Skillman}, E.~D., {Kennicutt}, R.~C., \& {Hodge}, P.~W. 1989, \apj, 347, 875,
  \dodoi{10.1086/168178}

\bibitem[{{Smith} {et~al.}(2012){Smith}, {Eales}, {Gomez}, {Roman-Duval},
  {Fritz}, {Braun}, {Baes}, {Bendo}, {Blommaert}, {Boquien}, {Boselli},
  {Clements}, {Cooray}, {Cortese}, {De Looze}, {Ford}, {Gear}, {Gentile},
  {Gordon}, {Kirk}, {Lebouteiller}, {Madden}, {Mentuch}, {O'Halloran}, {Page},
  {Schulz}, {Spinoglio}, {Verstappen}, {Wilson}, \& {Thilker}}]{smith2012}
{Smith}, M.~W.~L., {Eales}, S.~A., {Gomez}, H.~L., {et~al.} 2012, \apj, 756,
  40, \dodoi{10.1088/0004-637X/756/1/40}

\bibitem[{{Stepnik} {et~al.}(2003){Stepnik}, {Abergel}, {Bernard}, {Boulanger},
  {Cambr{\'e}sy}, {Giard}, {Jones}, {Lagache}, {Lamarre}, {Meny}, {Pajot}, {Le
  Peintre}, {Ristorcelli}, {Serra}, \& {Torre}}]{stepnik2003}
{Stepnik}, B., {Abergel}, A., {Bernard}, J.-P., {et~al.} 2003, \aap, 398, 551,
  \dodoi{10.1051/0004-6361:20021309}

\bibitem[{{Tamburro} {et~al.}(2009){Tamburro}, {Rix}, {Leroy}, {Mac Low},
  {Walter}, {Kennicutt}, {Brinks}, \& {de Blok}}]{tamburro2009}
{Tamburro}, D., {Rix}, H.~W., {Leroy}, A.~K., {et~al.} 2009, \aj, 137, 4424,
  \dodoi{10.1088/0004-6256/137/5/4424}

\bibitem[{{Tautvai{\v{s}}ien{\.{e}}} {et~al.}(2007){Tautvai{\v{s}}ien{\.{e}}},
  {Geisler}, {Wallerstein}, {Borissova}, {Bizyaev}, {Pagel}, {Charbonnel}, \&
  {Smith}}]{tautvaisiene2007}
{Tautvai{\v{s}}ien{\.{e}}}, G., {Geisler}, D., {Wallerstein}, G., {et~al.}
  2007, \aj, 134, 2318, \dodoi{10.1086/523630}

\bibitem[{{Tchernyshyov} {et~al.}(2015){Tchernyshyov}, {Meixner}, {Seale},
  {Fox}, {Friedman}, {Dwek}, \& {Galliano}}]{tchernyshyov2015}
{Tchernyshyov}, K., {Meixner}, M., {Seale}, J., {et~al.} 2015, \apj, 811, 78,
  \dodoi{10.1088/0004-637X/811/2/78}

\bibitem[{{Tramper} {et~al.}(2014){Tramper}, {Sana}, {de Koter}, {Kaper}, \&
  {Ram{\'\i}rez-Agudelo}}]{tramper2014}
{Tramper}, F., {Sana}, H., {de Koter}, A., {Kaper}, L., \&
  {Ram{\'\i}rez-Agudelo}, O.~H. 2014, \aap, 572, A36,
  \dodoi{10.1051/0004-6361/201424312}

\bibitem[{{Tumlinson} {et~al.}(2002){Tumlinson}, {Shull}, {Rachford},
  {Browning}, {Snow}, {Fullerton}, {Jenkins}, {Savage}, {Crowther}, {Moos},
  {Sembach}, {Sonneborn}, \& {York}}]{tumlinson2002}
{Tumlinson}, J., {Shull}, J.~M., {Rachford}, B.~L., {et~al.} 2002, \apj, 566,
  857, \dodoi{10.1086/338112}

\bibitem[{Vehtari {et~al.}(2021)Vehtari, Gelman, Simpson, Carpenter, \&
  B{\"u}rkner}]{vehtari2021ranknorm}
Vehtari, A., Gelman, A., Simpson, D., Carpenter, B., \& B{\"u}rkner, P.-C.
  2021, Bayesian Analysis, 16, 667 , \dodoi{10.1214/20-BA1221}

\bibitem[{{Wakker} \& {Mathis}(2000)}]{wakker2000}
{Wakker}, B.~P., \& {Mathis}, J.~S. 2000, \apjl, 544, L107,
  \dodoi{10.1086/317316}

\bibitem[{{Welty} {et~al.}(1997){Welty}, {Lauroesch}, {Blades}, {Hobbs}, \&
  {York}}]{welty1997}
{Welty}, D.~E., {Lauroesch}, J.~T., {Blades}, J.~C., {Hobbs}, L.~M., \& {York},
  D.~G. 1997, \apj, 489, 672, \dodoi{10.1086/304811}

\bibitem[{{Welty} {et~al.}(2012){Welty}, {Xue}, \& {Wong}}]{welty2012}
{Welty}, D.~E., {Xue}, R., \& {Wong}, T. 2012, \apj, 745, 173,
  \dodoi{10.1088/0004-637X/745/2/173}

\bibitem[{{Ysard} {et~al.}(2018){Ysard}, {Jones}, {Demyk}, {Bout{\'e}raon}, \&
  {Koehler}}]{ysard2018}
{Ysard}, N., {Jones}, A.~P., {Demyk}, K., {Bout{\'e}raon}, T., \& {Koehler}, M.
  2018, \aap, 617, A124, \dodoi{10.1051/0004-6361/201833386}

\bibitem[{{Zhukovska} {et~al.}(2016){Zhukovska}, {Dobbs}, {Jenkins}, \&
  {Klessen}}]{zhukovska2016}
{Zhukovska}, S., {Dobbs}, C., {Jenkins}, E.~B., \& {Klessen}, R.~S. 2016, \apj,
  831, 147, \dodoi{10.3847/0004-637X/831/2/147}

\end{thebibliography}
\bibliographystyle{aasjournal}



\end{document}